\documentclass[aps,prd,preprintnumbers,superscriptaddress,nofootinbib]{revtex4-2}
\usepackage[dvipsnames]{xcolor}
\definecolor{red}{rgb}{0.9, 0,0}
\definecolor{cerulean}{rgb}{0., 0.42,0.9}
\definecolor{navy}{rgb}{0.05, 0.05,0.8}

\usepackage[colorlinks]{hyperref}
\hypersetup{
    colorlinks = true,
    citecolor  = red,
	linkcolor  = navy
}

\usepackage{mathtools}
\usepackage{amsmath}
\usepackage{amssymb}
\usepackage{amsfonts}
\usepackage{txfonts}
\usepackage{graphics}
\usepackage{endnotes}

\def\vec#1{\boldsymbol{\mathbf{#1}}}
\newcommand{\unit}[1]{\vec{\hat{#1}}}

\begin{document}

\title{Proper Time Shifts in Pulsar Timing Arrays}

\author{Vincent S. H. Lee}
\email{vincentszehimlee@berkeley.edu}
\affiliation{Department of Physics, University of California, Berkeley, CA 94720, USA}
\affiliation{Department of Physics, University of California, San Diego, La Jolla, CA 92093, USA}

\preprint{N3AS-26-016}

\begin{abstract}
Pulsar timing arrays are a powerful tool for probing low-frequency gravitational phenomena and physics beyond the Standard Model. We establish a general, gauge-invariant framework for the timing residual by identifying the measurable quantity as the proper time interval between consecutive pulse arrivals at Earth. In linearized general relativity, we derive a general expression for this proper time shift and show that it decomposes into Doppler, Shapiro, and Einstein delay contributions whose sum is gauge invariant, similar to frameworks that have been developed for other gravitational-wave detectors such as laser and atom interferometers. We also provide a recipe for computing the resulting timing residual for any specified perturbation, and we illustrate the method with several physically well-motivated sources, recovering known results where they exist in the literature. This recipe can be directly applied to search for new physics from real pulsar timing data using existing numerical software.
\end{abstract}

\maketitle

\section{Introduction}
Pulsar timing arrays (PTAs) have long served as a leading probe of low-frequency gravitational waves (GWs) with numerous active collaborations, \textit{e.g.} NANOGrav~\cite{McLaughlin:2013ira}, EPTA~\cite{Ferdman:2010xq}, PPTA~\cite{2013PASA...30...17M}, InPTA~\cite{Tarafdar:2022toa}, CPTA~\cite{Xu:2023wog}, and MPTA~\cite{Spiewak:2022btk}. By precisely measuring the times of arrival (TOAs) of pulses from millisecond pulsars and their deviations from expected arrival times, PTAs can probe extremely small perturbations in spacetime~\cite{Taylor:2021yjx}. Beyond GWs, PTAs can also be used to probe gravitational effects predicted by beyond-the-Standard-Model (BSM) physics, such as primordial black holes (PBHs)~\cite{Seto:2007kj, 2012MNRAS.426.1369K, Schutz:2016khr, Dror:2019twh}, dark matter (DM) substructure~\cite{Siegel:2007fz, Baghram:2011is, Clark:2015sha, Clark:2015tha, Kashiyama:2018gsh, Ramani:2020hdo, Lee:2020wfn, Lee:2021zqw, Berghaus:2025kvn, Foster:2026kfg, Cherukupalli:2026cda}, and ultralight dark matter (ULDM)~\cite{Khmelnitsky:2013lxt, Porayko:2014rfa, Graham:2015ifn, Aoki:2016mtn, DeMartino:2017qsa, Porayko:2018sfa, Kato:2019bqz, Nomura:2019cvc, Kaplan:2022lmz, Unal:2022ooa, Kim:2023kyy, Eberhardt:2024ocm, Kim:2023pkx, Xia:2023hov, Luu:2023rgg, Hwang:2023odi, EuropeanPulsarTimingArray:2023egv, Boddy:2025oxn, Gan:2025icr, Dror:2025nvg, Foster:2026mvs}. For instance, NANOGrav reported compelling evidence for a stochastic gravitational-wave background (GWB) in their 15-year dataset~\cite{NANOGrav:2023gor}, and, using the same dataset, also investigated whether the observed signal could arise from a broad class of BSM scenarios~\cite{NANOGrav:2023hvm}. A numerical code, \texttt{PTArcade}, was later developed to read in user-provided signal templates from any BSM model of interest and perform Bayesian analyses on PTA data, yielding posterior distributions and upper limits on the corresponding model parameters~\cite{Mitridate:2023oar}.

While the power of PTAs to detect gravitational effects from BSM physics is widely appreciated, the timing deviation observable, $\delta t(t)$ \footnote{This is also known as the \textit{timing residual}, up to degeneracy with the pulsar's timing model, which will be discussed in detail in Sec.~\ref{subsec:timing_model}}, is usually only computed on a case-by-case basis in the literature. Existing works either restrict to GWs~\cite{Maggiore:2007ulw, Creighton:2008bu}, or treat each BSM source individually, frequently in a fixed gauge. For example, in the context of detecting DM with PTAs, the time shift due to ULDM have been derived in Refs.~\cite{Kim:2023kyy,Eberhardt:2024ocm}, and for DM substructure in Ref.~\cite{Dror:2019twh}. In these works the total time shift is broken into separate pieces, each attributed to a distinct way the metric perturbation influences the signal in the Newtonian gauge. While each such contribution is physically intuitive, it is not by itself a gauge-invariant quantity, and it is therefore unclear whether the sum represents a genuine physical observable. A general, source-agnostic expression for the residual is only derived recently in Ref.~\cite{Dror:2025nvg}, which follows from integrating the redshift sourced by the linearized Riemann tensor $R_{i0j0}$, giving a manifestly gauge-invariant quantity (see also Ref.~\cite{Magi:2026occ} for the equivalence of the pulse arrival rate and the photon redshift). This curvature form, however, does not decompose the observable as a sum of physical effects.

To our best knowledge, at the time of writing, for PTAs there was no single formulation that simultaneously applies to an arbitrary $h_{\mu\nu}$, resolves the observable into its separate physical contributions, and is manifestly gauge invariant (see, however, the Note Added on Ref.~\cite{Magi:2026upf}). However, such a formalism already exists for other forms of GW detectors as probes of BSM physics, such as laser interferometers~\cite{Lee:2024oxo} and atom interferometers~\cite{Badurina:2024rpp}. The idea is to identify the experimental observable as the \textit{proper} time shift, as measured along worldlines of massive bodies that act as proxies for the observer, between two physical events. For laser interferometers, it is the proper time measured by the beamsplitter between when the laser first gets split into two interferometer arms and when it then gets recombined. For single-photon atom interferometers, it is the proper time experienced by the atom cloud between when it is first split by a laser into a superposition of two spatially separate paths and when it is recombined by a laser at the end of the sequence. A proper time quantity is, by design, gauge invariant. Moreover, in linearized GR, it has been shown that the proper time observable in both laser and atom interferometers can be written as the sum of three distinct contributions: the Doppler effect, which accounts for the motion of the massive bodies in the experimental setup; the Shapiro delay, which accounts for the delay in the arrival time of the photons; and the Einstein delay, which is the gravitational redshift as measured by the massive bodies. This decomposition has been previously noted in the literature in the context of GWs~\cite{Rakhmanov:2004eh}, vacuum fluctuations in quantum gravity (geontropic fluctuations\footnote{See Refs.~\cite{Verlinde:2019xfb, Zurek:2020ukz, Zurek:2022xzl} for references})~\cite{Li:2022mvy}, and DM~\cite{Du:2023dhk, Badurina:2025xwl}. While each individual contribution takes a different value in each gauge, the sum is invariant under a gauge transformation.

In this work, we argue that an analogous formalism also exists for PTAs. Specifically, we identify the proper time, as measured by an observer on Earth, between the arrivals of two consecutive pulses from a pulsar, as a gauge-invariant observable for a general metric perturbation $h_{\mu\nu}$. This is illustrated in Fig.~\ref{fig:pulsar_proper_time}. In linearized gravity, we show that this observable can again be decomposed into three contributions: a Doppler term, accounting for the motion of the pulsar and Earth; a Shapiro delay term, accounting for the change in photon propagation time; and an Einstein delay term, accounting for the gravitational redshift of the clock. Beyond clarifying gauge-invariance, formulating the observable as a proper time and computing it by solving for the intersection of geodesics has an additional advantage, namely that the formalism could be generalized to modified theories of gravity, in which worldlines can instead follow modified geodesics, offering a route to formally construct tests of these theories with PTAs~\cite{Liang:2021bct,Domenech:2025ccu} (see, \textit{e.g.}, Refs.~\cite{Lee:2010cg, deRham:2016nuf, Qin:2020hfy, Wu:2023rib, Bi:2023ewq, Wu:2023pbt, Liang:2023ary, Liang:2024mex, Hu:2024wub, Cordes:2024oem, Liang:2025vji, Gree:2026nlt} for theoretical and observational works on using PTAs to test massive gravity, or gravity with modified dispersion relations). We provide a recipe for computing this proper time shift as a time series, $\delta t(t)$, for any specified $h_{\mu\nu}$, independent of its source, which agrees with the result of Ref.~\cite{Dror:2025nvg} in terms of $R_{i0j0}$. Readers interested in direct application can skip ahead to Eq.~\eqref{eqn:delta_t_t} and Eq.~\eqref{eqn:final_observable_simple} where the expression for $\delta t(t)$ is given. This can then be directly incorporated into numerical software such as \texttt{PTArcade} for data analysis with real PTA data.

The structure of this paper is as follows. In Sec.~\ref{sec:timeshift}, we consider a general metric perturbation $h_{\mu\nu}$ and compute the resulting proper time shift $\delta t(t)$, obtaining the general expressions in Eq.~\eqref{eqn:delta_t_t} and Eq.~\eqref{eqn:final_observable_simple} by solving for the intersection of geodesics, which are directly useful for data analysis. In Sec.~\ref{sec:worked_examples}, we consider a few worked examples by applying these expressions to metric perturbations arising in several well-motivated scenarios, including plane GWs, a stochastic GWB, ULDM, and the gravitational memory effect. We then compare our results with existing literature. Finally, we conclude in Sec.~\ref{sec:conclusion}.

Throughout this paper, we work in the linearized GR regime, where the metric can be written as $g_{\mu\nu}=\eta_{\mu\nu}+h_{\mu\nu}$, and we keep only terms to leading order in $h_{\mu\nu}$. Here $\eta_{\mu\nu}=\mathrm{diag}(-1,+1,+1,+1)$ is the Minkowski metric in the mostly positive convention. We also take $c=\hbar=1$.

\section{Time-shifts due to a General Metric Perturbation on an Ideal Pulsar}\label{sec:timeshift}

\subsection{single, idealized pulsar}\label{subsec:single_ideal}

We begin our analysis by considering an \textit{ideal pulsar} setup, in which we neglect the effects of noise sources and spin-down in the pulsar. Let $\vec{x}_E$ and $\vec{x}_P$ denote the positions of Earth and the pulsar, respectively, and let $\unit{n}$ be the unit vector pointing from Earth to the pulsar. The pulsar emits pulses at regular intervals $P$, which are subsequently detected by a telescope on Earth.

\begin{figure}[h]
	\includegraphics[width=0.5\textwidth]{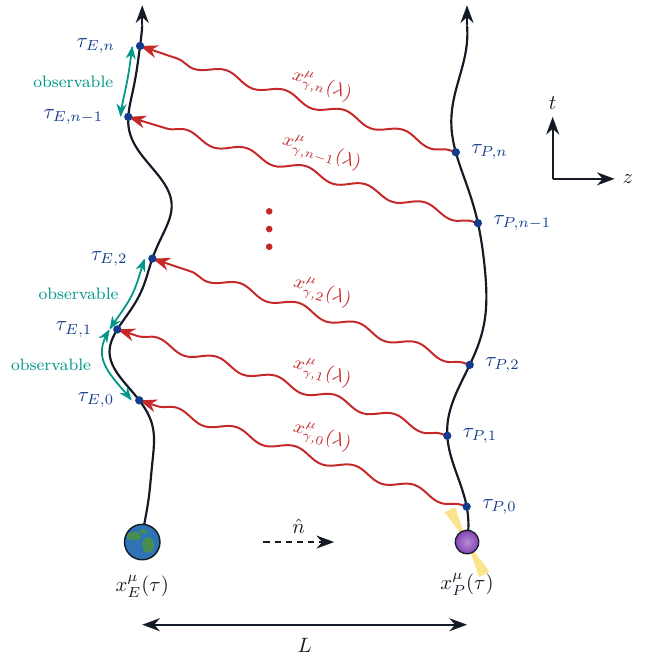}\caption{Schematic spacetime diagram of the Earth--pulsar system. Here coordinate time $t$ increases upward and the spatial coordinate $z$, aligned with the line of sight, to the right. Earth (left) and the pulsar (right) follow the timelike worldlines $x^{\mu}_{E}(\tau)$ and $x^{\mu}_{P}(\tau)$, parametrized by their proper time $\tau$ and are curvy due to their deflection by the metric perturbation
	$h_{\mu\nu}$. The two are separated by a distance $L$, and $\unit{n}$ is the unit
	vector pointing from Earth to the pulsar. The pulsar emits pulses at successive
	proper times $\tau_{P,0},\tau_{P,1},\dots,\tau_{P,n}$, spaced by a fixed proper
	interval $P$ along its worldline. Each pulse propagates to Earth along a null
	worldline $x^{\mu}_{\gamma,n}(\lambda)$ (red), with affine parameter $\lambda$,
	and is received at the Earth proper time $\tau_{E,n}$. Because the entire system
	is perturbed by $h_{\mu\nu}$, the arrival proper times are no longer evenly
	spaced. Their shifts $\delta\tau_{E,n}$ constitute the gauge-invariant observable
	derived in Sec.~\ref{sec:timeshift} (\textit{cf.} Eq.~\eqref{eqn:delta_t_t}). This figure was produced with assistance from Anthropic's \texttt{Claude Opus 5} model.}\label{fig:pulsar_proper_time}
\end{figure}

Under a weak metric perturbation, $h_{\mu\nu}$, the worldlines for the pulsar, Earth, and the photons traveling from the pulsar to the Earth will also be perturbed. This is illustrated in Fig.~\ref{fig:pulsar_proper_time}. We denote the pulsar and Earth worldlines as $x^{\mu}_P(\tau)$ and $x^{\mu}_E(\tau)$, respectively, and the photon worldline for each pulse as $x^{\mu}_{\gamma,0}(\lambda)$, $x^{\mu}_{\gamma,1}(\lambda)$, $\cdots$, where $\tau$ and $\lambda$ are the pulsar/Earth proper times, and the pulse's affine parameter, respectively. In the perturbed spacetime, the pulses are now emitted at regular \textit{proper} time intervals, \textit{i.e.} each pulse is emitted when proper time of $P$ has elapsed as measured by the pulsar's worldline, denoted as $\tau_{P,0}$, $\tau_{P,1}$, $\cdots$\footnote{Here we have neglected tidal deformation of the pulsar’s shape due to $h_{\mu\nu}$, which, for a pulsar with mass $M$ and radius $R$, would lead to a fractional change in the spin period of order $\delta P/P \sim (R^3/GM)\,f^2 h$, where $f$ is the characteristic frequency of variation of $h_{\mu\nu}$ at the pulsar, obtained by comparing the tidal acceleration across the star $\sim R \partial^2 h$ to the star’s self gravity $\sim GM/R^2$.}. The intervals of the time-of-arrivals of each pulse, as measured in the proper time of the detector's worldline on Earth, however, will not be regular, as the entire pulsar-Earth system is perturbed by $h_{\mu\nu}$. We denote these proper time-of-arrivals as $\tau_{E,0}$, $\tau_{E,1}$, $\cdots$. The shifts in these quantities constitute the observable of the experiment on $h_{\mu\nu}$, which, as we will later show, is invariant under a gauge transformation in $h_{\mu\nu}$ \footnote{Here we are working with a simplified setup where every single pulse is measured and used in the data analysis. In reality, each ``pulse" in Fig.~\ref{fig:pulsar_proper_time} should be understood as an \textit{epoch}, where multiple pulses are measured and averaged over a duration of minutes to hours in order to obtain a single robust measurement of the TOA~\cite{1975ApJ...198..661H}. This distinction is not important to us as long as the typical frequency of the time dependence of $h_{\mu\nu}$ is much smaller than the inverse of the duration of pulse averaging, which is $\gtrsim 10^{-4}$~Hz, and is thus always satisfied for a typical PTA observation with peak sensitivity at $\sim \mathcal{O}(\mathrm{nHz})$}. We also denote the pulse's affine parameter when it was emitted at the pulsar as $\lambda_{P,0}$, $\lambda_{P,1}$, $\cdots$, and when it arrived at Earth as $\lambda_{E,0}$, $\lambda_{E,1}$, $\cdots$.

We now examine more closely how $h_{\mu\nu}$ affects the proper time pulse measurements. Throughout this section, ``unperturbed” quantities are denoted by an overbar, while leading-order perturbations induced by $h_{\mu\nu}$ are denoted by $\delta$. The unperturbed solution corresponds to Minkowski spacetime, in which the pulsar and Earth follow worldlines at fixed spatial positions and pulses are emitted and received at perfectly regular intervals, while the perturbed quantities describe small deviations from these values due to $h_{\mu\nu}$. We label the first pulse, which marks the beginning of the measurement, with a 0 subscript, and define the measured proper-time shifts of subsequent pulses relative to this reference. As an illustration, we consider the $n$-th pulse, for which we can write

\begin{align}\label{eqn:perturbed_times}
	\tau_{P,n} &= \bar{\tau}_{P,n} \nonumber \\
	\tau_{E,n} &= \bar{\tau}_{E,n} + \delta \tau_{E,n} \nonumber \\
	\lambda_{P,n} &= \bar{\tau}_{P,n} \nonumber \\
	\lambda_{E,n} &= \bar{\lambda}_{E,n} + \delta \lambda_{E,n} \, .
\end{align}
As mentioned above, the parameters that mark the emission are not affected by $h_{\mu\nu}$ and are thus simply given by the unperturbed quantities. The worldlines themselves, however, will be perturbed
\begin{align}\label{eqn:worldlines}
	x^{\mu}_P(\tau) &= \bar{x}^{\mu}_P(\tau) + \delta x^{\mu}_P(\tau) \nonumber \\
	x^{\mu}_E(\tau) &= \bar{x}^{\mu}_E(\tau) + \delta x^{\mu}_E(\tau) \nonumber \\
	x^{\mu}_{\gamma,n}(\lambda) &= \bar{x}^{\mu}_{\gamma,n}(\lambda) + \delta x^{\mu}_{\gamma,n}(\lambda) \, ,
\end{align}
where the unperturbed worldlines are given by
\begin{align}\label{eqn:unperturbed_worldlines}
	\bar{x}^{\mu}_P(\tau) &= (\tau, \vec{x}_P) \nonumber \\
	\bar{x}^{\mu}_E(\tau) &= (\tau, \vec{x}_E) \nonumber \\
	\bar{x}^{\mu}_{\gamma,n}(\lambda) &= (\lambda, -(\lambda-\bar{\lambda}_{P,n})\unit{n}+\vec{x}_P) \, .
\end{align}
Our goal is to compute the proper TOAs, $\tau_{E,n}$. To achieve this, we use Fig.~\ref{fig:pulsar_proper_time} and write down the intersections of the worldlines at two instances: emission of pulse, and measurement of pulse.  
\begin{align}\label{eqn:worldline_intersection}
	x^{\mu}_P(\tau_{P,n}) &= x^{\mu}_{\gamma,n}(\lambda_{P,n}) \nonumber \\
	x^{\mu}_E(\tau_{E,n}) &= x^{\mu}_{\gamma,n}(\lambda_{E,n}) \, .
\end{align}
We then expand the worldlines in Eq.~\eqref{eqn:worldline_intersection} to linear order in $h_{\mu\nu}$ and isolate the leading-order contribution:
\begin{align}\label{eqn:worldline_expanded}
	x^{\mu}_P(\tau_{P,n}) &= \bar{x}^{\mu}_P(\bar{\tau}_{P,n}) + \delta x^{\mu}_P(\bar{\tau}_{P,n}) \nonumber \\
	x^{\mu}_E(\tau_{E,n}) &= \bar{x}^{\mu}_E(\bar{\tau}_{E,n}) + \delta x^{\mu}_E(\bar{\tau}_{E,n}) + \delta\tau_{E,n}\,\bar{v}^{\mu}_{E}(\bar{\tau}_{E,n}) \nonumber \\
	x^{\mu}_{\gamma,n}(\lambda_{P,n}) &= \bar{x}^{\mu}_{\gamma,n}(\bar{\lambda}_{P,n}) + \delta x^{\mu}_{\gamma,n}(\bar{\lambda}_{P,n}) \nonumber \\
	x^{\mu}_{\gamma,n}(\lambda_{E,n}) &= \bar{x}^{\mu}_{\gamma,n}(\bar{\lambda}_{E,n}) + \delta x^{\mu}_{\gamma,n}(\bar{\lambda}_{E,n})+ \delta\lambda_{E,n}\,\bar{v}^{\mu}_{\gamma,n}(\bar{\lambda}_{E,n}) \, ,
\end{align}
where we define the four-velocity of the pulsar and Earth as $v_{P/E}^{\mu}(\tau)\equiv (d/d\tau)x_{P/E}^{\mu}(\tau)$, and the affine tangent of the pulse as $v_{\gamma,n}^{\mu}(\lambda)\equiv (d/d\lambda)x_{\gamma,n}^{\mu}(\lambda)$. Solving Eq.~\eqref{eqn:worldline_intersection} using the expansions above, the zeroth-order terms immediately yield $\bar{\tau}_{P,n}=\bar{\lambda}_{P,n}$ and $\bar{\tau}_{E,n}=\bar{\lambda}_{E,n}$. At linear order, we obtain the leading contribution to the proper-time shift of the $n$-th pulse as
\begin{equation}\label{eqn:linear_equation_solved}
	\delta \tau_{E,n} = n_{\mu}\left[\delta x^{\mu}_E(\bar{\tau}_{E,n}) - \delta x^{\mu}_P(\bar{\tau}_{P,n})\right] - n_{\mu}\left[\delta x_{\gamma,n}^{\mu}(\bar{\lambda}_{E,n})-\delta x_{\gamma,n}^{\mu}(\bar{\lambda}_{P,n})\right] \, ,
\end{equation}
where we have defined the four-vector $n^{\mu}=(1,-n^i)$, which satisfies $n_{\mu}\bar{v}^{\mu}_E(\tau)=n_{\mu}\bar{v}^{\mu}_P(\tau)=-1$ and $n_{\mu}\bar{v}^{\mu}_{\gamma,n}(\lambda)=0$, as apparent from Eq.~\eqref{eqn:unperturbed_worldlines}.

The proper time shift derived for the $n$-th observed pulse in Eq.~\eqref{eqn:linear_equation_solved} can also be applied to the first pulse that is observed by substituting $n\to 0$. This allows us to easily compute the proper time shift relative to the first pulse, denoted by $\delta \tau_n\equiv \delta \tau_{E,n} - \delta \tau_{E,0}$, using Eq.~\eqref{eqn:linear_equation_solved}, which can be suggestively written as a sum of three distinct terms
\begin{align}\label{eqn:delta_tau_n}
	\delta \tau_{n} = \delta \tau_{n}^{(\mathcal{E})} + \delta \tau_{n}^{(\mathcal{D})} + \delta \tau_{n}^{(\mathcal{S})}   \, ,
\end{align}
with the individual contributions defined by
\begin{align}\label{eqn:delta_tau_n,EDS}
	\delta \tau_{n}^{(\mathcal{D})} &\equiv -n_i\left[\delta x^i_E(\bar{\tau}_{E,n})-\delta x^i_E(\bar{\tau}_{E,0})\right] + n_i\left[\delta x^i_P(\bar{\tau}_{P,n})-\delta x^i_P(\bar{\tau}_{P,0})\right] \nonumber \\
	\delta \tau_{n}^{(\mathcal{S})} &\equiv -\int_{\bar{\lambda}_{P,n}}^{\bar{\lambda}_{E,n}}d\lambda\, n_{\mu}\delta v^{\mu}_{\gamma,n}(\lambda) + \int_{\bar{\lambda}_{P,0}}^{\bar{\lambda}_{E,0}}d\lambda\, n_{\mu}\delta v^{\mu}_{\gamma,0}(\lambda) \nonumber \\
	\delta \tau_{n}^{(\mathcal{E})} &\equiv -\left[\delta x^0_E(\bar{\tau}_{E,n})-\delta x^0_E(\bar{\tau}_{E,0})\right] + \left[\delta x^0_P(\bar{\tau}_{P,n})-\delta x^0_P(\bar{\tau}_{P,0})\right]  \, .
\end{align}
Here, the three terms correspond to distinct physical effects: $\delta\tau^{(\mathcal{D})}_n$ is the Doppler effect due to the motion of the pulsar and Earth, $\delta\tau^{(\mathcal{S})}_n$ is the Shapiro effect, which is the time delay in the pulse as it propagates from the pulsar to Earth in a fluctuating background, and $\delta\tau^{(\mathcal{E})}_n$ is the Einstein delay, accounting for the time dilation of the clocks at the pulsar and Earth.

To further evaluate Eq.~\eqref{eqn:delta_tau_n}, we need to compute the perturbed worldlines. For massive bodies, including the pulsar and Earth, we consider the linear-order contribution to the geodesic equation
\begin{equation}\label{eqn:geodesic_equation}
	\frac{d^2}{d\tau^2}\delta  x^{\rho}_{P/E}(\tau)+\Gamma^{\rho}_{\mu\nu}\frac{d\bar{x}_{P/E}^{\mu}(\tau)}{d\tau}\frac{d\bar{x}_{P/E}^{\nu}(\tau)}{d\tau} = 0 \, ,
\end{equation}
where we used the fact that $\Gamma^{\rho}_{\mu\nu}\sim\mathcal{O}(h)$. Substituting the unperturbed worldline expressions from Eq.~\eqref{eqn:unperturbed_worldlines} into Eq.~\eqref{eqn:geodesic_equation}, we find $(d^2/d\tau^2)\delta x^{\mu}_{P/E}(\tau)=-\Gamma^{\mu}_{00}$, giving
\begin{align}\label{eqn:geodesic_equation_2}
	\frac{d^2}{d\tau^2}\delta  x^{0}_{P/E}(\tau) &= \frac{1}{2}\partial_0h_{00}\nonumber \\
	\frac{d^2}{d\tau^2}\delta  x^{i}_{P/E}(\tau) &= \frac{1}{2}\partial_ih_{00}-\partial_0h_{0i} \, .
\end{align}
To solve this differential equation, we choose initial conditions,
$\delta x^{\mu}_{P/E}(\tau_{P/E}^*)=0$ and
$\delta v^{i}_{P/E}(\tau_{P/E}^*)=0$, at some reference proper times
$\tau_{P/E}^*$. Since the worldlines are parametrized by proper time,
the normalization $g_{\mu\nu}v^\mu v^\nu=-1$ fixes the temporal
component to
$\delta v^{0}_{P/E}(\tau_{P/E}^*)=\tfrac12
h_{00}(\tau_{P/E}^*,\vec{x}_{P/E})$
at linear order. With these initial conditions, the solution to
Eq.~\eqref{eqn:geodesic_equation_2} is
\begin{align}\label{eqn:geodesic_equation_3}
	\delta  x^{0}_{P/E}(\tau) &= \frac{1}{2}\int_{\tau_{P/E}^*}^{\tau}d\tau'\, h_{00}(\tau',\vec{x}_{P/E}) \nonumber \\
	\delta  x^{i}_{P/E}(\tau) &= \frac{1}{2}\eta^{ij}\int_{\tau^*_{P/E}}^{\tau}d\tau'\int_{\tau^*_{P/E}}^{\tau'}d\tau''\,\partial_jh_{00}(\tau'',\vec{x}_{P/E}) -\eta^{ij} \int_{\tau^*_{P/E}}^{\tau}d\tau'\,h_{0j}(\tau',\vec{x}_{P/E}) + \eta^{ij} (\tau-\tau^*_{P/E})h_{0j}(\tau^*_{P/E},\vec{x}_{P/E}) \, .
\end{align}
For the photon trajectory, it is more convenient to work with the null geodesic condition, $g_{\mu\nu}v^{\mu}_{\gamma,n}(\lambda)v^{\nu}_{\gamma,n}(\lambda)=0$, which in linearized gravity simplifies to
\begin{align}\label{eqn:photon_geodesic}
	n_{\mu}\delta v^{\mu}_{\gamma,n}(\lambda) = -\frac{1}{2}h_{\mu\nu}n^{\mu}n^{\nu} = -\frac{1}{2}\left[h_{00}-2h_{0i}n^i+h_{ij}n^in^j\right]\Bigg|_{(\lambda,-(\lambda-\bar{\lambda}_{P,n})\unit{n}+\vec{x}_P)} \, .
\end{align}
The three contributions in the observable in Eq.~\eqref{eqn:delta_tau_n,EDS} can then be written in terms of derivatives and integrals of $h_{\mu\nu}$ readily using Eqs.~\eqref{eqn:geodesic_equation_3}--\eqref{eqn:photon_geodesic}.

Finally, we want to rewrite the observable in Eq.~\eqref{eqn:delta_tau_n,EDS} in a way that is easy to use in a realistic data analysis pipeline such as \texttt{PTArcade}. We first note that the proper times, $\tau$, in the arguments of Eq.~\eqref{eqn:delta_tau_n,EDS} can be replaced by just coordinate time, $t$, since they only differ in $\mathcal{O}(h)$ and they are arguments of quantities that are already $\mathcal{O}(h)$, making the error in replacing $\tau$ by $t$ order $\mathcal{O}(h^2)$. Accordingly, for a given pulse observed at proper time $\tau_{E,n}$, we can identify it with the coordinate time $t$. Then the proper TOA shift at time $t$ is written as $\delta t(t)\equiv \delta \tau_n\Big|_{n=(t-t_0)/P}$, where the symbol $\delta t$ is used instead of $\delta \tau$ to align with the notation commonly used in the community, even though these are still proper time quantities. The relevant time quantities for this pulse are
\begin{align}\label{eqn:tau_t_conversion}
	\bar{\tau}_{P,n} &= t-L \nonumber \\
	\bar{\tau}_{E,n} &= t \nonumber \\
	\bar{\lambda}_{P,n} &= t-L \nonumber \\
	\bar{\lambda}_{E,n} &= t \, ,
\end{align}
where $L$ is the pulsar-Earth distance, and the analogous quantities for the first pulse with $n$ replaced by $0$ are obtained by replacing $t$ with $t_0$, the coordinate time of the first pulse measured. Using Eq.~\eqref{eqn:tau_t_conversion}, we can rewrite everything in Eq.~\eqref{eqn:delta_tau_n} and Eq.~\eqref{eqn:delta_tau_n,EDS} in terms of coordinate time. The result can be neatly written as
\begin{equation}\label{eqn:delta_t_t}
	\delta t(t) = \delta  t^{(\mathcal{D})}_E(t) - \delta t^{(\mathcal{D})}_P(t) + \delta t^{(\mathcal{S})}(t) + \delta t^{(\mathcal{E})}_E(t) - \delta t^{(\mathcal{E})}_P(t) \, ,
\end{equation}
where we have separated the Doppler and the Einstein term as the difference between an Earth term and a pulsar term, and the individual terms are given by
\begin{align}\label{eqn:final_observable}
	\delta t^{(\mathcal{D})}_E(t) &= \left[-\frac{1}{2}\int_{t_E^*}^tdt'\int_{t_E^*}^{t'}dt''\,n^i\partial_ih_{00}(t'',\vec{x}_E)+\int_{t_E^*}^tdt'\,n^ih_{0i}(t',\vec{x}_E)-(t-t_E^*)n^ih_{0i}(t^*_E,\vec{x}_E)\right] - [t\to t_0] \nonumber \\
	\delta t^{(\mathcal{D})}_P(t) &= \left[-\frac{1}{2}\int_{t_P^*}^{t-L}dt'\int_{t_P^*}^{t'}dt''\,n^i\partial_ih_{00}(t'',\vec{x}_P)+\int_{t_P^*}^{t-L}dt'\,n^ih_{0i}(t',\vec{x}_P)-(t-L-t_P^*)n^ih_{0i}(t^*_P,\vec{x}_P)\right] - [t\to t_0] \nonumber \\
	\delta t^{(\mathcal{S})}(t) &= \left\{+\frac{1}{2}\int_{t-L}^t dt'\, \left[h_{00}-2n^ih_{0i}+n^in^jh_{ij}\right](t',\vec{x}_E-(t'-t)\unit{n})\right\} - \left\{t\to t_0\right\} \nonumber \\
	\delta t^{(\mathcal{E})}_E(t) &=  \left[-\frac{1}{2}\int_{t_{E}^*}^{t}dt'\, h_{00}(t',\vec{x}_{E})\right]  - \left[t\to t_0\right] \nonumber \\
	\delta t^{(\mathcal{E})}_P(t) &=  \left[-\frac{1}{2}\int_{t_{P}^*}^{t-L}dt'\, h_{00}(t',\vec{x}_{P})\right]  - \left[t\to t_0\right] \, ,
\end{align}
where $t_E^*$ and $t_P^*$ are the reference coordinate times set by the initial conditions of Earth and the pulsar, $\delta x^{\mu}_E$ and $\delta x^{\mu}_P$ are set to zero at those reference times, and we used $\vec{x}_P-L\unit{n}=\vec{x}_E$. 

The time shift quantity in Eqs.~\eqref{eqn:delta_t_t}--\eqref{eqn:final_observable} is the full observable under a general metric perturbation, $h_{\mu\nu}$. It is, however, not written in the most convenient form. In particular, the expression depends on $t_E^*$, $t_P^*$ and $t_0$, and it might not be immediately clear what values to use in a realistic analysis. On the other hand, the measured value of the time shifts in a realistic experiment has contributions from other sources as well. As we will show in Sec.~\ref{subsec:timing_model}, the \textit{timing model} also enters the measured time shifts to account for the intrinsic pulsar evolution that is unrelated to gravitational perturbations. The timing model generally includes a quadratic polynomial in $t$ with coefficients fitted to the data. Consequently, any contributions to Eqs.~\eqref{eqn:delta_t_t}--\eqref{eqn:final_observable} that are at most quadratic in $t$ are not observable: they are completely degenerate with the timing model and only shift its best-fit coefficients without affecting the likelihood or the timing residuals (\textit{i.e.}, the measured time shifts minus the fitted timing model). This has two implications for a realistic analysis. First, we observe from Eqs.~\eqref{eqn:delta_t_t}--\eqref{eqn:final_observable} that $t_E^*$, $t_P^*$ and $t_0$ can only contribute to at most a linear function in $t$, and thus can be conveniently set to any arbitrary value. Second, any other terms (particularly those originating from the boundaries of the time integrals) can be neglected. With this in mind, Eq.~\eqref{eqn:final_observable} can be rewritten in a form convenient for analysis
\begin{align}\label{eqn:final_observable_simple}
	\delta t^{(\mathcal{D})}_E(t) &= -\frac{1}{2}\int^tdt'\int^{t'}dt''\,n^i\partial_ih_{00}(t'',\vec{x}_E)+\int^tdt'\,n^ih_{0i}(t',\vec{x}_E) \nonumber \\
	\delta t^{(\mathcal{D})}_P(t) &= -\frac{1}{2}\int^{t-L}dt'\int^{t'}dt''\,n^i\partial_ih_{00}(t'',\vec{x}_P)+\int^{t-L}dt'\,n^ih_{0i}(t',\vec{x}_P)\nonumber \\
	\delta t^{(\mathcal{S})}(t) &= +\frac{1}{2}\int_{0}^L dz\, \left[h_{00}-2n^ih_{0i}+n^in^jh_{ij}\right](t-z,\vec{x}_E+z\unit{n}) \nonumber \\
	\delta t^{(\mathcal{E})}_E(t) &=  -\frac{1}{2}\int^{t}dt'\, h_{00}(t',\vec{x}_{E}) \nonumber \\
	\delta t^{(\mathcal{E})}_P(t) &=  -\frac{1}{2}\int^{t-L}dt'\, h_{00}(t',\vec{x}_{P}) \, ,
\end{align}
up to at most a linear polynomial in $t$, where any unspecified lower boundary of the integrals can be chosen arbitrarily, and we have redefined the integration variable using $z\equiv t-t'$ to parameterize the distance along the Earth--pulsar line-of-sight for the Shapiro term. This expression in Eq.~\eqref{eqn:final_observable_simple}, together with Eq.~\eqref{eqn:delta_t_t}, constitutes our main result.

\subsection{multiple pulsars}\label{subsec:multiple_pulsar}

The above analysis only considers the time shifts observed from a single pulsar. In a pulsar timing array analysis, one has access to timing measurements on multiple pulsars. The proper time shift observable for the $I$-th pulsar, located at $\vec{x}_{P,I}=\vec{x}_E+L_I\unit{n}_I$ where $L_I$ is the corresponding Earth-pulsar distance, can be written as
\begin{equation}\label{eqn:delta_t_t_I}
	\delta t_I(t) = \delta  t^{(\mathcal{D})}_{I,E}(t) - \delta t^{(\mathcal{D})}_{I,P}(t) + \delta t^{(\mathcal{S})}_I(t) + \delta t^{(\mathcal{E})}_{I,E}(t) - \delta t^{(\mathcal{E})}_{I,P}(t) \, ,
\end{equation}
where
\begin{align}\label{eqn:final_observable_simple_I}
	\delta t^{(\mathcal{D})}_{I,E}(t) &= -\frac{1}{2}\int^tdt'\int^{t'}dt''\,n_I^i\partial_ih_{00}(t'',\vec{x}_E)+\int^tdt'\,n_I^ih_{0i}(t',\vec{x}_E) \nonumber \\
	\delta t^{(\mathcal{D})}_{I,P}(t) &= -\frac{1}{2}\int^{t-L_I}dt'\int^{t'}dt''\,n_I^i\partial_ih_{00}(t'',\vec{x}_{P,I})+\int^{t-L_I}dt'\,n_I^ih_{0i}(t',\vec{x}_{P,I})\nonumber \\
	\delta t^{(\mathcal{S})}_I(t) &= +\frac{1}{2}\int_{0}^{L_I} dz\, \left[h_{00}-2n_I^ih_{0i}+n_I^in_I^jh_{ij}\right](t-z,\vec{x}_E+z\unit{n}_I) \nonumber \\
	\delta t^{(\mathcal{E})}_{I,E}(t) &=  -\frac{1}{2}\int^{t}dt'\, h_{00}(t',\vec{x}_{E}) \nonumber \\
	\delta t^{(\mathcal{E})}_{I,P}(t) &=  -\frac{1}{2}\int^{t-L_I}dt'\, h_{00}(t',\vec{x}_{P,I}) \, .
\end{align}

\subsection{additional contributions to timing residuals}\label{subsec:timing_model}

For completeness, we briefly mention other effects that can influence the measured TOAs independently of the metric perturbation considered above. In a realistic PTA observation, the observed TOAs and timing residuals are not determined solely by the gravitational perturbation derived in Sec.~\ref{subsec:single_ideal}. Schematically, the observed timing residuals can be written as
\begin{equation}\label{eqn:timing_model}
	\delta t_{I}^{\mathrm{obs}}(t)
	=\delta t_I(t;\boldsymbol{\theta})+\delta t_{I}^{\mathrm{timing}}(t;\boldsymbol{\theta}_{\mathrm{timing}})+	\delta t_{I}^{\mathrm{noise}}(t;\boldsymbol{\theta}_{\mathrm{noise}}) + \delta  t_{I}^{\mathrm{GWB}}(t;\boldsymbol{\theta}_{\mathrm{GWB}}) \, ,
\end{equation}
where $\delta t_I$ denotes the time shift induced by the metric perturbation given in Eq.~\eqref{eqn:final_observable_simple_I}, $\delta t_{I}^{\mathrm{timing}}$ represents the deterministic pulsar timing model, $\delta t_{I}^{\mathrm{noise}}$ accounts for instrumental and intrinsic pulsar noise, and $\delta t_{I}^{\mathrm{GWB}}$ represents contributions from a potential stochastic GWB. The symbol $\boldsymbol{\theta}$ denotes the parameters governing each contribution.

The timing model accounts for deterministic astrophysical and instrumental effects unrelated to the gravitational signal of interest, such as timing corrections back to the quasi-inertial reference frame of the Solar System Barycenter (SSB) and interstellar dispersion~\cite{Taylor:2021yjx}. It also includes the pulsar’s intrinsic spin-down, which introduces a quadratic polynomial in $t$ whose coefficients are not known \textit{a priori} and must be determined from the data~\cite{Hobbs:2006cd, Edwards:2006zg, 2009MNRAS.394.1945H, 2021ApJ...911...45L}. Pulsars also exhibit rotational irregularities, generally referred to as timing noise, which can often be modeled as white or red noise~\cite{Anderson:1975zze, 2010Sci...329..408L, 2010MNRAS.402.1027H, 2010ApJ...725.1607S, 2011MNRAS.414.1679E}. Finally, if a stochastic GWB exists, as suggested by NANOGrav’s recent evidence~\cite{NANOGrav:2023gor}, it also contributes to the observed time shifts. Both $\delta t_{I}^{\mathrm{noise}}$ and $\delta t_{I}^{\mathrm{GWB}}$ are stochastic processes best described by their power spectral density.

The timing model term $\delta t_{I}^{\mathrm{timing}}$ justifies dropping linear functions in $t$ in Eq.~\eqref{eqn:final_observable_simple} and Eq.~\eqref{eqn:final_observable_simple_I}. However, because the coefficients $\boldsymbol{\theta}_{\mathrm{timing}}$ in $\delta t_{I}^{\mathrm{timing}}(t;\boldsymbol{\theta}_{\mathrm{timing}})$ are not known a priori and must be inferred from the data, partial degeneracies between $\delta t_{I}^{\mathrm{timing}}$ and the signal shape $\delta t_I$ can further reduce the data’s sensitivity to the signal. This effect is well known in, \textit{e.g.}, searches for DM substructure~\cite{Ramani:2020hdo, Lee:2020wfn, Foster:2026kfg, Cherukupalli:2026cda}. This degeneracy complicates analytic estimates of the signal-to-noise ratio. In practice, however, data analysis software such as \texttt{enterprise}~\cite{2019ascl.soft12015E, enterprise} or \texttt{PTArcade}~\cite{Mitridate:2023oar} accounts for this degeneracy automatically by marginalizing over the timing-model parameters in the likelihood function and sampling the posterior distributions of the remaining model parameters. Consequently, one typically needs only to input the form of $\delta t_{I}$ derived from the relevant metric perturbation from Eqs.~\eqref{eqn:delta_t_t_I}--\eqref{eqn:final_observable_simple_I} and the other contributions in Eq.~\eqref{eqn:timing_model}, without manually treating the degeneracy.

\section{Explicit Demonstration of Gauge Invariance}\label{sec:gauge_invariance}

The observable defined in Eq.~\eqref{eqn:delta_t_t} and Eq.~\eqref{eqn:final_observable_simple} is a proper time quantity, and is thus automatically invariant under a general gauge transformation on the gravity field, defined as
\begin{equation}\label{eqn:gauge_transformation}
	h_{\mu\nu} \to h_{\mu\nu}' = h_{\mu\nu} - \partial_{\mu}\xi_{\nu} - \partial_{\nu}\xi_{\mu} \, ,
\end{equation}
for a general four-vector field $\xi^{\mu}$. In this section, we demonstrate this gauge invariance explicitly by directly computing $\delta t(t)$ in the new gauge for a single pulsar. Denoting the time shift quantities in the new gauge by a prime symbol, we compute the Doppler and Einstein contributions in Eq.~\eqref{eqn:final_observable_simple} in the new gauge and find
\begin{align}\label{eqn:new_gauge_Doppler_Einstein}
	\delta t^{(\mathcal{D})\prime}_E(t) &= \delta t^{(\mathcal{D})}_E(t) - n^i\xi_i(t,\vec{x}_E) \nonumber \\
	\delta t^{(\mathcal{D})\prime}_P(t) &= \delta t^{(\mathcal{D})}_P(t) - n^i\xi_i(t-L,\vec{x}_P)\nonumber \\
	\delta t^{(\mathcal{E})\prime}_E(t) &= 	\delta t^{(\mathcal{E})}_E(t)+ \xi_0(t,\vec{x}_{E}) \nonumber \\
	\delta t^{(\mathcal{E})\prime}_P(t) &= \delta t^{(\mathcal{E})}_P(t)+ \xi_0(t-L,\vec{x}_{P}) \, .
\end{align}
The Shapiro term in the new gauge is slightly more complicated to evaluate, since the integrand has to be evaluated along the photon trajectory. It is more convenient to write the Shapiro term Eq.~\eqref{eqn:final_observable_simple} as $\delta t^{(\mathcal{S})}(t)=(1/2)\int d\lambda\, h_{\mu\nu}n^{\mu}n^{\nu}$, and thus in the new gauge
\begin{align}\label{eqn:new_gauge_shapiro}
	\delta t^{(\mathcal{S})\prime}(t) &= \delta t^{(S)}(t)-\int d\lambda\,(n^{\mu}n^{\nu}\partial_{\mu}\xi_{\nu})\Big|_{\bar{x}_{\gamma}(\lambda)}=\delta t^{(S)}(t)-\int d\lambda\,\frac{d}{d\lambda}(n^{\mu}\xi_{\mu})\Big|_{\bar{x}_{\gamma}(\lambda)}\nonumber \\
	&=\delta t^{(\mathcal{S})}(t)-\left[\xi_0(t,\vec{x}_E)-n^i\xi_i(t,\vec{x}_E)\right] + \left[\xi_0(t-L,\vec{x}_P)-n^i\xi_i(t-L,\vec{x}_P)\right]\, .
\end{align}
Combining Eqs.~\eqref{eqn:new_gauge_Doppler_Einstein}--\eqref{eqn:new_gauge_shapiro}, we immediately see that the extra terms in the total proper time shift (\textit{cf.} Eq.~\eqref{eqn:delta_t_t}) in the new gauge exactly cancel with each other
\begin{equation}
	\delta t'(t) = \delta t(t) \, ,
\end{equation}
which concludes our proof that the proper time shift observable is gauge-invariant. We note here that, throughout this exercise, the initial conditions are understood to transform consistently under the gauge transformation, so that the same physical worldlines are compared in both gauges.

As an additional check, we differentiate Eq.~\eqref{eqn:delta_t_t} twice with respect to $t$, substitute Eq.~\eqref{eqn:final_observable_simple}, and integrate by parts using $d/dz=-\partial_0+n^i\partial_i$ along the unperturbed photon trajectory. We obtain
\begin{equation}\label{eqn:riemann_observable}
	\frac{d^2}{dt^2}\delta t(t) = -\int_0^L dz\,n^i n^j R_{i0j0}(t-z,\vec{x}_E+z\unit{n}) \, ,
\end{equation}
where
\begin{equation}\label{eqn:linearized_riemann}
	R_{i0j0}=-\frac{1}{2}\left(\partial_0^2h_{ij}-\partial_0\partial_ih_{0j}-\partial_0\partial_jh_{0i}+\partial_i\partial_jh_{00}\right) \, ,
\end{equation}
is the relevant component of the linearized Riemann tensor. This expression is manifestly gauge invariant and agrees with Ref.~\cite{Dror:2025nvg}.

\section{Worked Examples}\label{sec:worked_examples}

In Sec.~\ref{sec:timeshift}, we have worked out the proper time shift observable, $\delta t(t)$, for a general metric perturbation, $h_{\mu\nu}$, in Eq.~\eqref{eqn:delta_t_t} and Eq.~\eqref{eqn:final_observable_simple}. To illustrate how to use the formula, we now compute the proper time shifts by metric perturbations from a few well-motivated physics sources. We note here that theoretical and data analytical works on these sources have been performed in the literature before, often in a different framework, without putting all contributions in Eq.~\eqref{eqn:delta_t_t} in a unified manner. For the purpose of this work, we derive the observables for these sources and comment on the similarity and difference with the literature. Detailed analyses on the sources and implementation of searches in realistic datasets are beyond the scope of this work and are left for future investigations.

The rest of the section is organized as follows. In Sec.~\ref{subsec:GW}, we derive the proper time shift for a plane GW in two different gauges and confirm that they agree, and reproduce standard result from the literature. In addition, we compute the same observable for a stochastic GWB formed by superposition of numerous individual sources using the same formalism. In Sec.~\ref{subsec:ULDM} we consider the proper time shift due to a scalar ultralight dark matter field. Finally, in Sec.~\ref{subsec:memory}, we derive the proper time observable for the gravitational memory effect by considering two metrics arising from different configurations of an energetic null particle. We note that the proper time observable for dark matter substructure has been worked out in Ref.~\cite{Cherukupalli:2026cda}.

\subsection{gravitational waves}\label{subsec:GW}

Our first example is the benchmark scenario of a transient plane GW. Since GWs are the main focus for PTA experiments, the signal has already been worked out in a large number of works in the literature~\cite{Detweiler:1979wn, Anholm:2008wy, Mingarelli:2013dsa, Gair:2014rwa, Jenet:2014bea, Mingarelli:2014xfa, Maggiore:2018sht, Taylor:2021yjx}. We will show that the proper time shift expression produces the standard result in the literature, serving as a sanity check.

The example of GW is unique in that it is commonly described in two different gauges: the transverse-traceless (TT) gauge~\cite{Maggiore:2007ulw} and the proper detector (PD) frame~\cite{1922RendL..31...21F, Manasse:1963zz} (also known as Fermi-normal coordinates). We will show that although the individual contributions in Eq.~\eqref{eqn:final_observable_simple} differ in the two gauges, the total sum remains gauge-invariant.

We assume that the gravitational wave is traveling in the $\unit{z}$ direction with a frequency of $f_g$, angular frequency of $\omega_g\equiv 2\pi f_g$, phase $\phi_0$, and $h_+$ and $h_{\times}$ as its two polarizations. We also assume that the Earth--pulsar direction unit vector $\unit{n}=(\sin\theta\cos\varphi,\sin\theta\sin\varphi,\cos\theta)$ has spherical angular coordinates of $(\theta,\varphi)$. We set Earth to be located at the origin without loss of generality, and thus the pulsar is located at $\vec{x}_P=L\unit{n}$.

\subsubsection{plane GW in the transverse-traceless gauge}

In the TT-gauge, only the physical degrees of freedom are non-zero in the metric perturbation, which is written as~\cite{Maggiore:2007ulw}
\begin{equation}\label{eqn:TT_gauge}
	h_{\mu\nu} = \begin{pmatrix}
		0 & 0 & 0 & 0 \\
		0 & h_+ & h_{\times} & 0 \\
		0 & h_{\times} & -h_+ & 0 \\
		0 & 0 & 0 & 0 
	\end{pmatrix} \cos\left[\omega_g(t-z)+\phi_0\right] \, .
\end{equation}
We immediately see that out of the three terms in Eq.~\eqref{eqn:final_observable_simple}, only the Shapiro term contributes. Using $n^in^jh_{ij}=\sin^2\theta(h_+\cos2\varphi+h_{\times}\sin 2\varphi)\cos\left[\omega_g(t-z)+\phi_0\right]$, the proper time shift in Eq.~\eqref{eqn:delta_t_t} and Eq.~\eqref{eqn:final_observable_simple} is computed to be
\begin{align}\label{eqn:proper_time_TT_gauge}
	\delta t(t) &= \frac{2}{\omega_g}\sin^2\left(\frac{\theta}{2}\right)(h_+\cos2\varphi+h_{\times}\sin 2\varphi)\sin\left(\omega_g L\cos^2\left(\frac{\theta}{2}\right)\right)\cos\left[\omega_g\left(t-L\cos^2\left(\frac{\theta}{2}\right)\right)+\phi_0\right] \, .
\end{align}
This expression agrees with the results from the literature, which usually express the timing residual as a time integral of the redshift $z(t)$ of the pulse arrival rate. %

\subsubsection{plane GW in the proper detector frame}

The PD frame describes the spacetime perturbation as measured by an inertial observer who is experiencing free fall~\cite{Maggiore:2007ulw, Misner:1973prb}. Typically, the metric perturbation in the PD frame is written as a local expansion of the metric around the observer's worldline, and truncation of the expansion is valid for regions well within the typical wavelength of the perturbation~\cite{Manasse:1963zz, Ni:1978zz,  1979JMP....20.1473L, Marzlin:1994ia, Rakhmanov:2014noa}. However, even beyond this region, the expansion series can be resummed into an analytic form, as derived in Ref.~\cite{Berlin:2021txa}.

For algebraic simplicity, in the PD frame calculation, we will take a simplifying geometry by setting $\theta=\pi/2$ and $\varphi=0$, such that the Earth--pulsar line-of-sight is perpendicular to the direction of the GW propagation, and the entire Earth--pulsar system can be set at $y=z=0$. In this geometry, it is clear from Eq.~\eqref{eqn:final_observable_simple} that the only contributing components to the proper time shift are $h_{00}$, $h_{0x}$, and $h_{xx}$, which are given by~\cite{Berlin:2021txa}
\begin{align}\label{eqn:proper_detector_frame_2}
	h_{00}(t,x)\Bigg|_{y=0,z=0} &= -\frac{1}{2}\omega_g^{2}h_+x^2\cos(\omega_gt+\phi_0) \nonumber \\
	h_{0x}(t,x)\Bigg|_{y=0,z=0} &= 0 \nonumber \\
	h_{xx}(t,x)\Bigg|_{y=0,z=0} &= 0 \, .
\end{align}
We see that, in the PD frame, the Doppler, Shapiro and Einstein contributions in Eq.~\eqref{eqn:final_observable_simple} are all non-zero. Evaluating each contribution using Eq.~\eqref{eqn:proper_detector_frame_2}, we find (dropping constants and terms linear in $t$)
\begin{align}\label{eqn:PD_GW}
	\delta t^{(\mathcal{D})}_E(t) &= 0 \nonumber \\
	\delta t^{(\mathcal{D})}_P(t) &= -\frac{1}{2}h_+L\cos\left[\omega_g (t-L)+\phi_0\right] \nonumber \\
	\delta t^{(\mathcal{S})}(t) &= -\frac{1}{2}h_+L\cos\left[\omega_g (t-L)+\phi_0\right]+\frac{1}{4}h_+\omega_g L^2\sin\left[\omega_g (t-L)+\phi_0\right]+\frac{1}{\omega_g}h_+\sin\left(\frac{\omega_g L}{2}\right)\cos\left[\omega_g\left(t-\frac{L}{2}\right)+\phi_0\right] \nonumber \\
	\delta t^{(\mathcal{E})}_E(t) &= 0 \nonumber \\
	\delta t^{(\mathcal{E})}_P(t) &= \frac{1}{4}h_+\omega_g L^2\sin\left[\omega_g (t-L)+\phi_0\right] \, ,
\end{align}
and, from Eq.~\eqref{eqn:delta_t_t}
\begin{align}\label{eqn:PD_GW_2}
	\delta t(t) =\frac{1}{\omega_g}h_+\sin\left(\frac{\omega_g L}{2}\right)\cos\left[\omega_g\left(t-\frac{L}{2}\right)+\phi_0\right] \, .
\end{align}
It is thus clear from comparing Eq.~\eqref{eqn:proper_time_TT_gauge} and Eq.~\eqref{eqn:PD_GW_2} that $\delta t(t)$ in the PD frame gives the same answer as in the TT gauge in this simple geometry.

\subsubsection{stochastic gravitational-wave background}

Having verified that the proper time formalism reproduces the residual of a single plane GW in two gauges, we now consider a stochastic GWB formed by the superposition of many individual sources. The response of a pulsar timing array to such a background is a classical result that has been derived many times~\cite{1978SvA....22...36S, Detweiler:1979wn, Hellings:1983fr, Anholm:2008wy, Mingarelli:2013dsa, Jenet:2014bea, Gair:2014rwa, Mingarelli:2014xfa, Maggiore:2018sht, Taylor:2021yjx, Allen:2022dzg, Allen:2022ksj}. Here we compute the proper time observable for the stochastic GWB and show that it recovers results known in the standard literature. We work in the TT gauge, where only the Shapiro term of Eq.~\eqref{eqn:final_observable_simple_I} contributes, as noted above, so that the entire GWB observable sits in a single line-of-sight integral.

A general GW background can be expanded in plane waves as~\cite{Allen:1997ad, Maggiore:2007ulw, Romano:2016dpx}
\begin{equation}\label{eqn:GWB_plane_wave}
	h_{ij}(t,\vec{x}) = \sum_{A}\int_{-\infty}^{\infty} df \int_{S^2} d^2\hat{\Omega}\, \tilde{h}_A(f,\hat{\Omega})\, e^{A}_{ij}(\hat{\Omega})\, e^{2\pi i f\left(t-\hat{\Omega}\cdot\vec{x}\right)} \, ,
\end{equation}
where $\hat{\Omega}$ is the propagation direction, $f$ is the frequency of each individual source, and $e^{A}_{ij}(\hat{\Omega})$ with $A=+,\times$ are the transverse-traceless polarization tensors, normalized as $e^{A}_{ij}e^{A'\,ij}=2\delta_{AA'}$. Physically, $\tilde{h}_A(f,\hat{\Omega})$ is built from the incoherent sum of the individual waves emitted by a population of sources distributed across the sky.

Setting Earth at the origin, $\vec{x}_E=0$, we insert this expansion into the Shapiro term of Eq.~\eqref{eqn:final_observable_simple_I} and find
\begin{equation}\label{eqn:GWB_delta_t}
	\delta t_I(t) = \sum_{A}\int_{-\infty}^{\infty} df \int_{S^2} d^2\hat{\Omega}\, \tilde{h}_A(f,\hat{\Omega})\, F^{A}_I(\hat{\Omega})\, \frac{1-e^{-2\pi i f L_I\left(1+\hat{\Omega}\cdot\unit{n}_I\right)}}{2\pi i f}\, e^{2\pi i f t} \, ,
\end{equation}
where we have defined the antenna pattern function
\begin{equation}\label{eqn:GWB_antenna_pattern}
	F^{A}_I(\hat{\Omega}) \equiv \frac{n_I^i\, n_I^j\, e^{A}_{ij}(\hat{\Omega})}{2\left(1+\hat{\Omega}\cdot\unit{n}_I\right)} \, .
\end{equation}
The two terms in the numerator of Eq.~\eqref{eqn:GWB_delta_t} are known as the Earth and pulsar terms in standard PTA literature, since they arise from the GWB contribution to the Shapiro delay close to the Earth and the pulsar, respectively. In the standard treatment~\cite{Detweiler:1979wn, Anholm:2008wy, Taylor:2021yjx}, this structure is derived by integrating the redshift of the pulse arrival rate along the null geodesic, whereas here it emerges directly from the Shapiro contribution to a manifestly gauge-invariant proper time observable.

To connect with the literature, we define the redshift of the time shift measurement as $z_I(t)\equiv (d/dt)\,\delta t_I(t)$, whose Fourier transform is (using Eq.~\eqref{eqn:GWB_delta_t})
\begin{equation}\label{eqn:GWB_redshift}
	\tilde{z}_I(f) = \sum_{A}\int_{S^2} d^2\hat{\Omega}\, \tilde{h}_A(f,\hat{\Omega})\, F^{A}_I(\hat{\Omega})\left[1-e^{-2\pi i f L_I\left(1+\hat{\Omega}\cdot\unit{n}_I\right)}\right] \, .
\end{equation}
The cross-correlation between the redshifts of pulsars $I$ and $J$ is then~\cite{Taylor:2021yjx}
\begin{align}\label{eqn:GWB_correlator}
	\left\langle \tilde{z}_I(f)\,\tilde{z}^{*}_J(f')\right\rangle = \sum_{A}\sum_{A'}\int_{S^2}\int_{S^{2\prime}} d^2\hat{\Omega}\, d^2\hat{\Omega}'\, &\left[1-e^{-2\pi i f L_I\left(1+\hat{\Omega}\cdot\unit{n}_I\right)}\right]\left[1-e^{2\pi i f' L_J\left(1+\hat{\Omega}'\cdot\unit{n}_J\right)}\right] \nonumber \\
	\times &\left\langle \tilde{h}_A(f,\hat{\Omega})\,\tilde{h}^{*}_{A'}(f',\hat{\Omega}')\right\rangle F^{A}_I(\hat{\Omega})\, F^{A'}_J(\hat{\Omega}') \, .
\end{align}
We emphasize that this expression is exact in $f L_I$, since we have made no expansion in the GW wavelength relative to the Earth--pulsar distance. From this point the analysis is standard~\cite{Hellings:1983fr, Anholm:2008wy, Taylor:2021yjx}. We specify a stationary, isotropic, unpolarized, Gaussian ensemble for $\tilde{h}_A(f,\hat{\Omega})$, and we work in the short-wavelength limit $f L_I \gg 1$ appropriate to the PTA band, where we assume that the separation between distinct pulsars is much larger than the GW wavelength, $1/f \simeq 0.1\,\mathrm{pc}\left(100\,\mathrm{nHz}/f\right)$. The pulsar terms then oscillate rapidly over the sky and are suppressed upon integration for $I\neq J$, while contributing a factor of two for $I=J$, up to corrections of order $1/(f L_I)$~\cite{Mingarelli:2014xfa}. The remaining sky integral over the antenna patterns yields the one-sided cross-power spectral density of the proper time shifts. Writing
\begin{equation}\label{eqn:GWB_S_IJ_def}
	\left\langle \tilde{z}_I(f)\,\tilde{z}^{*}_J(f')\right\rangle = \frac{1}{2}\,\delta\left(f-f'\right)\left(2\pi f\right)^{2}S_{IJ}(f) \, .
\end{equation}
The result is~\cite{Taylor:2021yjx}
\begin{equation}\label{eqn:GWB_S_IJ}
	S_{IJ}(f) = \chi_{IJ}\,\frac{h_c^2(f)}{12\pi^2 f^3} \, , \qquad \chi_{IJ} = \frac{3}{2}\,x_{IJ}\ln x_{IJ} - \frac{x_{IJ}}{4} + \frac{1}{2} + \frac{1}{2}\,\delta_{IJ} \, , \qquad x_{IJ}\equiv \frac{1-\cos\gamma_{IJ}}{2} \, ,
\end{equation}
where $h_c(f)$ is the characteristic strain of the background, $\gamma_{IJ}$ is the angular separation between pulsars $I$ and $J$, and $\chi_{IJ}$ is the Hellings--Downs curve~\cite{Hellings:1983fr}. The $\delta_{IJ}$ term arises from the pulsar-term autocorrelation, which survives the sky integration only for $I=J$. This is the standard expression employed in PTA searches for a stochastic GWB.

We have thus recovered the full cross-correlation statistics of the GWB from the Shapiro term of Eq.~\eqref{eqn:final_observable_simple_I} alone. The antenna pattern, the Earth-term and pulsar-term structure, and the Hellings--Downs correlation all follow from a single contribution to the proper time observable, whereas in another gauge the same $S_{IJ}(f)$ would be distributed across nonvanishing Doppler, Shapiro, and Einstein terms.

\subsection{ultralight dark matter}\label{subsec:ULDM}

Ultralight dark matter with mass $\lesssim \mathrm{eV}$ has high occupation number, and thus behaves like a coherent field~\cite{Antypas:2022asj}. There is a large body of work in the literature discussing the prospects of using PTA for ULDM detection as well as actual implementation in data analysis~\cite{Khmelnitsky:2013lxt, Porayko:2014rfa, Graham:2015ifn, Aoki:2016mtn, DeMartino:2017qsa, Porayko:2018sfa, Kato:2019bqz, Nomura:2019cvc, Kaplan:2022lmz, Unal:2022ooa, Kim:2023kyy, Eberhardt:2024ocm, Kim:2023pkx, Xia:2023hov, Luu:2023rgg, Hwang:2023odi, EuropeanPulsarTimingArray:2023egv, Boddy:2025oxn, Gan:2025icr, Foster:2026mvs}. The metric perturbation due to scalar ULDM, in Newtonian gauge, can be written as
\begin{equation}\label{eqn:Newtonian_gauge}
	ds^2 = -(1+2\Phi)dt^2 + (1-2\Psi)dx_idx^i \, ,
\end{equation}
where the potentials $\Phi$ and $\Psi$ are stochastic quantities that depend on the value of the ULDM field. Putting this form of the metric perturbation into the proper time shift expression in Eq.~\eqref{eqn:final_observable_simple_I}, we find that the time shifts for the $I$-th pulsar are given by
\begin{align}\label{eqn:ULDM_proper_time}
	\delta t^{(\mathcal{D})}_{I,E}(t) &= \int^t dt'\int^{t'}dt''\,\unit{n}_I\cdot \nabla\Phi(t'',\vec{x}_E) \nonumber \\
	\delta t^{(\mathcal{D})}_{I,P}(t) &= \int^{t-L_I} dt'\int^{t'}dt''\,\unit{n}_I\cdot \nabla\Phi(t'',\vec{x}_{P,I}) \nonumber \\
	\delta t^{(\mathcal{S})}_{I}(t) &= -\int_0^{L_I} dz\,\left[\Phi(t-z,\vec{x}_E+z\unit{n}_I)+\Psi(t-z,\vec{x}_E+z\unit{n}_I)\right] \nonumber \\
	\delta t^{(\mathcal{E})}_{I,E}(t) &= \int^t dt'\,\Phi(t',\vec{x}_E) \nonumber \\
	\delta t^{(\mathcal{E})}_{I,P}(t) &= \int^{t-L_I} dt'\,\Phi(t',\vec{x}_{P,I}) \, .
\end{align}
To compare this to the literature, we can compute the fractional shift in frequency of the pulses, defined as $\delta \nu_I(t)/\nu_I\equiv - (d/dt)\delta t_I(t)$, and find the total as
\begin{align}\label{eqn:ULDM_redshift}
	\frac{\delta \nu_I(t)}{\nu_I} = &-\unit{n}_I\cdot \left[\int^{t}dt'\,\nabla\Phi(t',\vec{x}_E) - \int^{t-L_I} dt'\, \nabla\Phi(t',\vec{x}_{P,I})\right] + \int_0^{L_I}dz\,\unit{n}_I\cdot\nabla\left[\Phi(t-z,\vec{x}_E+z\unit{n}_I)+\Psi(t-z,\vec{x}_E+z\unit{n}_I)\right] \nonumber \\
	+ &\left[\Psi(t,\vec{x}_E) -  \,\Psi(t-L_I,\vec{x}_{P,I})\right] \, .
\end{align}
This mostly agrees with the result from Ref.~\cite{Kim:2023kyy} up to relative signs between terms. One can now derive the time shifts by integrating Eq.~\eqref{eqn:ULDM_redshift} over time.

\subsection{gravitational shock wave / memory effect}\label{subsec:memory}

Our final example is the gravitational memory effect, which is the prediction of general relativity that a burst of gravitational radiation or of unbound matter-energy leaves freely falling test masses {\em permanently} displaced from their initial configuration~\cite{Zeldovich:1974gvh, Braginsky:1985vlg, Braginsky:1987kwo, Christodoulou:1991cr, Wiseman:1991ss, 1992PhRvD..46.4304B, Thorne:1992sdb, Bieri:2013ada}. This can be due to, for example, mergers of supermassive black hole binaries~\cite{Favata:2009ii, Pollney:2010hs, Madison:2014vca}, or supernova neutrinos~\cite{1978ApJ...223.1037E, Turner:1978jj, 1996PhRvL..76..352B}. Memory has close connection to the soft graviton theorem and the BMS supertranslation symmetry at null infinity~\cite{Strominger:2014pwa, Strominger:2017zoo}, so a detection could probe the vacuum structure of asymptotically flat gravity~\cite{He:2023qha, Verlinde:2022hhs, He:2024vlp}. Works on detecting the memory effect using LIGO and NANOGrav include Refs.~\cite{Wang:2014zls, Lasky:2016knh, Hubner:2021amk, Cheung:2024zow} and Refs.~\cite{NANOGrav:2015xuc, NANOGrav:2019vto, NANOGrav:2023vfo, Agazie:2025oug}, while there are proposals of using future laser interferometers such as LISA and Einstein Telescope~\cite{Islo:2019qht, Johnson:2018xly, Grant:2022bla, Inchauspe:2024ibs}, (for reviews, see Refs.~\cite{Favata:2010zu, Mitman:2024uss}). To date, the memory effect has not been observed yet.

It has long been understood that the null memory effect is related to the metric perturbation sourced by highly energetic, null particles~\cite{1978ApJ...223.1037E, Turner:1978jj}. For instance, the geometry created by a null particle is described by the Aichelburg--Sexl (AS) metric~\cite{Aichelburg:1970dh, Dray:1984ha}, obtained originally as the ultrarelativistic boost of the Schwarzschild solution, which is flat everywhere except for a delta-function curvature concentrated on the null plane swept out by the particle. Its connection to the memory effect was clarified in Ref.~\cite{Tolish:2014bka}, which argued that the AS metric produces no \textit{displacement} memory, as its only permanent imprint is a relative \textit{velocity} kick imparted to bodies crossing the null plane~\cite{Tolish:2014bka, Zhang:2017rno, Zhang:2017jma}. Instead, the mechanism for producing a \textit{displacement} memory in free-falling test bodies needs to include the production of the null particle itself, which can be modeled as a massive particle decaying into an energetic light particle and a heavy remnant. In this case, the original massive body can be understood as a merging massive black hole binary, where the null particle is the GW burst emitted.

In this subsection, we apply our proper time shift formalism to verify this claim. We consider 1) the AS metric from an ``eternal'' null particle, as well as 2) the metric due to a massive body decaying into an energetic null particle and a heavy remnant, similar to a massive black hole merger. We compute the timing residual of a single Earth--pulsar pair for both metrics.

We first consider the AS metric which arises from a single null particle that was created in the infinite past~\cite{Aichelburg:1970dh, Dray:1984ha}
\begin{align}\label{eqn:AS_metric}
	ds^{2} &= -du\,dv + f(\rho)\,\delta(u)\,du^{2} + dx^{2} + dy^{2} \nonumber \\
	f(\rho) &= -8GE\ln\rho \, ,
\end{align}
where $u \equiv t-z$ and $v \equiv t+z$ are the lightcone coordinates, assuming that the particle is moving along the $\unit{z}$ direction, $E$ is the energy of the particle, and $\rho\equiv \sqrt{x^2+y^2}$ the transverse distance from its trajectory, so that the curvature sits on the null plane $u=0$. Reading the perturbation off the line element, we find $h_{00} = f\delta(u)$, $h_{0z} = -f\delta(u)$, and $h_{zz} = f\delta(u)$. Furthermore, we let Earth sit at transverse displacement $\vec{b}_{E}\equiv b_{E}\unit{x}$ from the shock axis, with components $\vec{x}_{E}=(b_{E},0,0)$, and the pulsar at $\vec{x}_{P} = \vec{x}_{E} + L\unit{n}$ with transverse position $\vec{b}_{P} = \vec{b}_{E} + L\vec{n}_{\perp}$, where $\vec{n}_{\perp} \equiv \unit{n} - (\unit{n}\cdot\unit{z})\unit{z}$ and $\cos\theta \equiv \unit{n}\cdot\unit{z}$. We denote $b_{P}\equiv |\vec{b}_{P}|$ to be the transverse displacement of the pulsar. The wavefront crosses Earth at $t=0$ and crosses the pulsar worldline at coordinate time $L\cos\theta$.

The evaluation of the five entries of Eq.~\eqref{eqn:final_observable_simple} for this metric is elementary, and we collect them in App.~\ref{app:details}. Combining them, we find
\begin{align}\label{eqn:AS_total}
	\delta t(t) = & -4GE\left(1+\cos\theta\right)\ln\left[\frac{\rho_{\gamma}(t)}{b_{E}}\right]\Theta(t) + \frac{4GE\,\vec{n}_{\perp}\cdot\vec{b}_{E}}{b_{E}^{2}}\,t\,\Theta(t) \nonumber \\
	& + 4GE\left(1+\cos\theta\right)\ln\left[\frac{\rho_{\gamma}(t)}{b_{P}}\right]\Theta\left(t-t_{P}\right) - \frac{4GE\,\vec{n}_{\perp}\cdot\vec{b}_{P}}{b_{P}^{2}}\left(t-t_{P}\right)\Theta\left(t-t_{P}\right) \, ,
\end{align}
where $\rho_{\gamma}(t) \equiv \left|\vec{b}_E + z_{*}(t)\,\vec{n}_{\perp}\right|$ with $z_{*}(t) = t/\left(1+\cos\theta\right)$ is the transverse distance from the shock axis of the point at which the wavefront intersects the line of sight at reception time $t$. Equation~\eqref{eqn:AS_total} represents instantaneous ``kicks" at times $t=0$ and $t=t_P$, at the Earth and pulsar locations, where the null plane produced by the energetic particle intersects with the Earth--pulsar system, generating the proper time shift. To understand whether this corresponds to the memory effect, we take the $t\to \infty$ limit in Eq.~\eqref{eqn:AS_total} to extract the late time behavior, and find
\begin{align}\label{eqn:AS_asymptote}
	\delta t(t) \to 4GE\left(\frac{\vec{n}_{\perp}\cdot\vec{b}_{E}}{b_{E}^{2}}-\frac{\vec{n}_{\perp}\cdot\vec{b}_{P}}{b_{P}^{2}}\right)t \, ,
\end{align}
where we drop a constant term that becomes subdominant for sufficiently large $t$. This is the structure of a \textit{velocity} memory effect~\cite{Tolish:2014bka, Zhang:2017rno, Zhang:2017jma}, where, long after the shock has passed, the Earth--pulsar system retains the memory of a velocity kick induced by the null particle, and continues to drift with a constant velocity.

The first calculation thus realizes the statement of Ref.~\cite{Tolish:2014bka} about an ``eternal" null particle leaving a velocity memory. In contrast, producing a displacement memory requires the consideration of the production of the null particle instead, which is absent in the above analysis. Conservation of four-momentum forbids creating null energy from nothing, so the elementary process is a particle of mass $M$ at rest which, at time $t_{0}$, emits a null particle of energy $E$ along $\unit{z}$ and recoils, and the metric is the retarded Lorenz-gauge solution of the linearized Einstein equation with this source. The corresponding form of the metric has been derived in Ref.~\cite{Tolish:2014bka}, which consists of the sum of three terms: a static piece (I) from the mass before the emission, a boosted piece (II) from the recoiling mass after, and a piece (III) sourced by the null particle itself,
\begin{align}\label{eqn:TW_created}
	h_{\mu\nu} &= h^{\mathrm{I}}_{\mu\nu} + h^{\mathrm{II}}_{\mu\nu} + h^{\mathrm{III}}_{\mu\nu} \nonumber \\
	h^{\mathrm{I}}_{\mu\nu} &= \frac{2GM}{\sqrt{x^{2}+y^{2}+\left(z-t_{0}\right)^{2}}}\left(\eta_{\mu\nu}+2t_{\mu}t_{\nu}\right)\Theta(-U) \nonumber \\
	h^{\mathrm{II}}_{\mu\nu} &= \frac{2GM'}{\sqrt{x^{2}+y^{2}+\left(z'-t_{0}\right)^{2}}}\left(\eta_{\mu\nu}+2t'_{\mu}t'_{\nu}\right)\Theta(U) \nonumber \\
	h^{\mathrm{III}}_{\mu\nu} &= \frac{4GE}{t-z}\,k_{\mu}k_{\nu}\,\Theta(U) \, ,
\end{align}
where $t^{\mu} = \left(1,0,0,0\right)$ is the four-velocity of static observers in the global inertial frame and $k^{\mu} = \left(1,\unit{z}\right)$ is tangent to the null geodesic. The decay occurs at the event $x^{\mu}_{d} = \left(t_{0},0,0,t_{0}\right)$. Four-momentum conservation fixes the rest mass, velocity, and Lorentz factor of the remnant, $M' = \sqrt{M^{2}-2ME}$, $v = E/\left(M-E\right)$, and $\gamma = \left(M-E\right)/M'$, so that its four-velocity is $t'^{\mu} = \gamma\left(1,0,0,-v\right)$, and $z' = t_{0} + \gamma\left[\left(z-t_{0}\right)+v\left(t-t_{0}\right)\right]$ is the longitudinal coordinate of its rest frame. Here $U$ is the retarded time with respect to the decay,
\begin{equation}\label{eqn:U_def}
	U \equiv \left(t-t_{0}\right) - \sqrt{x^{2}+y^{2}+\left(z-t_{0}\right)^{2}} \, .
\end{equation}
All three pieces in Eq.~\eqref{eqn:TW_created} are required, since they descend from a single conserved stress tensor, and only the sum is a solution to the linearized Einstein equation. It is recognized in Ref.~\cite{Tolish:2014bka} that the two metrics in Eq.~\eqref{eqn:AS_metric} and Eq.~\eqref{eqn:TW_created} are limits of one another: sending $t_{0}\to-\infty$ in Eq.~\eqref{eqn:TW_created}, the Riemann tensor becomes that of the AS metric of Eq.~\eqref{eqn:AS_metric}, where the information on the production of the null particle itself is forgotten. The two metrics therefore realize the two sides of the statement that memory is associated with the creation of null stress-energy and not with its passage.
\begin{figure}[tb]
	\centering
	\includegraphics[width=0.98\columnwidth]{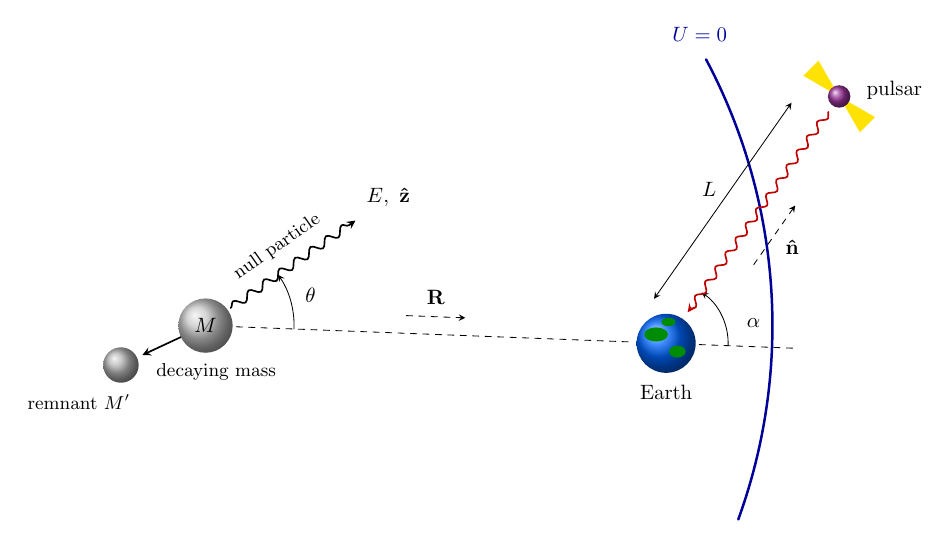}
	\caption{Geometry of the created null particle and the Earth--pulsar pair. A particle of mass $M$ at rest, the decaying mass, emits a null particle of energy $E$ along $\unit{z}$ and recoils as the remnant $M'$, conserving four-momentum. The pair lies at distance $R$ from the emission event along the direction $\unit{R}$, with $\unit{R}\cdot\unit{z} = \cos\theta$. We note that $\unit{z}$ points along the particle's trajectory while $\unit{R}$ points toward the observer, and the two coincide only for emission directly at the observer. The blue curve is the creation wavefront $U=0$, the future light cone of the emission event, expanding at the speed of light. Across the Earth--pulsar system it is plane up to corrections of $\mathcal{O}(L/R)$ and propagates along $\unit{R}$. The red wavy line is a pulse in flight from the pulsar to Earth: the snapshot shows the epoch after the front has crossed Earth and before the pulsar term arrives, during which pulses straddle the front and the residual of Eq.~\eqref{eqn:B_result} ramps. The pulsar lies at distance $L$ from Earth along $\unit{n}$. Separations are not to scale, $L \ll R$. This figure was produced with assistance from Anthropic's \texttt{Claude Opus 5} model.}
	\label{fig:memory_geometry}
\end{figure}

The proper time observable for this setup can be computed by putting the metric in Eq.~\eqref{eqn:TW_created} into Eq.~\eqref{eqn:final_observable_simple}. We carry out this calculation in App.~\ref{app:details} and find the approximate solution
\begin{equation}\label{eqn:B_result}
	\delta t(t) \approx \frac{1}{2}\cos\left(2\psi\right)\left(1-\cos\alpha\right)h_{0}\left[\left(t-t_{E}\right)\Theta\left(t-t_{E}\right) - \left(t-t_{E}-t_{P}\right)\Theta\left(t-t_{E}-t_{P}\right)\right] \, ,
\end{equation}
where $\cos\alpha \equiv \unit{R}\cdot\unit{n}$ and $\cos\theta \equiv \unit{R}\cdot\unit{z}$, with $\unit{R}$ the unit vector pointing from the decay toward Earth, $\psi$ is the azimuth of $\unit{n}$ about $\unit{R}$ measured from the sky projection of $\unit{z}$,
\begin{equation}\label{eqn:psi_def}
	\cos\psi = \frac{\unit{n}\cdot\unit{z} - \cos\theta\cos\alpha}{\sin\alpha\sin\theta}\, ,
\end{equation}
$t_{E} = t_{0} + R$ is the arrival time of the wavefront at Earth, $t_{P} = L\left(1+\cos\alpha\right)$ is the delay of the pulsar term, and $h_0$ is the amplitude given by
\begin{equation}\label{eqn:h0_mapping}
	h_{0} = \frac{2GE}{R}\left(1+\unit{R}\cdot\unit{z}\right) \, .
\end{equation}
Equation~\eqref{eqn:B_result} is precisely the burst-with-memory template employed by NANOGrav in their searches~\cite{Agazie:2025oug, vanHaasteren:2009fy, Madison:2014vca}. This can be interpreted as the memory wavefront passing over the Earth--pulsar system, inducing a ramp in the proper time shift measured from all pulsars in an angular pattern given by Eqs.~\eqref{eqn:B_result}--\eqref{eqn:psi_def}. Importantly, if we now take the limit $t\to\infty$, the proper time shift in Eq.~\eqref{eqn:B_result} saturates at the constant
\begin{equation}\label{eqn:memory_plateau}
	\delta t \to \frac{1}{2}\cos\left(2\psi\right)\left(1-\cos\alpha\right)h_{0}\,t_{P}  \, .
\end{equation}
We thus see that the Earth--pulsar system is left permanently displaced and at relative rest (as opposed to the AS metric coming from an ``eternal" null particle in Eq.~\eqref{eqn:AS_asymptote}), which is the displacement memory.

\section{Conclusion}\label{sec:conclusion}

We have derived the proper time shift measured by a PTA for an arbitrary metric perturbation, $h_{\mu\nu}$, at linear order, in analogy with previous work on other GW detectors such as laser~\cite{Lee:2024oxo} and atom interferometers~\cite{Badurina:2024rpp}. The response separates into Doppler, Shapiro, and Einstein contributions whose sum is gauge invariant. The expressions for the proper time shift, which can be used directly in software such as \texttt{enterprise}~\cite{2019ascl.soft12015E, enterprise} and \texttt{PTArcade}~\cite{Mitridate:2023oar} to study BSM physics with PTAs, are collected in Eqs.~\eqref{eqn:delta_t_t} and \eqref{eqn:final_observable_simple}. We illustrate the power of this formalism with physically motivated examples of $h_{\mu\nu}$, including GWs, ULDM, and the memory effect.

\textit{Note added.} This work is cited as ``in preparation" in the first arXiv version of Ref.~\cite{Cherukupalli:2026cda}, a paper on DM substructure coauthored by the author, which appeared on arXiv on July 3, 2026. There, Eq.~\eqref{eqn:delta_t_t} and Eq.~\eqref{eqn:final_observable_simple} are quoted in Sec.~III and App.~B, identified as a gauge-invariant proper time observable, and applied to a transiting DM subhalo. The present manuscript, which provides the derivation of these expressions, was first circulated for internal review within the NANOGrav collaboration on August 17, 2026. While it was under review, Ref.~\cite{Magi:2026upf} appeared on arXiv. Ref.~\cite{Magi:2026upf} defines the same observable as the modulation in proper time intervals between two consecutive pulses as measured on Earth, and computes it to second order in the metric perturbation in a general gauge. At linear order, the time derivative of Eq.~\eqref{eqn:delta_t_t} agrees with the timing modulation of Ref.~\cite{Magi:2026upf}, and the second derivative agrees with the curvature form of Ref.~\cite{Dror:2025nvg}. We note that the two works differ in approach and scope. Ref.~\cite{Magi:2026upf} computes the timing modulation in the limit of a vanishing emission interval, using the methods of cosmological perturbation theory, with the pulsar and the observer allowed to be non-geodesic. This work instead extends the proper time formalism for laser and atom interferometers in Refs.~\cite{Lee:2024oxo, Badurina:2024rpp}, coauthored by the author, to PTAs. We compute $\delta t(t)$ by solving the geodesic equations of the pulsar, Earth, and the pulses and locating their intersections, and write the observable as the sum of Doppler, Shapiro, and Einstein contributions, as in Refs.~\cite{Lee:2024oxo, Badurina:2024rpp}. The result is written in a form that accounts for the degeneracy with the timing model and can be used directly in existing PTA analysis software. The result is then applied to the sources in Sec.~\ref{sec:worked_examples}. We thank Matteo Magi and Jaiyul Yoo for bringing Ref.~\cite{Magi:2026upf} to our attention and for correspondence on the relation between the two works.

\acknowledgments

The author thanks Kim V. Berghaus, Abhiram Cherukupalli, Qiuyue Liang, Andrea Mitridate, Kai Schmitz, and Kathryn M. Zurek for helpful discussions. The author is a member of the NANOGrav collaboration. The author is supported by the Network for Neutrinos, Nuclear Astrophysics and Symmetries (N3AS) through the National Science Foundation Physics Frontier Center, Grant No. PHY-2020275. The author acknowledges the use of Anthropic's \texttt{Claude Opus 5} model, accessed through Lawrence Berkeley National Laboratory's CBorg AI platform, for text polishing, figure generation, and equation checking. This research used the CBorg AI platform and resources provided by the IT Division at the Lawrence Berkeley National Laboratory (Supported by the Director, Office of Science, Office of Basic Energy Sciences, of the U.S. Department of Energy under Contract No.\ DE-AC02-05CH11231).

\appendix

\section{Evaluation of the Proper Time Shifts from a Null Particle}\label{app:details}

In this appendix we evaluate the individual entries of Eq.~\eqref{eqn:final_observable_simple} for the two metrics of the main text and combine them into the residuals quoted there: the AS metric produced by an ``eternal" (\textit{i.e.} created in the infinite past) null particle, Eq.~\eqref{eqn:AS_metric}, leading to Eq.~\eqref{eqn:AS_total}, and a null particle (with a massive remnant) created from the decay of a heavy parent, Eq.~\eqref{eqn:TW_created}, leading to Eqs.~\eqref{eqn:B_result}--\eqref{eqn:h0_mapping}. 

\subsection{the AS metric}\label{app:AS}

With the perturbation $h_{00} = f\delta(u)$, $h_{0z} = -f\delta(u)$, $h_{zz} = f\delta(u)$ of Eq.~\eqref{eqn:AS_metric} and the geometry of the main text, the entries of Eq.~\eqref{eqn:final_observable_simple} evaluate as follows. The Einstein entries integrate $h_{00}$ once in time at fixed position, where $\delta(u) = \delta(t)$ at Earth and $\delta \left(t-L\cos\theta\right)$ at the pulsar, whose entries run to the emission time $t-L$, producing $t_{P} = L\left(1+\cos\theta\right)$:
\begin{align}\label{eqn:AS_Einstein}
	\delta t^{(\mathcal{E})}_{E}(t) &= -\frac{1}{2}\,f(b_E)\,\Theta(t) \nonumber \\
	\delta t^{(\mathcal{E})}_{P}(t) &= -\frac{1}{2}\,f(b_{P})\,\Theta \left(t-t_{P}\right) \, .
\end{align}
The Doppler entries comprise a double time integral of $n^{i}\partial_{i}h_{00} = \left(\vec{n}_{\perp} \cdot \nabla_{\perp}f\right)\delta(u) - \cos\theta\,f\,\delta'(u)$ and a single time integral of $n^{i}h_{0i} = -\cos\theta\,f\,\delta(u)$, which evaluates to
\begin{align}\label{eqn:AS_Doppler}
	\delta t^{(\mathcal{D})}_{E}(t) &= -\frac{1}{2}\left(\vec{n}_{\perp} \cdot \nabla_{\perp}f\right)\Big|_{\vec{b}_{E}}\,t\,\Theta(t) - \frac{\cos\theta}{2}\,f(b_{E})\,\Theta(t) \nonumber \\
	\delta t^{(\mathcal{D})}_{P}(t) &= -\frac{1}{2}\left(\vec{n}_{\perp} \cdot \nabla_{\perp}f\right)\Big|_{\vec{b}_{P}}\left(t-t_{P}\right)\Theta \left(t-t_{P}\right) - \frac{\cos\theta}{2}\,f(b_{P})\,\Theta \left(t-t_{P}\right) \, .
\end{align}
The Shapiro term is an integral over the combination $h_{00} - 2n^{i}h_{0i} + n^{i}n^{j}h_{ij} = f\,\delta(u)\left(1+\cos\theta\right)^{2}$ along the pulse path $u = t - z\left(1+\cos\theta\right)$, so the delta function selects the intersection $z_{*}(t)=t/(1+\cos\theta)$ of wavefront and line of sight
\begin{equation}\label{eqn:AS_Shapiro}
	\delta t^{(\mathcal{S})}(t) = \frac{1+\cos\theta}{2}\,f \left(\rho_{\gamma}(t)\right)\left[\Theta(t) - \Theta \left(t-t_{P}\right)\right] \, ,
\end{equation}
where $\rho_{\gamma}(t)=\left|\vec{b}_E + z_{*}(t)\vec{n}_{\perp}\right|$. Combining Eqs.~\eqref{eqn:AS_Einstein}--\eqref{eqn:AS_Shapiro} yields Eq.~\eqref{eqn:AS_total} of the main text.

\subsection{null particle created from decay}\label{app:memory}

We now turn our attention to evaluating the proper time shift for the second system, where an energetic particle with energy $E$ is produced at $t=t_0$ by a heavy particle with mass $M$ decaying and recoiling as a remnant with mass $M'$. As described in Sec.~\ref{subsec:memory}, the metric has been derived in Ref.~\cite{Tolish:2014bka} as Eq.~\eqref{eqn:TW_created}. To compute the proper time shift in Eq.~\eqref{eqn:final_observable_simple}, we first simplify the metric in Eq.~\eqref{eqn:TW_created} using two approximations: 1) small recoil ($E\ll M$), and 2) far field ($L\ll R$, where $\vec{R}$ is the vector pointing from the decaying mass to Earth and $L\unit{n}$ is the vector pointing from Earth to the pulsar, see Fig.~\ref{fig:memory_geometry}). In particular, denoting the spatial position of the decay by $\vec{x}_{d} = \left(0,0,t_{0}\right)$, so that $\vec{R} = \vec{x}_{E} - \vec{x}_{d}$ and $\left|\vec{x}-\vec{x}_{d}\right| \approx R + \unit{R}\cdot\left(\vec{x}-\vec{x}_{E}\right)$ to first order in $L/R$, we have $U\equiv t-t_0-|\vec{x}-\vec{x}_d| \approx \left(t-t_{0}-R\right) - \unit{R}\cdot\left(\vec{x}-\vec{x}_{E}\right)$. Piece I becomes
\begin{equation}\label{eqn:hI_expansion}
	h^{\mathrm{I}}_{\mu\nu} = \frac{2GM}{R}\left(\eta_{\mu\nu}+2t_{\mu}t_{\nu}\right)\Theta \left(-U\right) \, .
\end{equation}
For piece II, to first order in $E/M$ we have $M' \approx M-E$, $v\approx E/M$, $\gamma \approx 1+(E^2/2M^2)$ and $t'_{\mu} \approx t_{\mu} - \left(E/M\right)\unit{z}_{\mu}$ with $\unit{z}_{\mu} = \left(0,\unit{z}\right)$, so that $\eta_{\mu\nu}+2t'_{\mu}t'_{\nu} \approx \eta_{\mu\nu}+2t_{\mu}t_{\nu} - \left(2E/M\right)\left(t_{\mu}\unit{z}_{\nu}+t_{\nu}\unit{z}_{\mu}\right)$, while $z' \approx z + \left(E/M\right)\left(t-t_{0}\right)$ with $t-t_{0} \approx R$ behind the wavefront, so that the distance to the remnant is $R\left(1 + \left(E/M\right)\unit{R}\cdot\unit{z}\right)$. Multiplying the three factors and keeping terms of order $E$,
\begin{equation}\label{eqn:hII_expansion}
	h^{\mathrm{II}}_{\mu\nu} = \left\{\frac{2GM}{R}\left(\eta_{\mu\nu}+2t_{\mu}t_{\nu}\right) - \frac{2GE}{R}\left[\left(1+\unit{R}\cdot\unit{z}\right)\left(\eta_{\mu\nu}+2t_{\mu}t_{\nu}\right) + 2\left(t_{\mu}\unit{z}_{\nu}+t_{\nu}\unit{z}_{\mu}\right)\right]\right\}\Theta \left(U\right) \, .
\end{equation}
For piece III, the denominator is $t-z = \left(t-t_{0}\right)-\left(z-t_{0}\right) \approx R\left(1-\unit{R}\cdot\unit{z}\right)$ at the pair behind the wavefront, so that
\begin{equation}\label{eqn:hIII_expansion}
	h^{\mathrm{III}}_{\mu\nu} = \frac{4GE}{R\left(1-\unit{R}\cdot\unit{z}\right)}\,k_{\mu}k_{\nu}\,\Theta \left(U\right) \, .
\end{equation}
Collecting all terms, the two step-function pieces combine into a constant piece plus a step whose amplitude is linear in $E$,
\begin{equation}\label{eqn:memory_step_0}
	h_{\mu\nu} = \frac{2GM}{R}\,\mathrm{diag}(1,1,1,1) + \Delta h_{\mu\nu}\,\Theta \left(t-t_{E}-\unit{R}\cdot\left(\vec{x}-\vec{x}_{E}\right)\right) \, ,
\end{equation}
where $t_{E} = t_{0}+R$ and we used the expressions $\eta_{\mu\nu}+2t_{\mu}t_{\nu}=\mathrm{diag}(1,1,1,1)$ and $\left(t_{\mu}\unit{z}_{\nu}+t_{\nu}\unit{z}_{\mu}\right)_{0i}=\left(t_{\mu}\unit{z}_{\nu}+t_{\nu}\unit{z}_{\mu}\right)_{i0}=-\unit{z}_{i}$, all other components vanishing, and
\begin{align}\label{eqn:memory_step_amplitudes}
	\Delta h_{00} &= -\frac{2GE}{R}\left(1+\unit{R}\cdot\unit{z}\right) + \frac{4GE}{R\left(1-\unit{R}\cdot\unit{z}\right)} \nonumber \\
	\Delta h_{0i} &= \left[\frac{4GE}{R} - \frac{4GE}{R\left(1-\unit{R}\cdot\unit{z}\right)}\right] \unit{z}_{i} \nonumber \\
	\Delta h_{ij} &= -\frac{2GE}{R}\left(1+\unit{R}\cdot\unit{z}\right)\delta_{ij} + \frac{4GE}{R\left(1-\unit{R}\cdot\unit{z}\right)}\,\unit{z}_{i}\unit{z}_{j} \, .
\end{align}
The advantage of writing the metric in the form of Eqs.~\eqref{eqn:memory_step_0}--\eqref{eqn:memory_step_amplitudes} is that, since $\Delta h_{\mu\nu}$ is approximately a constant tensor (its spatial gradient introduces extra suppression in $1/R$), the integrals in Eq.~\eqref{eqn:final_observable} can be readily evaluated. The result is
\begin{align}\label{eqn:memory_entries}
	\delta t^{(\mathcal{E})}_{E} &= -\frac{GM}{R}\,t - \frac{\Delta h_{00}}{2}\left(t-t_{E}\right)\Theta \left(t-t_{E}\right) \nonumber \\
	\delta t^{(\mathcal{E})}_{P} &= -\frac{GM}{R}\left(t-L\right) - \frac{\Delta h_{00}}{2}\left(t-t_{E}-t_{P}\right)\Theta \left(t-t_{E}-t_{P}\right) \nonumber \\
	\delta t^{(\mathcal{D})}_{E} &= \left[\frac{\cos\alpha}{2}\,\Delta h_{00} + n^{i}\Delta h_{0i}\right]\left(t-t_{E}\right)\Theta \left(t-t_{E}\right) \nonumber \\
	\delta t^{(\mathcal{D})}_{P} &= \left[\frac{\cos\alpha}{2}\,\Delta h_{00} + n^{i}\Delta h_{0i}\right]\left(t-t_{E}-t_{P}\right)\Theta \left(t-t_{E}-t_{P}\right) \nonumber \\
	\delta t^{(\mathcal{S})} &= \frac{2GML}{R} + \frac{\Delta h_{00} - 2n^{i}\Delta h_{0i} + n^{i}n^{j}\Delta h_{ij}}{2\left(1+\cos\alpha\right)}\left[\left(t-t_{E}\right)\Theta \left(t-t_{E}\right) - \left(t-t_{E}-t_{P}\right)\Theta \left(t-t_{E}-t_{P}\right)\right] \, ,
\end{align}
where $\cos\alpha\equiv \unit{R}\cdot \unit{n}$, and $t_{P} = L\left(1+\cos\alpha\right)$ corresponds to the delay of the pulsar term relative
to the Earth term. The combination of Eq.~\eqref{eqn:delta_t_t} is
\begin{equation}\label{eqn:plane_step_slope}
	\delta t(t) = \frac{\cos^{2}\alpha\Delta h_{00} + 2\cos\alpha n^{i}\Delta h_{0i} + n^{i}n^{j}\Delta h_{ij}}{2\left(1+\cos\alpha\right)}\left[\left(t-t_{E}\right)\Theta \left(t-t_{E}\right) - \left(t-t_{E}-t_{P}\right)\Theta \left(t-t_{E}-t_{P}\right)\right] \, ,
\end{equation}
where we dropped a constant piece of $GML/R$ which is degenerate with the timing model. This recovers the time signature of the memory template written as Eq.~\eqref{eqn:B_result} in Sec.~\ref{subsec:memory} commonly employed in realistic searches with PTA data~\cite{Agazie:2025oug}. The angular structure and the amplitude in Eq.~\eqref{eqn:psi_def} and Eq.~\eqref{eqn:h0_mapping} can be readily derived by putting Eq.~\eqref{eqn:memory_step_amplitudes} into Eq.~\eqref{eqn:plane_step_slope}.

\bibliography{bibliography}

\begin{thebibliography}{146}%
\makeatletter
\providecommand \@ifxundefined [1]{%
 \@ifx{#1\undefined}
}%
\providecommand \@ifnum [1]{%
 \ifnum #1\expandafter \@firstoftwo
 \else \expandafter \@secondoftwo
 \fi
}%
\providecommand \@ifx [1]{%
 \ifx #1\expandafter \@firstoftwo
 \else \expandafter \@secondoftwo
 \fi
}%
\providecommand \natexlab [1]{#1}%
\providecommand \enquote  [1]{``#1''}%
\providecommand \bibnamefont  [1]{#1}%
\providecommand \bibfnamefont [1]{#1}%
\providecommand \citenamefont [1]{#1}%
\providecommand \href@noop [0]{\@secondoftwo}%
\providecommand \href [0]{\begingroup \@sanitize@url \@href}%
\providecommand \@href[1]{\@@startlink{#1}\@@href}%
\providecommand \@@href[1]{\endgroup#1\@@endlink}%
\providecommand \@sanitize@url [0]{\catcode `\\12\catcode `\$12\catcode
  `\&12\catcode `\#12\catcode `\^12\catcode `\_12\catcode `\%12\relax}%
\providecommand \@@startlink[1]{}%
\providecommand \@@endlink[0]{}%
\providecommand \url  [0]{\begingroup\@sanitize@url \@url }%
\providecommand \@url [1]{\endgroup\@href {#1}{\urlprefix }}%
\providecommand \urlprefix  [0]{URL }%
\providecommand \Eprint [0]{\href }%
\providecommand \doibase [0]{https://doi.org/}%
\providecommand \selectlanguage [0]{\@gobble}%
\providecommand \bibinfo  [0]{\@secondoftwo}%
\providecommand \bibfield  [0]{\@secondoftwo}%
\providecommand \translation [1]{[#1]}%
\providecommand \BibitemOpen [0]{}%
\providecommand \bibitemStop [0]{}%
\providecommand \bibitemNoStop [0]{.\EOS\space}%
\providecommand \EOS [0]{\spacefactor3000\relax}%
\providecommand \BibitemShut  [1]{\csname bibitem#1\endcsname}%
\let\auto@bib@innerbib\@empty
\bibitem [{\citenamefont {McLaughlin}(2013)}]{McLaughlin:2013ira}%
  \BibitemOpen
  \bibfield  {author} {\bibinfo {author} {\bibfnamefont {M.~A.}\ \bibnamefont
  {McLaughlin}},\ }\bibfield  {title} {\bibinfo {title} {{The North American
  Nanohertz Observatory for Gravitational Waves}},\ }\href
  {https://doi.org/10.1088/0264-9381/30/22/224008} {\bibfield  {journal}
  {\bibinfo  {journal} {Class. Quant. Grav.}\ }\textbf {\bibinfo {volume}
  {30}},\ \bibinfo {pages} {224008} (\bibinfo {year} {2013})},\ \Eprint
  {https://arxiv.org/abs/1310.0758} {arXiv:1310.0758 [astro-ph.IM]}
  \BibitemShut {NoStop}%
\bibitem [{\citenamefont {Ferdman}\ \emph {et~al.}(2010)\citenamefont {Ferdman}
  \emph {et~al.}}]{Ferdman:2010xq}%
  \BibitemOpen
  \bibfield  {author} {\bibinfo {author} {\bibfnamefont {R.~D.}\ \bibnamefont
  {Ferdman}} \emph {et~al.},\ }\bibfield  {title} {\bibinfo {title} {{The
  European Pulsar Timing Array: current efforts and a LEAP toward the
  future}},\ }\href {https://doi.org/10.1088/0264-9381/27/8/084014} {\bibfield
  {journal} {\bibinfo  {journal} {Class. Quant. Grav.}\ }\textbf {\bibinfo
  {volume} {27}},\ \bibinfo {pages} {084014} (\bibinfo {year} {2010})},\
  \Eprint {https://arxiv.org/abs/1003.3405} {arXiv:1003.3405 [astro-ph.HE]}
  \BibitemShut {NoStop}%
\bibitem [{\citenamefont {{Manchester}}\ \emph {et~al.}(2013)\citenamefont
  {{Manchester}}, \citenamefont {{Hobbs}}, \citenamefont {{Bailes}},
  \citenamefont {{Coles}}, \citenamefont {{van Straten}}, \citenamefont
  {{Keith}}, \citenamefont {{Shannon}}, \citenamefont {{Bhat}}, \citenamefont
  {{Brown}}, \citenamefont {{Burke-Spolaor}}, \citenamefont {{Champion}},
  \citenamefont {{Chaudhary}}, \citenamefont {{Edwards}}, \citenamefont
  {{Hampson}}, \citenamefont {{Hotan}}, \citenamefont {{Jameson}},
  \citenamefont {{Jenet}}, \citenamefont {{Kesteven}}, \citenamefont {{Khoo}},
  \citenamefont {{Kocz}}, \citenamefont {{Maciesiak}}, \citenamefont
  {{Oslowski}}, \citenamefont {{Ravi}}, \citenamefont {{Reynolds}},
  \citenamefont {{Sarkissian}}, \citenamefont {{Verbiest}}, \citenamefont
  {{Wen}}, \citenamefont {{Wilson}}, \citenamefont {{Yardley}}, \citenamefont
  {{Yan}},\ and\ \citenamefont {{You}}}]{2013PASA...30...17M}%
  \BibitemOpen
  \bibfield  {author} {\bibinfo {author} {\bibfnamefont {R.~N.}\ \bibnamefont
  {{Manchester}}}, \bibinfo {author} {\bibfnamefont {G.}~\bibnamefont
  {{Hobbs}}}, \bibinfo {author} {\bibfnamefont {M.}~\bibnamefont {{Bailes}}},
  \bibinfo {author} {\bibfnamefont {W.~A.}\ \bibnamefont {{Coles}}}, \bibinfo
  {author} {\bibfnamefont {W.}~\bibnamefont {{van Straten}}}, \bibinfo {author}
  {\bibfnamefont {M.~J.}\ \bibnamefont {{Keith}}}, \bibinfo {author}
  {\bibfnamefont {R.~M.}\ \bibnamefont {{Shannon}}}, \bibinfo {author}
  {\bibfnamefont {N.~D.~R.}\ \bibnamefont {{Bhat}}}, \bibinfo {author}
  {\bibfnamefont {A.}~\bibnamefont {{Brown}}}, \bibinfo {author} {\bibfnamefont
  {S.~G.}\ \bibnamefont {{Burke-Spolaor}}}, \bibinfo {author} {\bibfnamefont
  {D.~J.}\ \bibnamefont {{Champion}}}, \bibinfo {author} {\bibfnamefont
  {A.}~\bibnamefont {{Chaudhary}}}, \bibinfo {author} {\bibfnamefont {R.~T.}\
  \bibnamefont {{Edwards}}}, \bibinfo {author} {\bibfnamefont {G.}~\bibnamefont
  {{Hampson}}}, \bibinfo {author} {\bibfnamefont {A.~W.}\ \bibnamefont
  {{Hotan}}}, \bibinfo {author} {\bibfnamefont {A.}~\bibnamefont {{Jameson}}},
  \bibinfo {author} {\bibfnamefont {F.~A.}\ \bibnamefont {{Jenet}}}, \bibinfo
  {author} {\bibfnamefont {M.~J.}\ \bibnamefont {{Kesteven}}}, \bibinfo
  {author} {\bibfnamefont {J.}~\bibnamefont {{Khoo}}}, \bibinfo {author}
  {\bibfnamefont {J.}~\bibnamefont {{Kocz}}}, \bibinfo {author} {\bibfnamefont
  {K.}~\bibnamefont {{Maciesiak}}}, \bibinfo {author} {\bibfnamefont
  {S.}~\bibnamefont {{Oslowski}}}, \bibinfo {author} {\bibfnamefont
  {V.}~\bibnamefont {{Ravi}}}, \bibinfo {author} {\bibfnamefont {J.~R.}\
  \bibnamefont {{Reynolds}}}, \bibinfo {author} {\bibfnamefont {J.~M.}\
  \bibnamefont {{Sarkissian}}}, \bibinfo {author} {\bibfnamefont {J.~P.~W.}\
  \bibnamefont {{Verbiest}}}, \bibinfo {author} {\bibfnamefont {Z.~L.}\
  \bibnamefont {{Wen}}}, \bibinfo {author} {\bibfnamefont {W.~E.}\ \bibnamefont
  {{Wilson}}}, \bibinfo {author} {\bibfnamefont {D.}~\bibnamefont {{Yardley}}},
  \bibinfo {author} {\bibfnamefont {W.~M.}\ \bibnamefont {{Yan}}},\ and\
  \bibinfo {author} {\bibfnamefont {X.~P.}\ \bibnamefont {{You}}},\ }\bibfield
  {title} {\bibinfo {title} {{The Parkes Pulsar Timing Array Project}},\ }\href
  {https://doi.org/10.1017/pasa.2012.017} {\bibfield  {journal} {\bibinfo
  {journal} {Publ. Astron. Soc. Aust.}\ }\textbf {\bibinfo {volume} {30}},\
  \bibinfo {eid} {e017} (\bibinfo {year} {2013})},\ \Eprint
  {https://arxiv.org/abs/1210.6130} {arXiv:1210.6130 [astro-ph.IM]}
  \BibitemShut {NoStop}%
\bibitem [{\citenamefont {Tarafdar}\ \emph {et~al.}(2022)\citenamefont
  {Tarafdar} \emph {et~al.}}]{Tarafdar:2022toa}%
  \BibitemOpen
  \bibfield  {author} {\bibinfo {author} {\bibfnamefont {P.}~\bibnamefont
  {Tarafdar}} \emph {et~al.},\ }\bibfield  {title} {\bibinfo {title} {{The
  Indian Pulsar Timing Array: First data release}},\ }\href
  {https://doi.org/10.1017/pasa.2022.46} {\bibfield  {journal} {\bibinfo
  {journal} {Publ. Astron. Soc. Austral.}\ }\textbf {\bibinfo {volume} {39}},\
  \bibinfo {pages} {e053} (\bibinfo {year} {2022})},\ \Eprint
  {https://arxiv.org/abs/2206.09289} {arXiv:2206.09289 [astro-ph.IM]}
  \BibitemShut {NoStop}%
\bibitem [{\citenamefont {Xu}\ \emph {et~al.}(2023)\citenamefont {Xu} \emph
  {et~al.}}]{Xu:2023wog}%
  \BibitemOpen
  \bibfield  {author} {\bibinfo {author} {\bibfnamefont {H.}~\bibnamefont {Xu}}
  \emph {et~al.},\ }\bibfield  {title} {\bibinfo {title} {{Searching for the
  Nano-Hertz Stochastic Gravitational Wave Background with the Chinese Pulsar
  Timing Array Data Release I}},\ }\href
  {https://doi.org/10.1088/1674-4527/acdfa5} {\bibfield  {journal} {\bibinfo
  {journal} {Res. Astron. Astrophys.}\ }\textbf {\bibinfo {volume} {23}},\
  \bibinfo {pages} {075024} (\bibinfo {year} {2023})},\ \Eprint
  {https://arxiv.org/abs/2306.16216} {arXiv:2306.16216 [astro-ph.HE]}
  \BibitemShut {NoStop}%
\bibitem [{\citenamefont {Spiewak}\ \emph {et~al.}(2022)\citenamefont {Spiewak}
  \emph {et~al.}}]{Spiewak:2022btk}%
  \BibitemOpen
  \bibfield  {author} {\bibinfo {author} {\bibfnamefont {R.}~\bibnamefont
  {Spiewak}} \emph {et~al.},\ }\bibfield  {title} {\bibinfo {title} {{The
  MeerTime Pulsar Timing Array: A census of emission properties and timing
  potential}},\ }\href {https://doi.org/10.1017/pasa.2022.19} {\bibfield
  {journal} {\bibinfo  {journal} {Publ. Astron. Soc. Austral.}\ }\textbf
  {\bibinfo {volume} {39}},\ \bibinfo {pages} {e027} (\bibinfo {year}
  {2022})},\ \Eprint {https://arxiv.org/abs/2204.04115} {arXiv:2204.04115
  [astro-ph.HE]} \BibitemShut {NoStop}%
\bibitem [{\citenamefont {Taylor}(2021)}]{Taylor:2021yjx}%
  \BibitemOpen
  \bibfield  {author} {\bibinfo {author} {\bibfnamefont {S.~R.}\ \bibnamefont
  {Taylor}},\ }\bibfield  {title} {\bibinfo {title} {{The Nanohertz
  Gravitational Wave Astronomer}},\ }\href@noop {} {\  (\bibinfo {year}
  {2021})},\ \Eprint {https://arxiv.org/abs/2105.13270} {arXiv:2105.13270
  [astro-ph.HE]} \BibitemShut {NoStop}%
\bibitem [{\citenamefont {Seto}\ and\ \citenamefont
  {Cooray}(2007)}]{Seto:2007kj}%
  \BibitemOpen
  \bibfield  {author} {\bibinfo {author} {\bibfnamefont {N.}~\bibnamefont
  {Seto}}\ and\ \bibinfo {author} {\bibfnamefont {A.}~\bibnamefont {Cooray}},\
  }\bibfield  {title} {\bibinfo {title} {{Searching for primordial black hole
  dark matter with pulsar timing arrays}},\ }\href
  {https://doi.org/10.1086/516570} {\bibfield  {journal} {\bibinfo  {journal}
  {Astrophys. J. Lett.}\ }\textbf {\bibinfo {volume} {659}},\ \bibinfo {pages}
  {L33} (\bibinfo {year} {2007})},\ \Eprint
  {https://arxiv.org/abs/astro-ph/0702586} {arXiv:astro-ph/0702586}
  \BibitemShut {NoStop}%
\bibitem [{\citenamefont {{Kashiyama}}\ and\ \citenamefont
  {{Seto}}(2012)}]{2012MNRAS.426.1369K}%
  \BibitemOpen
  \bibfield  {author} {\bibinfo {author} {\bibfnamefont {K.}~\bibnamefont
  {{Kashiyama}}}\ and\ \bibinfo {author} {\bibfnamefont {N.}~\bibnamefont
  {{Seto}}},\ }\bibfield  {title} {\bibinfo {title} {{Enhanced exploration for
  primordial black holes using pulsar timing arrays}},\ }\href
  {https://doi.org/10.1111/j.1365-2966.2012.21935.x} {\bibfield  {journal}
  {\bibinfo  {journal} {Mon. Not. R. Astron. Soc.}\ }\textbf {\bibinfo {volume}
  {426}},\ \bibinfo {pages} {1369} (\bibinfo {year} {2012})},\ \Eprint
  {https://arxiv.org/abs/1208.4101} {arXiv:1208.4101 [astro-ph.CO]}
  \BibitemShut {NoStop}%
\bibitem [{\citenamefont {Schutz}\ and\ \citenamefont
  {Liu}(2017)}]{Schutz:2016khr}%
  \BibitemOpen
  \bibfield  {author} {\bibinfo {author} {\bibfnamefont {K.}~\bibnamefont
  {Schutz}}\ and\ \bibinfo {author} {\bibfnamefont {A.}~\bibnamefont {Liu}},\
  }\bibfield  {title} {\bibinfo {title} {{Pulsar timing can constrain
  primordial black holes in the LIGO mass window}},\ }\href
  {https://doi.org/10.1103/PhysRevD.95.023002} {\bibfield  {journal} {\bibinfo
  {journal} {Phys. Rev. D}\ }\textbf {\bibinfo {volume} {95}},\ \bibinfo
  {pages} {023002} (\bibinfo {year} {2017})},\ \Eprint
  {https://arxiv.org/abs/1610.04234} {arXiv:1610.04234 [astro-ph.CO]}
  \BibitemShut {NoStop}%
\bibitem [{\citenamefont {Dror}\ \emph {et~al.}(2019)\citenamefont {Dror},
  \citenamefont {Ramani}, \citenamefont {Trickle},\ and\ \citenamefont
  {Zurek}}]{Dror:2019twh}%
  \BibitemOpen
  \bibfield  {author} {\bibinfo {author} {\bibfnamefont {J.~A.}\ \bibnamefont
  {Dror}}, \bibinfo {author} {\bibfnamefont {H.}~\bibnamefont {Ramani}},
  \bibinfo {author} {\bibfnamefont {T.}~\bibnamefont {Trickle}},\ and\ \bibinfo
  {author} {\bibfnamefont {K.~M.}\ \bibnamefont {Zurek}},\ }\bibfield  {title}
  {\bibinfo {title} {{Pulsar Timing Probes of Primordial Black Holes and
  Subhalos}},\ }\href {https://doi.org/10.1103/PhysRevD.100.023003} {\bibfield
  {journal} {\bibinfo  {journal} {Phys. Rev. D}\ }\textbf {\bibinfo {volume}
  {100}},\ \bibinfo {pages} {023003} (\bibinfo {year} {2019})},\ \Eprint
  {https://arxiv.org/abs/1901.04490} {arXiv:1901.04490 [astro-ph.CO]}
  \BibitemShut {NoStop}%
\bibitem [{\citenamefont {Siegel}\ \emph {et~al.}(2007)\citenamefont {Siegel},
  \citenamefont {Hertzberg},\ and\ \citenamefont {Fry}}]{Siegel:2007fz}%
  \BibitemOpen
  \bibfield  {author} {\bibinfo {author} {\bibfnamefont {E.~R.}\ \bibnamefont
  {Siegel}}, \bibinfo {author} {\bibfnamefont {M.~P.}\ \bibnamefont
  {Hertzberg}},\ and\ \bibinfo {author} {\bibfnamefont {J.~N.}\ \bibnamefont
  {Fry}},\ }\bibfield  {title} {\bibinfo {title} {{Probing Dark Matter
  Substructure with Pulsar Timing}},\ }\href
  {https://doi.org/10.1111/j.1365-2966.2007.12435.x} {\bibfield  {journal}
  {\bibinfo  {journal} {Mon. Not. Roy. Astron. Soc.}\ }\textbf {\bibinfo
  {volume} {382}},\ \bibinfo {pages} {879} (\bibinfo {year} {2007})},\ \Eprint
  {https://arxiv.org/abs/astro-ph/0702546} {arXiv:astro-ph/0702546}
  \BibitemShut {NoStop}%
\bibitem [{\citenamefont {Baghram}\ \emph {et~al.}(2011)\citenamefont
  {Baghram}, \citenamefont {Afshordi},\ and\ \citenamefont
  {Zurek}}]{Baghram:2011is}%
  \BibitemOpen
  \bibfield  {author} {\bibinfo {author} {\bibfnamefont {S.}~\bibnamefont
  {Baghram}}, \bibinfo {author} {\bibfnamefont {N.}~\bibnamefont {Afshordi}},\
  and\ \bibinfo {author} {\bibfnamefont {K.~M.}\ \bibnamefont {Zurek}},\
  }\bibfield  {title} {\bibinfo {title} {{Prospects for Detecting Dark Matter
  Halo Substructure with Pulsar Timing}},\ }\href
  {https://doi.org/10.1103/PhysRevD.84.043511} {\bibfield  {journal} {\bibinfo
  {journal} {Phys. Rev. D}\ }\textbf {\bibinfo {volume} {84}},\ \bibinfo
  {pages} {043511} (\bibinfo {year} {2011})},\ \Eprint
  {https://arxiv.org/abs/1101.5487} {arXiv:1101.5487 [astro-ph.CO]}
  \BibitemShut {NoStop}%
\bibitem [{\citenamefont {Clark}\ \emph
  {et~al.}(2016{\natexlab{a}})\citenamefont {Clark}, \citenamefont {Lewis},\
  and\ \citenamefont {Scott}}]{Clark:2015sha}%
  \BibitemOpen
  \bibfield  {author} {\bibinfo {author} {\bibfnamefont {H.~A.}\ \bibnamefont
  {Clark}}, \bibinfo {author} {\bibfnamefont {G.~F.}\ \bibnamefont {Lewis}},\
  and\ \bibinfo {author} {\bibfnamefont {P.}~\bibnamefont {Scott}},\ }\bibfield
   {title} {\bibinfo {title} {{Investigating dark matter substructure with
  pulsar timing {\textendash} I. Constraints on ultracompact minihaloes}},\
  }\href {https://doi.org/10.1093/mnras/stv2743} {\bibfield  {journal}
  {\bibinfo  {journal} {Mon. Not. Roy. Astron. Soc.}\ }\textbf {\bibinfo
  {volume} {456}},\ \bibinfo {pages} {1394} (\bibinfo {year}
  {2016}{\natexlab{a}})},\ \bibinfo {note} {[Erratum: Mon.Not.Roy.Astron.Soc.
  464, 2468 (2017)]},\ \Eprint {https://arxiv.org/abs/1509.02938}
  {arXiv:1509.02938 [astro-ph.CO]} \BibitemShut {NoStop}%
\bibitem [{\citenamefont {Clark}\ \emph
  {et~al.}(2016{\natexlab{b}})\citenamefont {Clark}, \citenamefont {Lewis},\
  and\ \citenamefont {Scott}}]{Clark:2015tha}%
  \BibitemOpen
  \bibfield  {author} {\bibinfo {author} {\bibfnamefont {H.~A.}\ \bibnamefont
  {Clark}}, \bibinfo {author} {\bibfnamefont {G.~F.}\ \bibnamefont {Lewis}},\
  and\ \bibinfo {author} {\bibfnamefont {P.}~\bibnamefont {Scott}},\ }\bibfield
   {title} {\bibinfo {title} {{Investigating dark matter substructure with
  pulsar timing {\textendash} II. Improved limits on small-scale cosmology}},\
  }\href {https://doi.org/10.1093/mnras/stv2529} {\bibfield  {journal}
  {\bibinfo  {journal} {Mon. Not. Roy. Astron. Soc.}\ }\textbf {\bibinfo
  {volume} {456}},\ \bibinfo {pages} {1402} (\bibinfo {year}
  {2016}{\natexlab{b}})},\ \bibinfo {note} {[Erratum: Mon.Not.Roy.Astron.Soc.
  464, 955--956 (2017)]},\ \Eprint {https://arxiv.org/abs/1509.02941}
  {arXiv:1509.02941 [astro-ph.CO]} \BibitemShut {NoStop}%
\bibitem [{\citenamefont {Kashiyama}\ and\ \citenamefont
  {Oguri}(2018)}]{Kashiyama:2018gsh}%
  \BibitemOpen
  \bibfield  {author} {\bibinfo {author} {\bibfnamefont {K.}~\bibnamefont
  {Kashiyama}}\ and\ \bibinfo {author} {\bibfnamefont {M.}~\bibnamefont
  {Oguri}},\ }\bibfield  {title} {\bibinfo {title} {{Detectability of
  Small-Scale Dark Matter Clumps with Pulsar Timing Arrays}},\ }\href@noop {}
  {\  (\bibinfo {year} {2018})},\ \Eprint {https://arxiv.org/abs/1801.07847}
  {arXiv:1801.07847 [astro-ph.CO]} \BibitemShut {NoStop}%
\bibitem [{\citenamefont {Ramani}\ \emph {et~al.}(2020)\citenamefont {Ramani},
  \citenamefont {Trickle},\ and\ \citenamefont {Zurek}}]{Ramani:2020hdo}%
  \BibitemOpen
  \bibfield  {author} {\bibinfo {author} {\bibfnamefont {H.}~\bibnamefont
  {Ramani}}, \bibinfo {author} {\bibfnamefont {T.}~\bibnamefont {Trickle}},\
  and\ \bibinfo {author} {\bibfnamefont {K.~M.}\ \bibnamefont {Zurek}},\
  }\bibfield  {title} {\bibinfo {title} {{Observability of Dark Matter
  Substructure with Pulsar Timing Correlations}},\ }\href
  {https://doi.org/10.1088/1475-7516/2020/12/033} {\bibfield  {journal}
  {\bibinfo  {journal} {JCAP}\ }\textbf {\bibinfo {volume} {12}},\ \bibinfo
  {pages} {033}},\ \Eprint {https://arxiv.org/abs/2005.03030} {arXiv:2005.03030
  [astro-ph.CO]} \BibitemShut {NoStop}%
\bibitem [{\citenamefont {Lee}\ \emph {et~al.}(2021{\natexlab{a}})\citenamefont
  {Lee}, \citenamefont {Mitridate}, \citenamefont {Trickle},\ and\
  \citenamefont {Zurek}}]{Lee:2020wfn}%
  \BibitemOpen
  \bibfield  {author} {\bibinfo {author} {\bibfnamefont {V.~S.~H.}\
  \bibnamefont {Lee}}, \bibinfo {author} {\bibfnamefont {A.}~\bibnamefont
  {Mitridate}}, \bibinfo {author} {\bibfnamefont {T.}~\bibnamefont {Trickle}},\
  and\ \bibinfo {author} {\bibfnamefont {K.~M.}\ \bibnamefont {Zurek}},\
  }\bibfield  {title} {\bibinfo {title} {{Probing Small-Scale Power Spectra
  with Pulsar Timing Arrays}},\ }\href
  {https://doi.org/10.1007/JHEP06(2021)028} {\bibfield  {journal} {\bibinfo
  {journal} {JHEP}\ }\textbf {\bibinfo {volume} {06}},\ \bibinfo {pages}
  {028}},\ \Eprint {https://arxiv.org/abs/2012.09857} {arXiv:2012.09857
  [astro-ph.CO]} \BibitemShut {NoStop}%
\bibitem [{\citenamefont {Lee}\ \emph {et~al.}(2021{\natexlab{b}})\citenamefont
  {Lee}, \citenamefont {Taylor}, \citenamefont {Trickle},\ and\ \citenamefont
  {Zurek}}]{Lee:2021zqw}%
  \BibitemOpen
  \bibfield  {author} {\bibinfo {author} {\bibfnamefont {V.~S.~H.}\
  \bibnamefont {Lee}}, \bibinfo {author} {\bibfnamefont {S.~R.}\ \bibnamefont
  {Taylor}}, \bibinfo {author} {\bibfnamefont {T.}~\bibnamefont {Trickle}},\
  and\ \bibinfo {author} {\bibfnamefont {K.~M.}\ \bibnamefont {Zurek}},\
  }\bibfield  {title} {\bibinfo {title} {{Bayesian Forecasts for Dark Matter
  Substructure Searches with Mock Pulsar Timing Data}},\ }\href
  {https://doi.org/10.1088/1475-7516/2021/08/025} {\bibfield  {journal}
  {\bibinfo  {journal} {JCAP}\ }\textbf {\bibinfo {volume} {08}},\ \bibinfo
  {pages} {025}},\ \Eprint {https://arxiv.org/abs/2104.05717} {arXiv:2104.05717
  [astro-ph.CO]} \BibitemShut {NoStop}%
\bibitem [{\citenamefont {Berghaus}\ \emph {et~al.}(2025)\citenamefont
  {Berghaus}, \citenamefont {Du}, \citenamefont {Lee}, \citenamefont {Prabhu},
  \citenamefont {Reischke}, \citenamefont {Connor},\ and\ \citenamefont
  {Zurek}}]{Berghaus:2025kvn}%
  \BibitemOpen
  \bibfield  {author} {\bibinfo {author} {\bibfnamefont {K.~V.}\ \bibnamefont
  {Berghaus}}, \bibinfo {author} {\bibfnamefont {Y.}~\bibnamefont {Du}},
  \bibinfo {author} {\bibfnamefont {V.~S.~H.}\ \bibnamefont {Lee}}, \bibinfo
  {author} {\bibfnamefont {A.}~\bibnamefont {Prabhu}}, \bibinfo {author}
  {\bibfnamefont {R.}~\bibnamefont {Reischke}}, \bibinfo {author}
  {\bibfnamefont {L.}~\bibnamefont {Connor}},\ and\ \bibinfo {author}
  {\bibfnamefont {K.~M.}\ \bibnamefont {Zurek}},\ }\bibfield  {title} {\bibinfo
  {title} {{Physics beyond the Standard Model with the DSA-2000}},\ }\href
  {https://doi.org/10.1088/1475-7516/2025/12/035} {\bibfield  {journal}
  {\bibinfo  {journal} {JCAP}\ }\textbf {\bibinfo {volume} {12}},\ \bibinfo
  {pages} {035}},\ \Eprint {https://arxiv.org/abs/2505.23892} {arXiv:2505.23892
  [hep-ph]} \BibitemShut {NoStop}%
\bibitem [{\citenamefont {Foster}\ \emph
  {et~al.}(2026{\natexlab{a}})\citenamefont {Foster}, \citenamefont {Trickle},\
  and\ \citenamefont {Vassallo}}]{Foster:2026kfg}%
  \BibitemOpen
  \bibfield  {author} {\bibinfo {author} {\bibfnamefont {J.~W.}\ \bibnamefont
  {Foster}}, \bibinfo {author} {\bibfnamefont {T.}~\bibnamefont {Trickle}},\
  and\ \bibinfo {author} {\bibfnamefont {F.}~\bibnamefont {Vassallo}},\
  }\bibfield  {title} {\bibinfo {title} {{Projecting the ultimate pulsar timing
  sensitivity to dark matter substructure in a stochastic gravitational wave
  background}},\ }\href@noop {} {\  (\bibinfo {year} {2026}{\natexlab{a}})},\
  \Eprint {https://arxiv.org/abs/2606.18329} {arXiv:2606.18329 [astro-ph.CO]}
  \BibitemShut {NoStop}%
\bibitem [{\citenamefont {Cherukupalli}\ \emph {et~al.}(2026)\citenamefont
  {Cherukupalli}, \citenamefont {Lee}, \citenamefont {Berghaus},\ and\
  \citenamefont {Zurek}}]{Cherukupalli:2026cda}%
  \BibitemOpen
  \bibfield  {author} {\bibinfo {author} {\bibfnamefont {A.}~\bibnamefont
  {Cherukupalli}}, \bibinfo {author} {\bibfnamefont {V.~S.~H.}\ \bibnamefont
  {Lee}}, \bibinfo {author} {\bibfnamefont {K.}~\bibnamefont {Berghaus}},\ and\
  \bibinfo {author} {\bibfnamefont {K.~M.}\ \bibnamefont {Zurek}},\ }\bibfield
  {title} {\bibinfo {title} {{Pulsar Timing Sensitivity to Dark Matter
  Substructure in the Presence of a Stochastic Gravitational-Wave
  Background}},\ }\href@noop {} {\  (\bibinfo {year} {2026})},\ \Eprint
  {https://arxiv.org/abs/2607.03533} {arXiv:2607.03533 [astro-ph.CO]}
  \BibitemShut {NoStop}%
\bibitem [{\citenamefont {Khmelnitsky}\ and\ \citenamefont
  {Rubakov}(2014)}]{Khmelnitsky:2013lxt}%
  \BibitemOpen
  \bibfield  {author} {\bibinfo {author} {\bibfnamefont {A.}~\bibnamefont
  {Khmelnitsky}}\ and\ \bibinfo {author} {\bibfnamefont {V.}~\bibnamefont
  {Rubakov}},\ }\bibfield  {title} {\bibinfo {title} {{Pulsar timing signal
  from ultralight scalar dark matter}},\ }\href
  {https://doi.org/10.1088/1475-7516/2014/02/019} {\bibfield  {journal}
  {\bibinfo  {journal} {JCAP}\ }\textbf {\bibinfo {volume} {02}},\ \bibinfo
  {pages} {019}},\ \Eprint {https://arxiv.org/abs/1309.5888} {arXiv:1309.5888
  [astro-ph.CO]} \BibitemShut {NoStop}%
\bibitem [{\citenamefont {Porayko}\ and\ \citenamefont
  {Postnov}(2014)}]{Porayko:2014rfa}%
  \BibitemOpen
  \bibfield  {author} {\bibinfo {author} {\bibfnamefont {N.~K.}\ \bibnamefont
  {Porayko}}\ and\ \bibinfo {author} {\bibfnamefont {K.~A.}\ \bibnamefont
  {Postnov}},\ }\bibfield  {title} {\bibinfo {title} {{Constraints on
  ultralight scalar dark matter from pulsar timing}},\ }\href
  {https://doi.org/10.1103/PhysRevD.90.062008} {\bibfield  {journal} {\bibinfo
  {journal} {Phys. Rev. D}\ }\textbf {\bibinfo {volume} {90}},\ \bibinfo
  {pages} {062008} (\bibinfo {year} {2014})},\ \Eprint
  {https://arxiv.org/abs/1408.4670} {arXiv:1408.4670 [astro-ph.CO]}
  \BibitemShut {NoStop}%
\bibitem [{\citenamefont {Graham}\ \emph {et~al.}(2016)\citenamefont {Graham},
  \citenamefont {Kaplan}, \citenamefont {Mardon}, \citenamefont {Rajendran},\
  and\ \citenamefont {Terrano}}]{Graham:2015ifn}%
  \BibitemOpen
  \bibfield  {author} {\bibinfo {author} {\bibfnamefont {P.~W.}\ \bibnamefont
  {Graham}}, \bibinfo {author} {\bibfnamefont {D.~E.}\ \bibnamefont {Kaplan}},
  \bibinfo {author} {\bibfnamefont {J.}~\bibnamefont {Mardon}}, \bibinfo
  {author} {\bibfnamefont {S.}~\bibnamefont {Rajendran}},\ and\ \bibinfo
  {author} {\bibfnamefont {W.~A.}\ \bibnamefont {Terrano}},\ }\bibfield
  {title} {\bibinfo {title} {{Dark Matter Direct Detection with
  Accelerometers}},\ }\href {https://doi.org/10.1103/PhysRevD.93.075029}
  {\bibfield  {journal} {\bibinfo  {journal} {Phys. Rev. D}\ }\textbf {\bibinfo
  {volume} {93}},\ \bibinfo {pages} {075029} (\bibinfo {year} {2016})},\
  \Eprint {https://arxiv.org/abs/1512.06165} {arXiv:1512.06165 [hep-ph]}
  \BibitemShut {NoStop}%
\bibitem [{\citenamefont {Aoki}\ and\ \citenamefont
  {Soda}(2016)}]{Aoki:2016mtn}%
  \BibitemOpen
  \bibfield  {author} {\bibinfo {author} {\bibfnamefont {A.}~\bibnamefont
  {Aoki}}\ and\ \bibinfo {author} {\bibfnamefont {J.}~\bibnamefont {Soda}},\
  }\bibfield  {title} {\bibinfo {title} {{Pulsar timing signal from ultralight
  axion in $f(R)$ theory}},\ }\href
  {https://doi.org/10.1103/PhysRevD.93.083503} {\bibfield  {journal} {\bibinfo
  {journal} {Phys. Rev. D}\ }\textbf {\bibinfo {volume} {93}},\ \bibinfo
  {pages} {083503} (\bibinfo {year} {2016})},\ \Eprint
  {https://arxiv.org/abs/1601.03904} {arXiv:1601.03904 [hep-ph]} \BibitemShut
  {NoStop}%
\bibitem [{\citenamefont {De~Martino}\ \emph {et~al.}(2017)\citenamefont
  {De~Martino}, \citenamefont {Broadhurst}, \citenamefont {Henry~Tye},
  \citenamefont {Chiueh}, \citenamefont {Schive},\ and\ \citenamefont
  {Lazkoz}}]{DeMartino:2017qsa}%
  \BibitemOpen
  \bibfield  {author} {\bibinfo {author} {\bibfnamefont {I.}~\bibnamefont
  {De~Martino}}, \bibinfo {author} {\bibfnamefont {T.}~\bibnamefont
  {Broadhurst}}, \bibinfo {author} {\bibfnamefont {S.~H.}\ \bibnamefont
  {Henry~Tye}}, \bibinfo {author} {\bibfnamefont {T.}~\bibnamefont {Chiueh}},
  \bibinfo {author} {\bibfnamefont {H.-Y.}\ \bibnamefont {Schive}},\ and\
  \bibinfo {author} {\bibfnamefont {R.}~\bibnamefont {Lazkoz}},\ }\bibfield
  {title} {\bibinfo {title} {{Recognizing Axionic Dark Matter by Compton and de
  Broglie Scale Modulation of Pulsar Timing}},\ }\href
  {https://doi.org/10.1103/PhysRevLett.119.221103} {\bibfield  {journal}
  {\bibinfo  {journal} {Phys. Rev. Lett.}\ }\textbf {\bibinfo {volume} {119}},\
  \bibinfo {pages} {221103} (\bibinfo {year} {2017})},\ \Eprint
  {https://arxiv.org/abs/1705.04367} {arXiv:1705.04367 [astro-ph.CO]}
  \BibitemShut {NoStop}%
\bibitem [{\citenamefont {Porayko}\ \emph {et~al.}(2018)\citenamefont {Porayko}
  \emph {et~al.}}]{Porayko:2018sfa}%
  \BibitemOpen
  \bibfield  {author} {\bibinfo {author} {\bibfnamefont {N.~K.}\ \bibnamefont
  {Porayko}} \emph {et~al.},\ }\bibfield  {title} {\bibinfo {title} {{Parkes
  Pulsar Timing Array constraints on ultralight scalar-field dark matter}},\
  }\href {https://doi.org/10.1103/PhysRevD.98.102002} {\bibfield  {journal}
  {\bibinfo  {journal} {Phys. Rev. D}\ }\textbf {\bibinfo {volume} {98}},\
  \bibinfo {pages} {102002} (\bibinfo {year} {2018})},\ \Eprint
  {https://arxiv.org/abs/1810.03227} {arXiv:1810.03227 [astro-ph.CO]}
  \BibitemShut {NoStop}%
\bibitem [{\citenamefont {Kato}\ and\ \citenamefont
  {Soda}(2020)}]{Kato:2019bqz}%
  \BibitemOpen
  \bibfield  {author} {\bibinfo {author} {\bibfnamefont {R.}~\bibnamefont
  {Kato}}\ and\ \bibinfo {author} {\bibfnamefont {J.}~\bibnamefont {Soda}},\
  }\bibfield  {title} {\bibinfo {title} {{Search for ultralight scalar dark
  matter with pulsar timing arrays}},\ }\href
  {https://doi.org/10.1088/1475-7516/2020/09/036} {\bibfield  {journal}
  {\bibinfo  {journal} {JCAP}\ }\textbf {\bibinfo {volume} {09}},\ \bibinfo
  {pages} {036}},\ \Eprint {https://arxiv.org/abs/1904.09143} {arXiv:1904.09143
  [astro-ph.HE]} \BibitemShut {NoStop}%
\bibitem [{\citenamefont {Nomura}\ \emph {et~al.}(2020)\citenamefont {Nomura},
  \citenamefont {Ito},\ and\ \citenamefont {Soda}}]{Nomura:2019cvc}%
  \BibitemOpen
  \bibfield  {author} {\bibinfo {author} {\bibfnamefont {K.}~\bibnamefont
  {Nomura}}, \bibinfo {author} {\bibfnamefont {A.}~\bibnamefont {Ito}},\ and\
  \bibinfo {author} {\bibfnamefont {J.}~\bibnamefont {Soda}},\ }\bibfield
  {title} {\bibinfo {title} {{Pulsar timing residual induced by ultralight
  vector dark matter}},\ }\href
  {https://doi.org/10.1140/epjc/s10052-020-7990-y} {\bibfield  {journal}
  {\bibinfo  {journal} {Eur. Phys. J. C}\ }\textbf {\bibinfo {volume} {80}},\
  \bibinfo {pages} {419} (\bibinfo {year} {2020})},\ \Eprint
  {https://arxiv.org/abs/1912.10210} {arXiv:1912.10210 [gr-qc]} \BibitemShut
  {NoStop}%
\bibitem [{\citenamefont {Kaplan}\ \emph {et~al.}(2022)\citenamefont {Kaplan},
  \citenamefont {Mitridate},\ and\ \citenamefont {Trickle}}]{Kaplan:2022lmz}%
  \BibitemOpen
  \bibfield  {author} {\bibinfo {author} {\bibfnamefont {D.~E.}\ \bibnamefont
  {Kaplan}}, \bibinfo {author} {\bibfnamefont {A.}~\bibnamefont {Mitridate}},\
  and\ \bibinfo {author} {\bibfnamefont {T.}~\bibnamefont {Trickle}},\
  }\bibfield  {title} {\bibinfo {title} {{Constraining fundamental constant
  variations from ultralight dark matter with pulsar timing arrays}},\ }\href
  {https://doi.org/10.1103/PhysRevD.106.035032} {\bibfield  {journal} {\bibinfo
   {journal} {Phys. Rev. D}\ }\textbf {\bibinfo {volume} {106}},\ \bibinfo
  {pages} {035032} (\bibinfo {year} {2022})},\ \Eprint
  {https://arxiv.org/abs/2205.06817} {arXiv:2205.06817 [hep-ph]} \BibitemShut
  {NoStop}%
\bibitem [{\citenamefont {Unal}\ \emph {et~al.}(2024)\citenamefont {Unal},
  \citenamefont {Urban},\ and\ \citenamefont {Kovetz}}]{Unal:2022ooa}%
  \BibitemOpen
  \bibfield  {author} {\bibinfo {author} {\bibfnamefont {C.}~\bibnamefont
  {Unal}}, \bibinfo {author} {\bibfnamefont {F.~R.}\ \bibnamefont {Urban}},\
  and\ \bibinfo {author} {\bibfnamefont {E.~D.}\ \bibnamefont {Kovetz}},\
  }\bibfield  {title} {\bibinfo {title} {{Probing ultralight scalar, vector and
  tensor dark matter with pulsar timing arrays}},\ }\href
  {https://doi.org/10.1016/j.physletb.2024.138830} {\bibfield  {journal}
  {\bibinfo  {journal} {Phys. Lett. B}\ }\textbf {\bibinfo {volume} {855}},\
  \bibinfo {pages} {138830} (\bibinfo {year} {2024})},\ \Eprint
  {https://arxiv.org/abs/2209.02741} {arXiv:2209.02741 [astro-ph.CO]}
  \BibitemShut {NoStop}%
\bibitem [{\citenamefont {Kim}\ and\ \citenamefont
  {Mitridate}(2024)}]{Kim:2023kyy}%
  \BibitemOpen
  \bibfield  {author} {\bibinfo {author} {\bibfnamefont {H.}~\bibnamefont
  {Kim}}\ and\ \bibinfo {author} {\bibfnamefont {A.}~\bibnamefont
  {Mitridate}},\ }\bibfield  {title} {\bibinfo {title} {{Stochastic ultralight
  dark matter fluctuations in pulsar timing arrays}},\ }\href
  {https://doi.org/10.1103/PhysRevD.109.055017} {\bibfield  {journal} {\bibinfo
   {journal} {Phys. Rev. D}\ }\textbf {\bibinfo {volume} {109}},\ \bibinfo
  {pages} {055017} (\bibinfo {year} {2024})},\ \Eprint
  {https://arxiv.org/abs/2312.12225} {arXiv:2312.12225 [hep-ph]} \BibitemShut
  {NoStop}%
\bibitem [{\citenamefont {Eberhardt}\ \emph {et~al.}(2025)\citenamefont
  {Eberhardt}, \citenamefont {Liang},\ and\ \citenamefont
  {Ferreira}}]{Eberhardt:2024ocm}%
  \BibitemOpen
  \bibfield  {author} {\bibinfo {author} {\bibfnamefont {A.}~\bibnamefont
  {Eberhardt}}, \bibinfo {author} {\bibfnamefont {Q.}~\bibnamefont {Liang}},\
  and\ \bibinfo {author} {\bibfnamefont {E.~G.~M.}\ \bibnamefont {Ferreira}},\
  }\bibfield  {title} {\bibinfo {title} {{Simulations of Shapiro,
  Gravitational, and Doppler time delays in pulsar networks for ultralight dark
  matter}},\ }\href {https://doi.org/10.1103/jwy2-lpd8} {\bibfield  {journal}
  {\bibinfo  {journal} {Phys. Rev. D}\ }\textbf {\bibinfo {volume} {112}},\
  \bibinfo {pages} {123036} (\bibinfo {year} {2025})},\ \Eprint
  {https://arxiv.org/abs/2411.18051} {arXiv:2411.18051 [astro-ph.CO]}
  \BibitemShut {NoStop}%
\bibitem [{\citenamefont {Kim}(2023)}]{Kim:2023pkx}%
  \BibitemOpen
  \bibfield  {author} {\bibinfo {author} {\bibfnamefont {H.}~\bibnamefont
  {Kim}},\ }\bibfield  {title} {\bibinfo {title} {{Gravitational interaction of
  ultralight dark matter with interferometers}},\ }\href
  {https://doi.org/10.1088/1475-7516/2023/12/018} {\bibfield  {journal}
  {\bibinfo  {journal} {JCAP}\ }\textbf {\bibinfo {volume} {12}},\ \bibinfo
  {pages} {018}},\ \Eprint {https://arxiv.org/abs/2306.13348} {arXiv:2306.13348
  [hep-ph]} \BibitemShut {NoStop}%
\bibitem [{\citenamefont {Xia}\ \emph {et~al.}(2023)\citenamefont {Xia},
  \citenamefont {Tang}, \citenamefont {Huang}, \citenamefont {Yuan},\ and\
  \citenamefont {Fan}}]{Xia:2023hov}%
  \BibitemOpen
  \bibfield  {author} {\bibinfo {author} {\bibfnamefont {Z.-Q.}\ \bibnamefont
  {Xia}}, \bibinfo {author} {\bibfnamefont {T.-P.}\ \bibnamefont {Tang}},
  \bibinfo {author} {\bibfnamefont {X.}~\bibnamefont {Huang}}, \bibinfo
  {author} {\bibfnamefont {Q.}~\bibnamefont {Yuan}},\ and\ \bibinfo {author}
  {\bibfnamefont {Y.-Z.}\ \bibnamefont {Fan}},\ }\bibfield  {title} {\bibinfo
  {title} {{Constraining ultralight dark matter using the Fermi-LAT pulsar
  timing array}},\ }\href {https://doi.org/10.1103/PhysRevD.107.L121302}
  {\bibfield  {journal} {\bibinfo  {journal} {Phys. Rev. D}\ }\textbf {\bibinfo
  {volume} {107}},\ \bibinfo {pages} {L121302} (\bibinfo {year} {2023})},\
  \Eprint {https://arxiv.org/abs/2303.17545} {arXiv:2303.17545 [astro-ph.HE]}
  \BibitemShut {NoStop}%
\bibitem [{\citenamefont {Luu}\ \emph {et~al.}(2024)\citenamefont {Luu},
  \citenamefont {Liu}, \citenamefont {Ren}, \citenamefont {Broadhurst},
  \citenamefont {Yang}, \citenamefont {Wang},\ and\ \citenamefont
  {Xie}}]{Luu:2023rgg}%
  \BibitemOpen
  \bibfield  {author} {\bibinfo {author} {\bibfnamefont {H.~N.}\ \bibnamefont
  {Luu}}, \bibinfo {author} {\bibfnamefont {T.}~\bibnamefont {Liu}}, \bibinfo
  {author} {\bibfnamefont {J.}~\bibnamefont {Ren}}, \bibinfo {author}
  {\bibfnamefont {T.}~\bibnamefont {Broadhurst}}, \bibinfo {author}
  {\bibfnamefont {R.}~\bibnamefont {Yang}}, \bibinfo {author} {\bibfnamefont
  {J.-S.}\ \bibnamefont {Wang}},\ and\ \bibinfo {author} {\bibfnamefont
  {Z.}~\bibnamefont {Xie}},\ }\bibfield  {title} {\bibinfo {title} {{Stochastic
  Wave Dark Matter with Fermi-LAT {\ensuremath{\gamma}}-Ray Pulsar Timing
  Array}},\ }\href {https://doi.org/10.3847/2041-8213/ad2ae2} {\bibfield
  {journal} {\bibinfo  {journal} {Astrophys. J. Lett.}\ }\textbf {\bibinfo
  {volume} {963}},\ \bibinfo {pages} {L46} (\bibinfo {year} {2024})},\ \Eprint
  {https://arxiv.org/abs/2304.04735} {arXiv:2304.04735 [astro-ph.HE]}
  \BibitemShut {NoStop}%
\bibitem [{\citenamefont {Hwang}\ \emph {et~al.}(2024)\citenamefont {Hwang},
  \citenamefont {Jeong}, \citenamefont {Noh},\ and\ \citenamefont
  {Smarra}}]{Hwang:2023odi}%
  \BibitemOpen
  \bibfield  {author} {\bibinfo {author} {\bibfnamefont {J.-c.}\ \bibnamefont
  {Hwang}}, \bibinfo {author} {\bibfnamefont {D.}~\bibnamefont {Jeong}},
  \bibinfo {author} {\bibfnamefont {H.}~\bibnamefont {Noh}},\ and\ \bibinfo
  {author} {\bibfnamefont {C.}~\bibnamefont {Smarra}},\ }\bibfield  {title}
  {\bibinfo {title} {{Pulsar Timing Array signature from oscillating metric
  perturbations due to ultra-light axion}},\ }\href
  {https://doi.org/10.1088/1475-7516/2024/02/014} {\bibfield  {journal}
  {\bibinfo  {journal} {JCAP}\ }\textbf {\bibinfo {volume} {02}},\ \bibinfo
  {pages} {014}},\ \Eprint {https://arxiv.org/abs/2311.00234} {arXiv:2311.00234
  [astro-ph.CO]} \BibitemShut {NoStop}%
\bibitem [{\citenamefont {Smarra}\ \emph {et~al.}(2023)\citenamefont {Smarra}
  \emph {et~al.}}]{EuropeanPulsarTimingArray:2023egv}%
  \BibitemOpen
  \bibfield  {author} {\bibinfo {author} {\bibfnamefont {C.}~\bibnamefont
  {Smarra}} \emph {et~al.} (\bibinfo {collaboration} {European Pulsar Timing
  Array}),\ }\bibfield  {title} {\bibinfo {title} {{Second Data Release from
  the European Pulsar Timing Array: Challenging the Ultralight Dark Matter
  Paradigm}},\ }\href {https://doi.org/10.1103/PhysRevLett.131.171001}
  {\bibfield  {journal} {\bibinfo  {journal} {Phys. Rev. Lett.}\ }\textbf
  {\bibinfo {volume} {131}},\ \bibinfo {pages} {171001} (\bibinfo {year}
  {2023})},\ \Eprint {https://arxiv.org/abs/2306.16228} {arXiv:2306.16228
  [astro-ph.HE]} \BibitemShut {NoStop}%
\bibitem [{\citenamefont {Boddy}\ \emph {et~al.}(2025)\citenamefont {Boddy},
  \citenamefont {Dror},\ and\ \citenamefont {Lam}}]{Boddy:2025oxn}%
  \BibitemOpen
  \bibfield  {author} {\bibinfo {author} {\bibfnamefont {K.~K.}\ \bibnamefont
  {Boddy}}, \bibinfo {author} {\bibfnamefont {J.~A.}\ \bibnamefont {Dror}},\
  and\ \bibinfo {author} {\bibfnamefont {A.}~\bibnamefont {Lam}},\ }\bibfield
  {title} {\bibinfo {title} {{Ultralight Dark Matter Statistics for Pulsar
  Timing Detection}},\ }\href {https://doi.org/10.1103/hgnx-w1dn} {\bibfield
  {journal} {\bibinfo  {journal} {Phys. Rev. Lett.}\ }\textbf {\bibinfo
  {volume} {135}},\ \bibinfo {pages} {101001} (\bibinfo {year} {2025})},\
  \Eprint {https://arxiv.org/abs/2502.15874} {arXiv:2502.15874 [hep-ph]}
  \BibitemShut {NoStop}%
\bibitem [{\citenamefont {Gan}\ \emph {et~al.}(2025)\citenamefont {Gan},
  \citenamefont {Kim},\ and\ \citenamefont {Mitridate}}]{Gan:2025icr}%
  \BibitemOpen
  \bibfield  {author} {\bibinfo {author} {\bibfnamefont {X.}~\bibnamefont
  {Gan}}, \bibinfo {author} {\bibfnamefont {H.}~\bibnamefont {Kim}},\ and\
  \bibinfo {author} {\bibfnamefont {A.}~\bibnamefont {Mitridate}},\ }\bibfield
  {title} {\bibinfo {title} {{Probing Quadratically Coupled Ultralight Dark
  Matter with Pulsar Timing Arrays}},\ }\href@noop {} {\  (\bibinfo {year}
  {2025})},\ \Eprint {https://arxiv.org/abs/2510.13945} {arXiv:2510.13945
  [hep-ph]} \BibitemShut {NoStop}%
\bibitem [{\citenamefont {Dror}\ and\ \citenamefont
  {Wei}(2025)}]{Dror:2025nvg}%
  \BibitemOpen
  \bibfield  {author} {\bibinfo {author} {\bibfnamefont {J.~A.}\ \bibnamefont
  {Dror}}\ and\ \bibinfo {author} {\bibfnamefont {Q.}~\bibnamefont {Wei}},\
  }\bibfield  {title} {\bibinfo {title} {{Pulsar timing detection of ultralight
  vector dark matter}},\ }\href {https://doi.org/10.1103/hh8p-gmxl} {\bibfield
  {journal} {\bibinfo  {journal} {Phys. Rev. D}\ }\textbf {\bibinfo {volume}
  {112}},\ \bibinfo {pages} {075024} (\bibinfo {year} {2025})},\ \Eprint
  {https://arxiv.org/abs/2505.22719} {arXiv:2505.22719 [hep-ph]} \BibitemShut
  {NoStop}%
\bibitem [{\citenamefont {Foster}\ \emph
  {et~al.}(2026{\natexlab{b}})\citenamefont {Foster}, \citenamefont {Boddy},
  \citenamefont {Dror}, \citenamefont {Lee}, \citenamefont {Mitridate},
  \citenamefont {Smith}, \citenamefont {Taylor},\ and\ \citenamefont
  {Trickle}}]{Foster:2026mvs}%
  \BibitemOpen
  \bibfield  {author} {\bibinfo {author} {\bibfnamefont {J.~W.}\ \bibnamefont
  {Foster}}, \bibinfo {author} {\bibfnamefont {K.~K.}\ \bibnamefont {Boddy}},
  \bibinfo {author} {\bibfnamefont {J.~A.}\ \bibnamefont {Dror}}, \bibinfo
  {author} {\bibfnamefont {V.~S.~H.}\ \bibnamefont {Lee}}, \bibinfo {author}
  {\bibfnamefont {A.}~\bibnamefont {Mitridate}}, \bibinfo {author}
  {\bibfnamefont {T.~L.}\ \bibnamefont {Smith}}, \bibinfo {author}
  {\bibfnamefont {K.}~\bibnamefont {Taylor}},\ and\ \bibinfo {author}
  {\bibfnamefont {T.}~\bibnamefont {Trickle}},\ }\bibfield  {title} {\bibinfo
  {title} {{Correlated signals of ultralight scalar dark matter in pulsar
  timing}},\ }\href@noop {} {\  (\bibinfo {year} {2026}{\natexlab{b}})},\
  \Eprint {https://arxiv.org/abs/2607.24912} {arXiv:2607.24912 [astro-ph.CO]}
  \BibitemShut {NoStop}%
\bibitem [{\citenamefont {Agazie}\ \emph {et~al.}(2023)\citenamefont {Agazie}
  \emph {et~al.}}]{NANOGrav:2023gor}%
  \BibitemOpen
  \bibfield  {author} {\bibinfo {author} {\bibfnamefont {G.}~\bibnamefont
  {Agazie}} \emph {et~al.} (\bibinfo {collaboration} {NANOGrav}),\ }\bibfield
  {title} {\bibinfo {title} {{The NANOGrav 15 yr Data Set: Evidence for a
  Gravitational-wave Background}},\ }\href
  {https://doi.org/10.3847/2041-8213/acdac6} {\bibfield  {journal} {\bibinfo
  {journal} {Astrophys. J. Lett.}\ }\textbf {\bibinfo {volume} {951}},\
  \bibinfo {pages} {L8} (\bibinfo {year} {2023})},\ \Eprint
  {https://arxiv.org/abs/2306.16213} {arXiv:2306.16213 [astro-ph.HE]}
  \BibitemShut {NoStop}%
\bibitem [{\citenamefont {Afzal}\ \emph {et~al.}(2023)\citenamefont {Afzal}
  \emph {et~al.}}]{NANOGrav:2023hvm}%
  \BibitemOpen
  \bibfield  {author} {\bibinfo {author} {\bibfnamefont {A.}~\bibnamefont
  {Afzal}} \emph {et~al.} (\bibinfo {collaboration} {NANOGrav}),\ }\bibfield
  {title} {\bibinfo {title} {{The NANOGrav 15 yr Data Set: Search for Signals
  from New Physics}},\ }\href {https://doi.org/10.3847/2041-8213/acdc91}
  {\bibfield  {journal} {\bibinfo  {journal} {Astrophys. J. Lett.}\ }\textbf
  {\bibinfo {volume} {951}},\ \bibinfo {pages} {L11} (\bibinfo {year}
  {2023})},\ \Eprint {https://arxiv.org/abs/2306.16219} {arXiv:2306.16219
  [astro-ph.HE]} \BibitemShut {NoStop}%
\bibitem [{\citenamefont {Mitridate}\ \emph {et~al.}(2023)\citenamefont
  {Mitridate}, \citenamefont {Wright}, \citenamefont {von Eckardstein},
  \citenamefont {Schr{\"o}der}, \citenamefont {Nay}, \citenamefont {Olum},
  \citenamefont {Schmitz},\ and\ \citenamefont {Trickle}}]{Mitridate:2023oar}%
  \BibitemOpen
  \bibfield  {author} {\bibinfo {author} {\bibfnamefont {A.}~\bibnamefont
  {Mitridate}}, \bibinfo {author} {\bibfnamefont {D.}~\bibnamefont {Wright}},
  \bibinfo {author} {\bibfnamefont {R.}~\bibnamefont {von Eckardstein}},
  \bibinfo {author} {\bibfnamefont {T.}~\bibnamefont {Schr{\"o}der}}, \bibinfo
  {author} {\bibfnamefont {J.}~\bibnamefont {Nay}}, \bibinfo {author}
  {\bibfnamefont {K.}~\bibnamefont {Olum}}, \bibinfo {author} {\bibfnamefont
  {K.}~\bibnamefont {Schmitz}},\ and\ \bibinfo {author} {\bibfnamefont
  {T.}~\bibnamefont {Trickle}},\ }\bibfield  {title} {\bibinfo {title}
  {{PTArcade}},\ }\href@noop {} {\  (\bibinfo {year} {2023})},\ \Eprint
  {https://arxiv.org/abs/2306.16377} {arXiv:2306.16377 [hep-ph]} \BibitemShut
  {NoStop}%
\bibitem [{\citenamefont {Maggiore}(2007)}]{Maggiore:2007ulw}%
  \BibitemOpen
  \bibfield  {author} {\bibinfo {author} {\bibfnamefont {M.}~\bibnamefont
  {Maggiore}},\ }\href
  {https://doi.org/10.1093/acprof:oso/9780198570745.001.0001} {\emph {\bibinfo
  {title} {{Gravitational Waves. Vol. 1: Theory and Experiments}}}}\ (\bibinfo
  {publisher} {Oxford University Press},\ \bibinfo {year} {2007})\BibitemShut
  {NoStop}%
\bibitem [{\citenamefont {Creighton}\ \emph {et~al.}(2009)\citenamefont
  {Creighton}, \citenamefont {Jenet},\ and\ \citenamefont
  {Price}}]{Creighton:2008bu}%
  \BibitemOpen
  \bibfield  {author} {\bibinfo {author} {\bibfnamefont {T.}~\bibnamefont
  {Creighton}}, \bibinfo {author} {\bibfnamefont {F.~A.}\ \bibnamefont
  {Jenet}},\ and\ \bibinfo {author} {\bibfnamefont {R.~H.}\ \bibnamefont
  {Price}},\ }\bibfield  {title} {\bibinfo {title} {{Pulsar timing and
  spacetime curvature}},\ }\href {https://doi.org/10.1088/0004-637X/693/2/1113}
  {\bibfield  {journal} {\bibinfo  {journal} {Astrophys. J.}\ }\textbf
  {\bibinfo {volume} {693}},\ \bibinfo {pages} {1113} (\bibinfo {year}
  {2009})},\ \Eprint {https://arxiv.org/abs/0812.3941} {arXiv:0812.3941
  [astro-ph]} \BibitemShut {NoStop}%
\bibitem [{\citenamefont {Magi}\ and\ \citenamefont
  {Yoo}(2026{\natexlab{a}})}]{Magi:2026occ}%
  \BibitemOpen
  \bibfield  {author} {\bibinfo {author} {\bibfnamefont {M.}~\bibnamefont
  {Magi}}\ and\ \bibinfo {author} {\bibfnamefont {J.}~\bibnamefont {Yoo}},\
  }\bibfield  {title} {\bibinfo {title} {{Exact Equivalence of the Observed
  Redshift and the Pulsar Timing Modulation in the Infinitesimal-Pulse
  Limit}},\ }\href@noop {} {\  (\bibinfo {year} {2026}{\natexlab{a}})},\
  \Eprint {https://arxiv.org/abs/2608.23684} {arXiv:2608.23684 [astro-ph.CO]}
  \BibitemShut {NoStop}%
\bibitem [{\citenamefont {Magi}\ and\ \citenamefont
  {Yoo}(2026{\natexlab{b}})}]{Magi:2026upf}%
  \BibitemOpen
  \bibfield  {author} {\bibinfo {author} {\bibfnamefont {M.}~\bibnamefont
  {Magi}}\ and\ \bibinfo {author} {\bibfnamefont {J.}~\bibnamefont {Yoo}},\
  }\bibfield  {title} {\bibinfo {title} {{Complete Second-Order Relativistic
  Derivation of the Observed Pulsar Timing Modulations}},\ }\href@noop {} {\
  (\bibinfo {year} {2026}{\natexlab{b}})},\ \Eprint
  {https://arxiv.org/abs/2609.13364} {arXiv:2609.13364 [astro-ph.CO]}
  \BibitemShut {NoStop}%
\bibitem [{\citenamefont {Lee}\ and\ \citenamefont
  {Zurek}(2025)}]{Lee:2024oxo}%
  \BibitemOpen
  \bibfield  {author} {\bibinfo {author} {\bibfnamefont {V.~S.~H.}\
  \bibnamefont {Lee}}\ and\ \bibinfo {author} {\bibfnamefont {K.~M.}\
  \bibnamefont {Zurek}},\ }\bibfield  {title} {\bibinfo {title} {{Proper time
  observables of general gravitational perturbations in laser
  interferometry-based gravitational wave detectors}},\ }\href
  {https://doi.org/10.1103/6q7d-jz26} {\bibfield  {journal} {\bibinfo
  {journal} {Phys. Rev. D}\ }\textbf {\bibinfo {volume} {111}},\ \bibinfo
  {pages} {124037} (\bibinfo {year} {2025})},\ \Eprint
  {https://arxiv.org/abs/2408.03363} {arXiv:2408.03363 [hep-ph]} \BibitemShut
  {NoStop}%
\bibitem [{\citenamefont {Badurina}\ \emph
  {et~al.}(2025{\natexlab{a}})\citenamefont {Badurina}, \citenamefont {Du},
  \citenamefont {Lee}, \citenamefont {Wang},\ and\ \citenamefont
  {Zurek}}]{Badurina:2024rpp}%
  \BibitemOpen
  \bibfield  {author} {\bibinfo {author} {\bibfnamefont {L.}~\bibnamefont
  {Badurina}}, \bibinfo {author} {\bibfnamefont {Y.}~\bibnamefont {Du}},
  \bibinfo {author} {\bibfnamefont {V.~S.~H.}\ \bibnamefont {Lee}}, \bibinfo
  {author} {\bibfnamefont {Y.}~\bibnamefont {Wang}},\ and\ \bibinfo {author}
  {\bibfnamefont {K.~M.}\ \bibnamefont {Zurek}},\ }\bibfield  {title} {\bibinfo
  {title} {{Signatures of linearized gravity in atom interferometers: A
  simplified computational framework}},\ }\href
  {https://doi.org/10.1103/PhysRevD.111.042002} {\bibfield  {journal} {\bibinfo
   {journal} {Phys. Rev. D}\ }\textbf {\bibinfo {volume} {111}},\ \bibinfo
  {pages} {042002} (\bibinfo {year} {2025}{\natexlab{a}})},\ \Eprint
  {https://arxiv.org/abs/2409.03828} {arXiv:2409.03828 [gr-qc]} \BibitemShut
  {NoStop}%
\bibitem [{\citenamefont {Rakhmanov}(2005)}]{Rakhmanov:2004eh}%
  \BibitemOpen
  \bibfield  {author} {\bibinfo {author} {\bibfnamefont {M.}~\bibnamefont
  {Rakhmanov}},\ }\bibfield  {title} {\bibinfo {title} {{Response of test
  masses to gravitational waves in the local Lorentz gauge}},\ }\href
  {https://doi.org/10.1103/PhysRevD.71.084003} {\bibfield  {journal} {\bibinfo
  {journal} {Phys. Rev. D}\ }\textbf {\bibinfo {volume} {71}},\ \bibinfo
  {pages} {084003} (\bibinfo {year} {2005})},\ \Eprint
  {https://arxiv.org/abs/gr-qc/0406009} {arXiv:gr-qc/0406009} \BibitemShut
  {NoStop}%
\bibitem [{\citenamefont {Verlinde}\ and\ \citenamefont
  {Zurek}(2021)}]{Verlinde:2019xfb}%
  \BibitemOpen
  \bibfield  {author} {\bibinfo {author} {\bibfnamefont {E.~P.}\ \bibnamefont
  {Verlinde}}\ and\ \bibinfo {author} {\bibfnamefont {K.~M.}\ \bibnamefont
  {Zurek}},\ }\bibfield  {title} {\bibinfo {title} {{Observational signatures
  of quantum gravity in interferometers}},\ }\href
  {https://doi.org/10.1016/j.physletb.2021.136663} {\bibfield  {journal}
  {\bibinfo  {journal} {Phys. Lett. B}\ }\textbf {\bibinfo {volume} {822}},\
  \bibinfo {pages} {136663} (\bibinfo {year} {2021})},\ \Eprint
  {https://arxiv.org/abs/1902.08207} {arXiv:1902.08207 [gr-qc]} \BibitemShut
  {NoStop}%
\bibitem [{\citenamefont {Zurek}(2022{\natexlab{a}})}]{Zurek:2020ukz}%
  \BibitemOpen
  \bibfield  {author} {\bibinfo {author} {\bibfnamefont {K.~M.}\ \bibnamefont
  {Zurek}},\ }\bibfield  {title} {\bibinfo {title} {{On vacuum fluctuations in
  quantum gravity and interferometer arm fluctuations}},\ }\href
  {https://doi.org/10.1016/j.physletb.2022.136910} {\bibfield  {journal}
  {\bibinfo  {journal} {Phys. Lett. B}\ }\textbf {\bibinfo {volume} {826}},\
  \bibinfo {pages} {136910} (\bibinfo {year} {2022}{\natexlab{a}})},\ \Eprint
  {https://arxiv.org/abs/2012.05870} {arXiv:2012.05870 [hep-th]} \BibitemShut
  {NoStop}%
\bibitem [{\citenamefont {Zurek}(2022{\natexlab{b}})}]{Zurek:2022xzl}%
  \BibitemOpen
  \bibfield  {author} {\bibinfo {author} {\bibfnamefont {K.~M.}\ \bibnamefont
  {Zurek}},\ }\bibfield  {title} {\bibinfo {title} {{Snowmass 2021 White Paper:
  Observational Signatures of Quantum Gravity}},\ }\href@noop {} {\  (\bibinfo
  {year} {2022}{\natexlab{b}})},\ \Eprint {https://arxiv.org/abs/2205.01799}
  {arXiv:2205.01799 [gr-qc]} \BibitemShut {NoStop}%
\bibitem [{\citenamefont {Li}\ \emph {et~al.}(2023)\citenamefont {Li},
  \citenamefont {Lee}, \citenamefont {Chen},\ and\ \citenamefont
  {Zurek}}]{Li:2022mvy}%
  \BibitemOpen
  \bibfield  {author} {\bibinfo {author} {\bibfnamefont {D.}~\bibnamefont
  {Li}}, \bibinfo {author} {\bibfnamefont {V.~S.~H.}\ \bibnamefont {Lee}},
  \bibinfo {author} {\bibfnamefont {Y.}~\bibnamefont {Chen}},\ and\ \bibinfo
  {author} {\bibfnamefont {K.~M.}\ \bibnamefont {Zurek}},\ }\bibfield  {title}
  {\bibinfo {title} {{Interferometer response to geontropic fluctuations}},\
  }\href {https://doi.org/10.1103/PhysRevD.107.024002} {\bibfield  {journal}
  {\bibinfo  {journal} {Phys. Rev. D}\ }\textbf {\bibinfo {volume} {107}},\
  \bibinfo {pages} {024002} (\bibinfo {year} {2023})},\ \Eprint
  {https://arxiv.org/abs/2209.07543} {arXiv:2209.07543 [gr-qc]} \BibitemShut
  {NoStop}%
\bibitem [{\citenamefont {Du}\ \emph {et~al.}(2023)\citenamefont {Du},
  \citenamefont {Lee}, \citenamefont {Wang},\ and\ \citenamefont
  {Zurek}}]{Du:2023dhk}%
  \BibitemOpen
  \bibfield  {author} {\bibinfo {author} {\bibfnamefont {Y.}~\bibnamefont
  {Du}}, \bibinfo {author} {\bibfnamefont {V.~S.~H.}\ \bibnamefont {Lee}},
  \bibinfo {author} {\bibfnamefont {Y.}~\bibnamefont {Wang}},\ and\ \bibinfo
  {author} {\bibfnamefont {K.~M.}\ \bibnamefont {Zurek}},\ }\bibfield  {title}
  {\bibinfo {title} {{Macroscopic dark matter detection with gravitational wave
  experiments}},\ }\href {https://doi.org/10.1103/PhysRevD.108.122003}
  {\bibfield  {journal} {\bibinfo  {journal} {Phys. Rev. D}\ }\textbf {\bibinfo
  {volume} {108}},\ \bibinfo {pages} {122003} (\bibinfo {year} {2023})},\
  \Eprint {https://arxiv.org/abs/2306.13122} {arXiv:2306.13122 [astro-ph.CO]}
  \BibitemShut {NoStop}%
\bibitem [{\citenamefont {Badurina}\ \emph
  {et~al.}(2025{\natexlab{b}})\citenamefont {Badurina}, \citenamefont {Du},
  \citenamefont {Lee}, \citenamefont {Wang},\ and\ \citenamefont
  {Zurek}}]{Badurina:2025xwl}%
  \BibitemOpen
  \bibfield  {author} {\bibinfo {author} {\bibfnamefont {L.}~\bibnamefont
  {Badurina}}, \bibinfo {author} {\bibfnamefont {Y.}~\bibnamefont {Du}},
  \bibinfo {author} {\bibfnamefont {V.~S.~H.}\ \bibnamefont {Lee}}, \bibinfo
  {author} {\bibfnamefont {Y.}~\bibnamefont {Wang}},\ and\ \bibinfo {author}
  {\bibfnamefont {K.~M.}\ \bibnamefont {Zurek}},\ }\bibfield  {title} {\bibinfo
  {title} {{Detecting gravitational signatures of dark matter with atom
  gradiometers}},\ }\href {https://doi.org/10.1103/xs7b-zgtj} {\bibfield
  {journal} {\bibinfo  {journal} {Phys. Rev. D}\ }\textbf {\bibinfo {volume}
  {112}},\ \bibinfo {pages} {063014} (\bibinfo {year} {2025}{\natexlab{b}})},\
  \Eprint {https://arxiv.org/abs/2505.00781} {arXiv:2505.00781 [hep-ph]}
  \BibitemShut {NoStop}%
\bibitem [{\citenamefont {Liang}\ and\ \citenamefont
  {Trodden}(2021)}]{Liang:2021bct}%
  \BibitemOpen
  \bibfield  {author} {\bibinfo {author} {\bibfnamefont {Q.}~\bibnamefont
  {Liang}}\ and\ \bibinfo {author} {\bibfnamefont {M.}~\bibnamefont
  {Trodden}},\ }\bibfield  {title} {\bibinfo {title} {{Detecting the stochastic
  gravitational wave background from massive gravity with pulsar timing
  arrays}},\ }\href {https://doi.org/10.1103/PhysRevD.104.084052} {\bibfield
  {journal} {\bibinfo  {journal} {Phys. Rev. D}\ }\textbf {\bibinfo {volume}
  {104}},\ \bibinfo {pages} {084052} (\bibinfo {year} {2021})},\ \Eprint
  {https://arxiv.org/abs/2108.05344} {arXiv:2108.05344 [astro-ph.CO]}
  \BibitemShut {NoStop}%
\bibitem [{\citenamefont {Dom{\`e}nech}\ \emph {et~al.}(2026)\citenamefont
  {Dom{\`e}nech}, \citenamefont {Pi},\ and\ \citenamefont
  {Wang}}]{Domenech:2025ccu}%
  \BibitemOpen
  \bibfield  {author} {\bibinfo {author} {\bibfnamefont {G.}~\bibnamefont
  {Dom{\`e}nech}}, \bibinfo {author} {\bibfnamefont {S.}~\bibnamefont {Pi}},\
  and\ \bibinfo {author} {\bibfnamefont {A.}~\bibnamefont {Wang}},\ }\bibfield
  {title} {\bibinfo {title} {{Observable Gravitational Wave Strain at Second
  Order}},\ }\href {https://doi.org/10.1103/pwbs-xwrh} {\bibfield  {journal}
  {\bibinfo  {journal} {Phys. Rev. Lett.}\ }\textbf {\bibinfo {volume} {136}},\
  \bibinfo {pages} {221402} (\bibinfo {year} {2026})},\ \Eprint
  {https://arxiv.org/abs/2512.15704} {arXiv:2512.15704 [gr-qc]} \BibitemShut
  {NoStop}%
\bibitem [{\citenamefont {Lee}\ \emph {et~al.}(2010)\citenamefont {Lee},
  \citenamefont {Jenet}, \citenamefont {Price}, \citenamefont {Wex},\ and\
  \citenamefont {Kramer}}]{Lee:2010cg}%
  \BibitemOpen
  \bibfield  {author} {\bibinfo {author} {\bibfnamefont {K.}~\bibnamefont
  {Lee}}, \bibinfo {author} {\bibfnamefont {F.~A.}\ \bibnamefont {Jenet}},
  \bibinfo {author} {\bibfnamefont {R.~H.}\ \bibnamefont {Price}}, \bibinfo
  {author} {\bibfnamefont {N.}~\bibnamefont {Wex}},\ and\ \bibinfo {author}
  {\bibfnamefont {M.}~\bibnamefont {Kramer}},\ }\bibfield  {title} {\bibinfo
  {title} {{Detecting massive gravitons using pulsar timing arrays}},\ }\href
  {https://doi.org/10.1088/0004-637X/722/2/1589} {\bibfield  {journal}
  {\bibinfo  {journal} {Astrophys. J.}\ }\textbf {\bibinfo {volume} {722}},\
  \bibinfo {pages} {1589} (\bibinfo {year} {2010})},\ \Eprint
  {https://arxiv.org/abs/1008.2561} {arXiv:1008.2561 [astro-ph.HE]}
  \BibitemShut {NoStop}%
\bibitem [{\citenamefont {de~Rham}\ \emph {et~al.}(2017)\citenamefont
  {de~Rham}, \citenamefont {Deskins}, \citenamefont {Tolley},\ and\
  \citenamefont {Zhou}}]{deRham:2016nuf}%
  \BibitemOpen
  \bibfield  {author} {\bibinfo {author} {\bibfnamefont {C.}~\bibnamefont
  {de~Rham}}, \bibinfo {author} {\bibfnamefont {J.~T.}\ \bibnamefont
  {Deskins}}, \bibinfo {author} {\bibfnamefont {A.~J.}\ \bibnamefont
  {Tolley}},\ and\ \bibinfo {author} {\bibfnamefont {S.-Y.}\ \bibnamefont
  {Zhou}},\ }\bibfield  {title} {\bibinfo {title} {{Graviton Mass Bounds}},\
  }\href {https://doi.org/10.1103/RevModPhys.89.025004} {\bibfield  {journal}
  {\bibinfo  {journal} {Rev. Mod. Phys.}\ }\textbf {\bibinfo {volume} {89}},\
  \bibinfo {pages} {025004} (\bibinfo {year} {2017})},\ \Eprint
  {https://arxiv.org/abs/1606.08462} {arXiv:1606.08462 [astro-ph.CO]}
  \BibitemShut {NoStop}%
\bibitem [{\citenamefont {Qin}\ \emph {et~al.}(2021)\citenamefont {Qin},
  \citenamefont {Boddy},\ and\ \citenamefont {Kamionkowski}}]{Qin:2020hfy}%
  \BibitemOpen
  \bibfield  {author} {\bibinfo {author} {\bibfnamefont {W.}~\bibnamefont
  {Qin}}, \bibinfo {author} {\bibfnamefont {K.~K.}\ \bibnamefont {Boddy}},\
  and\ \bibinfo {author} {\bibfnamefont {M.}~\bibnamefont {Kamionkowski}},\
  }\bibfield  {title} {\bibinfo {title} {{Subluminal stochastic gravitational
  waves in pulsar-timing arrays and astrometry}},\ }\href
  {https://doi.org/10.1103/PhysRevD.103.024045} {\bibfield  {journal} {\bibinfo
   {journal} {Phys. Rev. D}\ }\textbf {\bibinfo {volume} {103}},\ \bibinfo
  {pages} {024045} (\bibinfo {year} {2021})},\ \Eprint
  {https://arxiv.org/abs/2007.11009} {arXiv:2007.11009 [gr-qc]} \BibitemShut
  {NoStop}%
\bibitem [{\citenamefont {Wu}\ \emph {et~al.}(2024)\citenamefont {Wu},
  \citenamefont {Chen}, \citenamefont {Bi},\ and\ \citenamefont
  {Huang}}]{Wu:2023rib}%
  \BibitemOpen
  \bibfield  {author} {\bibinfo {author} {\bibfnamefont {Y.-M.}\ \bibnamefont
  {Wu}}, \bibinfo {author} {\bibfnamefont {Z.-C.}\ \bibnamefont {Chen}},
  \bibinfo {author} {\bibfnamefont {Y.-C.}\ \bibnamefont {Bi}},\ and\ \bibinfo
  {author} {\bibfnamefont {Q.-G.}\ \bibnamefont {Huang}},\ }\bibfield  {title}
  {\bibinfo {title} {{Constraining the graviton mass with the NANOGrav
  15{\,}year data set}},\ }\href {https://doi.org/10.1088/1361-6382/ad2a9b}
  {\bibfield  {journal} {\bibinfo  {journal} {Class. Quant. Grav.}\ }\textbf
  {\bibinfo {volume} {41}},\ \bibinfo {pages} {075002} (\bibinfo {year}
  {2024})},\ \Eprint {https://arxiv.org/abs/2310.07469} {arXiv:2310.07469
  [astro-ph.CO]} \BibitemShut {NoStop}%
\bibitem [{\citenamefont {Bi}\ \emph {et~al.}(2024)\citenamefont {Bi},
  \citenamefont {Wu}, \citenamefont {Chen},\ and\ \citenamefont
  {Huang}}]{Bi:2023ewq}%
  \BibitemOpen
  \bibfield  {author} {\bibinfo {author} {\bibfnamefont {Y.-C.}\ \bibnamefont
  {Bi}}, \bibinfo {author} {\bibfnamefont {Y.-M.}\ \bibnamefont {Wu}}, \bibinfo
  {author} {\bibfnamefont {Z.-C.}\ \bibnamefont {Chen}},\ and\ \bibinfo
  {author} {\bibfnamefont {Q.-G.}\ \bibnamefont {Huang}},\ }\bibfield  {title}
  {\bibinfo {title} {{Constraints on the velocity of gravitational waves from
  the NANOGrav 15-year data set}},\ }\href
  {https://doi.org/10.1103/PhysRevD.109.L061101} {\bibfield  {journal}
  {\bibinfo  {journal} {Phys. Rev. D}\ }\textbf {\bibinfo {volume} {109}},\
  \bibinfo {pages} {L061101} (\bibinfo {year} {2024})},\ \Eprint
  {https://arxiv.org/abs/2310.08366} {arXiv:2310.08366 [astro-ph.CO]}
  \BibitemShut {NoStop}%
\bibitem [{\citenamefont {Wu}\ \emph {et~al.}(2023)\citenamefont {Wu},
  \citenamefont {Chen},\ and\ \citenamefont {Huang}}]{Wu:2023pbt}%
  \BibitemOpen
  \bibfield  {author} {\bibinfo {author} {\bibfnamefont {Y.-M.}\ \bibnamefont
  {Wu}}, \bibinfo {author} {\bibfnamefont {Z.-C.}\ \bibnamefont {Chen}},\ and\
  \bibinfo {author} {\bibfnamefont {Q.-G.}\ \bibnamefont {Huang}},\ }\bibfield
  {title} {\bibinfo {title} {{Search for stochastic gravitational-wave
  background from massive gravity in the NANOGrav 12.5-year dataset}},\ }\href
  {https://doi.org/10.1103/PhysRevD.107.042003} {\bibfield  {journal} {\bibinfo
   {journal} {Phys. Rev. D}\ }\textbf {\bibinfo {volume} {107}},\ \bibinfo
  {pages} {042003} (\bibinfo {year} {2023})},\ \Eprint
  {https://arxiv.org/abs/2302.00229} {arXiv:2302.00229 [gr-qc]} \BibitemShut
  {NoStop}%
\bibitem [{\citenamefont {Liang}\ \emph {et~al.}(2023)\citenamefont {Liang},
  \citenamefont {Lin},\ and\ \citenamefont {Trodden}}]{Liang:2023ary}%
  \BibitemOpen
  \bibfield  {author} {\bibinfo {author} {\bibfnamefont {Q.}~\bibnamefont
  {Liang}}, \bibinfo {author} {\bibfnamefont {M.-X.}\ \bibnamefont {Lin}},\
  and\ \bibinfo {author} {\bibfnamefont {M.}~\bibnamefont {Trodden}},\
  }\bibfield  {title} {\bibinfo {title} {{A test of gravity with Pulsar Timing
  Arrays}},\ }\href {https://doi.org/10.1088/1475-7516/2023/11/042} {\bibfield
  {journal} {\bibinfo  {journal} {JCAP}\ }\textbf {\bibinfo {volume} {11}},\
  \bibinfo {pages} {042}},\ \Eprint {https://arxiv.org/abs/2304.02640}
  {arXiv:2304.02640 [astro-ph.CO]} \BibitemShut {NoStop}%
\bibitem [{\citenamefont {Liang}\ \emph {et~al.}(2024)\citenamefont {Liang},
  \citenamefont {Obata},\ and\ \citenamefont {Sasaki}}]{Liang:2024mex}%
  \BibitemOpen
  \bibfield  {author} {\bibinfo {author} {\bibfnamefont {Q.}~\bibnamefont
  {Liang}}, \bibinfo {author} {\bibfnamefont {I.}~\bibnamefont {Obata}},\ and\
  \bibinfo {author} {\bibfnamefont {M.}~\bibnamefont {Sasaki}},\ }\bibfield
  {title} {\bibinfo {title} {{Testing gravity with frequency-dependent overlap
  reduction function in Pulsar Timing Array}},\ }\href
  {https://doi.org/10.1088/1475-7516/2024/10/097} {\bibfield  {journal}
  {\bibinfo  {journal} {JCAP}\ }\textbf {\bibinfo {volume} {10}},\ \bibinfo
  {pages} {097}},\ \Eprint {https://arxiv.org/abs/2405.11755} {arXiv:2405.11755
  [astro-ph.CO]} \BibitemShut {NoStop}%
\bibitem [{\citenamefont {Hu}\ \emph {et~al.}(2024)\citenamefont {Hu},
  \citenamefont {Liang}, \citenamefont {Lin},\ and\ \citenamefont
  {Trodden}}]{Hu:2024wub}%
  \BibitemOpen
  \bibfield  {author} {\bibinfo {author} {\bibfnamefont {W.}~\bibnamefont
  {Hu}}, \bibinfo {author} {\bibfnamefont {Q.}~\bibnamefont {Liang}}, \bibinfo
  {author} {\bibfnamefont {M.-X.}\ \bibnamefont {Lin}},\ and\ \bibinfo {author}
  {\bibfnamefont {M.}~\bibnamefont {Trodden}},\ }\bibfield  {title} {\bibinfo
  {title} {{Testing gravity with realistic gravitational waveforms in Pulsar
  Timing Arrays}},\ }\href {https://doi.org/10.1088/1475-7516/2024/12/054}
  {\bibfield  {journal} {\bibinfo  {journal} {JCAP}\ }\textbf {\bibinfo
  {volume} {12}},\ \bibinfo {pages} {054}},\ \Eprint
  {https://arxiv.org/abs/2408.11774} {arXiv:2408.11774 [astro-ph.CO]}
  \BibitemShut {NoStop}%
\bibitem [{\citenamefont {Cordes}\ \emph {et~al.}(2025)\citenamefont {Cordes},
  \citenamefont {Mitridate}, \citenamefont {Schmitz}, \citenamefont
  {Schr{\"o}der},\ and\ \citenamefont {Wassner}}]{Cordes:2024oem}%
  \BibitemOpen
  \bibfield  {author} {\bibinfo {author} {\bibfnamefont {N.}~\bibnamefont
  {Cordes}}, \bibinfo {author} {\bibfnamefont {A.}~\bibnamefont {Mitridate}},
  \bibinfo {author} {\bibfnamefont {K.}~\bibnamefont {Schmitz}}, \bibinfo
  {author} {\bibfnamefont {T.}~\bibnamefont {Schr{\"o}der}},\ and\ \bibinfo
  {author} {\bibfnamefont {K.}~\bibnamefont {Wassner}},\ }\bibfield  {title}
  {\bibinfo {title} {{On the overlap reduction function of pulsar timing array
  searches for gravitational waves in modified gravity}},\ }\href
  {https://doi.org/10.1088/1361-6382/ad9881} {\bibfield  {journal} {\bibinfo
  {journal} {Class. Quant. Grav.}\ }\textbf {\bibinfo {volume} {42}},\ \bibinfo
  {pages} {015003} (\bibinfo {year} {2025})},\ \Eprint
  {https://arxiv.org/abs/2407.04464} {arXiv:2407.04464 [gr-qc]} \BibitemShut
  {NoStop}%
\bibitem [{\citenamefont {Liang}\ \emph {et~al.}(2026)\citenamefont {Liang},
  \citenamefont {Nomura},\ and\ \citenamefont {Omiya}}]{Liang:2025vji}%
  \BibitemOpen
  \bibfield  {author} {\bibinfo {author} {\bibfnamefont {Q.}~\bibnamefont
  {Liang}}, \bibinfo {author} {\bibfnamefont {K.}~\bibnamefont {Nomura}},\ and\
  \bibinfo {author} {\bibfnamefont {H.}~\bibnamefont {Omiya}},\ }\bibfield
  {title} {\bibinfo {title} {{Detecting parity-violating gravitational wave
  backgrounds with pulsar polarization arrays}},\ }\href
  {https://doi.org/10.1103/m6sq-s53x} {\bibfield  {journal} {\bibinfo
  {journal} {Phys. Rev. D}\ }\textbf {\bibinfo {volume} {113}},\ \bibinfo
  {pages} {064003} (\bibinfo {year} {2026})},\ \Eprint
  {https://arxiv.org/abs/2511.07956} {arXiv:2511.07956 [gr-qc]} \BibitemShut
  {NoStop}%
\bibitem [{\citenamefont {Gr{\'e}e}\ \emph {et~al.}(2026)\citenamefont
  {Gr{\'e}e}, \citenamefont {Liang},\ and\ \citenamefont
  {Ferreira}}]{Gree:2026nlt}%
  \BibitemOpen
  \bibfield  {author} {\bibinfo {author} {\bibfnamefont {J.}~\bibnamefont
  {Gr{\'e}e}}, \bibinfo {author} {\bibfnamefont {Q.}~\bibnamefont {Liang}},\
  and\ \bibinfo {author} {\bibfnamefont {E.~G.~M.}\ \bibnamefont {Ferreira}},\
  }\bibfield  {title} {\bibinfo {title} {{Forecasting Sensitivity to Modified
  Dispersion Effects in Pulsar Timing Arrays}},\ }\href@noop {} {\  (\bibinfo
  {year} {2026})},\ \Eprint {https://arxiv.org/abs/2603.17563}
  {arXiv:2603.17563 [astro-ph.CO]} \BibitemShut {NoStop}%
\bibitem [{\citenamefont {{Helfand}}\ \emph {et~al.}(1975)\citenamefont
  {{Helfand}}, \citenamefont {{Manchester}},\ and\ \citenamefont
  {{Taylor}}}]{1975ApJ...198..661H}%
  \BibitemOpen
  \bibfield  {author} {\bibinfo {author} {\bibfnamefont {D.~J.}\ \bibnamefont
  {{Helfand}}}, \bibinfo {author} {\bibfnamefont {R.~N.}\ \bibnamefont
  {{Manchester}}},\ and\ \bibinfo {author} {\bibfnamefont {J.~H.}\ \bibnamefont
  {{Taylor}}},\ }\bibfield  {title} {\bibinfo {title} {{Observations of pulsar
  radio emission. III. Stability of integrated profiles.}},\ }\href
  {https://doi.org/10.1086/153644} {\bibfield  {journal} {\bibinfo  {journal}
  {\apj}\ }\textbf {\bibinfo {volume} {198}},\ \bibinfo {pages} {661} (\bibinfo
  {year} {1975})}\BibitemShut {NoStop}%
\bibitem [{\citenamefont {Hobbs}\ \emph {et~al.}(2006)\citenamefont {Hobbs},
  \citenamefont {Edwards},\ and\ \citenamefont {Manchester}}]{Hobbs:2006cd}%
  \BibitemOpen
  \bibfield  {author} {\bibinfo {author} {\bibfnamefont {G.}~\bibnamefont
  {Hobbs}}, \bibinfo {author} {\bibfnamefont {R.}~\bibnamefont {Edwards}},\
  and\ \bibinfo {author} {\bibfnamefont {R.}~\bibnamefont {Manchester}},\
  }\bibfield  {title} {\bibinfo {title} {{Tempo2, a new pulsar timing package.
  1. overview}},\ }\href {https://doi.org/10.1111/j.1365-2966.2006.10302.x}
  {\bibfield  {journal} {\bibinfo  {journal} {Mon. Not. Roy. Astron. Soc.}\
  }\textbf {\bibinfo {volume} {369}},\ \bibinfo {pages} {655} (\bibinfo {year}
  {2006})},\ \Eprint {https://arxiv.org/abs/astro-ph/0603381}
  {arXiv:astro-ph/0603381} \BibitemShut {NoStop}%
\bibitem [{\citenamefont {Edwards}\ \emph {et~al.}(2006)\citenamefont
  {Edwards}, \citenamefont {Hobbs},\ and\ \citenamefont
  {Manchester}}]{Edwards:2006zg}%
  \BibitemOpen
  \bibfield  {author} {\bibinfo {author} {\bibfnamefont {R.~T.}\ \bibnamefont
  {Edwards}}, \bibinfo {author} {\bibfnamefont {G.~B.}\ \bibnamefont {Hobbs}},\
  and\ \bibinfo {author} {\bibfnamefont {R.~N.}\ \bibnamefont {Manchester}},\
  }\bibfield  {title} {\bibinfo {title} {{Tempo2, a new pulsar timing package.
  2. The timing model and precision estimates}},\ }\href
  {https://doi.org/10.1111/j.1365-2966.2006.10870.x} {\bibfield  {journal}
  {\bibinfo  {journal} {Mon. Not. Roy. Astron. Soc.}\ }\textbf {\bibinfo
  {volume} {372}},\ \bibinfo {pages} {1549} (\bibinfo {year} {2006})},\ \Eprint
  {https://arxiv.org/abs/astro-ph/0607664} {arXiv:astro-ph/0607664}
  \BibitemShut {NoStop}%
\bibitem [{\citenamefont {{Hobbs}}\ \emph {et~al.}(2009)\citenamefont
  {{Hobbs}}, \citenamefont {{Jenet}}, \citenamefont {{Lee}}, \citenamefont
  {{Verbiest}}, \citenamefont {{Yardley}}, \citenamefont {{Manchester}},
  \citenamefont {{Lommen}}, \citenamefont {{Coles}}, \citenamefont
  {{Edwards}},\ and\ \citenamefont {{Shettigara}}}]{2009MNRAS.394.1945H}%
  \BibitemOpen
  \bibfield  {author} {\bibinfo {author} {\bibfnamefont {G.}~\bibnamefont
  {{Hobbs}}}, \bibinfo {author} {\bibfnamefont {F.}~\bibnamefont {{Jenet}}},
  \bibinfo {author} {\bibfnamefont {K.~J.}\ \bibnamefont {{Lee}}}, \bibinfo
  {author} {\bibfnamefont {J.~P.~W.}\ \bibnamefont {{Verbiest}}}, \bibinfo
  {author} {\bibfnamefont {D.}~\bibnamefont {{Yardley}}}, \bibinfo {author}
  {\bibfnamefont {R.}~\bibnamefont {{Manchester}}}, \bibinfo {author}
  {\bibfnamefont {A.}~\bibnamefont {{Lommen}}}, \bibinfo {author}
  {\bibfnamefont {W.}~\bibnamefont {{Coles}}}, \bibinfo {author} {\bibfnamefont
  {R.}~\bibnamefont {{Edwards}}},\ and\ \bibinfo {author} {\bibfnamefont
  {C.}~\bibnamefont {{Shettigara}}},\ }\bibfield  {title} {\bibinfo {title}
  {{TEMPO2: a new pulsar timing package - III. Gravitational wave
  simulation}},\ }\href {https://doi.org/10.1111/j.1365-2966.2009.14391.x}
  {\bibfield  {journal} {\bibinfo  {journal} {Mon. Not. R. Astron. Soc.}\
  }\textbf {\bibinfo {volume} {394}},\ \bibinfo {pages} {1945} (\bibinfo {year}
  {2009})},\ \Eprint {https://arxiv.org/abs/0901.0592} {arXiv:0901.0592
  [astro-ph.SR]} \BibitemShut {NoStop}%
\bibitem [{\citenamefont {{Luo}}\ \emph {et~al.}(2021)\citenamefont {{Luo}},
  \citenamefont {{Ransom}}, \citenamefont {{Demorest}}, \citenamefont {{Ray}},
  \citenamefont {{Archibald}}, \citenamefont {{Kerr}}, \citenamefont
  {{Jennings}}, \citenamefont {{Bachetti}}, \citenamefont {{van Haasteren}},
  \citenamefont {{Champagne}}, \citenamefont {{Colen}}, \citenamefont
  {{Phillips}}, \citenamefont {{Zimmerman}}, \citenamefont {{Stovall}},
  \citenamefont {{Lam}},\ and\ \citenamefont {{Jenet}}}]{2021ApJ...911...45L}%
  \BibitemOpen
  \bibfield  {author} {\bibinfo {author} {\bibfnamefont {J.}~\bibnamefont
  {{Luo}}}, \bibinfo {author} {\bibfnamefont {S.}~\bibnamefont {{Ransom}}},
  \bibinfo {author} {\bibfnamefont {P.}~\bibnamefont {{Demorest}}}, \bibinfo
  {author} {\bibfnamefont {P.~S.}\ \bibnamefont {{Ray}}}, \bibinfo {author}
  {\bibfnamefont {A.}~\bibnamefont {{Archibald}}}, \bibinfo {author}
  {\bibfnamefont {M.}~\bibnamefont {{Kerr}}}, \bibinfo {author} {\bibfnamefont
  {R.~J.}\ \bibnamefont {{Jennings}}}, \bibinfo {author} {\bibfnamefont
  {M.}~\bibnamefont {{Bachetti}}}, \bibinfo {author} {\bibfnamefont
  {R.}~\bibnamefont {{van Haasteren}}}, \bibinfo {author} {\bibfnamefont
  {C.~A.}\ \bibnamefont {{Champagne}}}, \bibinfo {author} {\bibfnamefont
  {J.}~\bibnamefont {{Colen}}}, \bibinfo {author} {\bibfnamefont
  {C.}~\bibnamefont {{Phillips}}}, \bibinfo {author} {\bibfnamefont
  {J.}~\bibnamefont {{Zimmerman}}}, \bibinfo {author} {\bibfnamefont
  {K.}~\bibnamefont {{Stovall}}}, \bibinfo {author} {\bibfnamefont {M.~T.}\
  \bibnamefont {{Lam}}},\ and\ \bibinfo {author} {\bibfnamefont {F.~A.}\
  \bibnamefont {{Jenet}}},\ }\bibfield  {title} {\bibinfo {title} {{PINT: A
  Modern Software Package for Pulsar Timing}},\ }\href
  {https://doi.org/10.3847/1538-4357/abe62f} {\bibfield  {journal} {\bibinfo
  {journal} {\apj}\ }\textbf {\bibinfo {volume} {911}},\ \bibinfo {eid} {45}
  (\bibinfo {year} {2021})},\ \Eprint {https://arxiv.org/abs/2012.00074}
  {arXiv:2012.00074 [astro-ph.IM]} \BibitemShut {NoStop}%
\bibitem [{\citenamefont {Anderson}\ and\ \citenamefont
  {Itoh}(1975)}]{Anderson:1975zze}%
  \BibitemOpen
  \bibfield  {author} {\bibinfo {author} {\bibfnamefont {P.~W.}\ \bibnamefont
  {Anderson}}\ and\ \bibinfo {author} {\bibfnamefont {N.}~\bibnamefont
  {Itoh}},\ }\bibfield  {title} {\bibinfo {title} {{Pulsar glitches and
  restlessness as a hard superfluidity phenomenon}},\ }\href
  {https://doi.org/10.1038/256025a0} {\bibfield  {journal} {\bibinfo  {journal}
  {Nature}\ }\textbf {\bibinfo {volume} {256}},\ \bibinfo {pages} {25}
  (\bibinfo {year} {1975})}\BibitemShut {NoStop}%
\bibitem [{\citenamefont {{Lyne}}\ \emph {et~al.}(2010)\citenamefont {{Lyne}},
  \citenamefont {{Hobbs}}, \citenamefont {{Kramer}}, \citenamefont {{Stairs}},\
  and\ \citenamefont {{Stappers}}}]{2010Sci...329..408L}%
  \BibitemOpen
  \bibfield  {author} {\bibinfo {author} {\bibfnamefont {A.}~\bibnamefont
  {{Lyne}}}, \bibinfo {author} {\bibfnamefont {G.}~\bibnamefont {{Hobbs}}},
  \bibinfo {author} {\bibfnamefont {M.}~\bibnamefont {{Kramer}}}, \bibinfo
  {author} {\bibfnamefont {I.}~\bibnamefont {{Stairs}}},\ and\ \bibinfo
  {author} {\bibfnamefont {B.}~\bibnamefont {{Stappers}}},\ }\bibfield  {title}
  {\bibinfo {title} {{Switched Magnetospheric Regulation of Pulsar
  Spin-Down}},\ }\href {https://doi.org/10.1126/science.1186683} {\bibfield
  {journal} {\bibinfo  {journal} {Science}\ }\textbf {\bibinfo {volume}
  {329}},\ \bibinfo {pages} {408} (\bibinfo {year} {2010})},\ \Eprint
  {https://arxiv.org/abs/1006.5184} {arXiv:1006.5184 [astro-ph.GA]}
  \BibitemShut {NoStop}%
\bibitem [{\citenamefont {{Hobbs}}\ \emph {et~al.}(2010)\citenamefont
  {{Hobbs}}, \citenamefont {{Lyne}},\ and\ \citenamefont
  {{Kramer}}}]{2010MNRAS.402.1027H}%
  \BibitemOpen
  \bibfield  {author} {\bibinfo {author} {\bibfnamefont {G.}~\bibnamefont
  {{Hobbs}}}, \bibinfo {author} {\bibfnamefont {A.~G.}\ \bibnamefont
  {{Lyne}}},\ and\ \bibinfo {author} {\bibfnamefont {M.}~\bibnamefont
  {{Kramer}}},\ }\bibfield  {title} {\bibinfo {title} {{An analysis of the
  timing irregularities for 366 pulsars}},\ }\href
  {https://doi.org/10.1111/j.1365-2966.2009.15938.x} {\bibfield  {journal}
  {\bibinfo  {journal} {Mon. Not. R. Astron. Soc.}\ }\textbf {\bibinfo {volume}
  {402}},\ \bibinfo {pages} {1027} (\bibinfo {year} {2010})},\ \Eprint
  {https://arxiv.org/abs/0912.4537} {arXiv:0912.4537 [astro-ph.GA]}
  \BibitemShut {NoStop}%
\bibitem [{\citenamefont {{Shannon}}\ and\ \citenamefont
  {{Cordes}}(2010)}]{2010ApJ...725.1607S}%
  \BibitemOpen
  \bibfield  {author} {\bibinfo {author} {\bibfnamefont {R.~M.}\ \bibnamefont
  {{Shannon}}}\ and\ \bibinfo {author} {\bibfnamefont {J.~M.}\ \bibnamefont
  {{Cordes}}},\ }\bibfield  {title} {\bibinfo {title} {{Assessing the Role of
  Spin Noise in the Precision Timing of Millisecond Pulsars}},\ }\href
  {https://doi.org/10.1088/0004-637X/725/2/1607} {\bibfield  {journal}
  {\bibinfo  {journal} {\apj}\ }\textbf {\bibinfo {volume} {725}},\ \bibinfo
  {pages} {1607} (\bibinfo {year} {2010})},\ \Eprint
  {https://arxiv.org/abs/1010.4794} {arXiv:1010.4794 [astro-ph.SR]}
  \BibitemShut {NoStop}%
\bibitem [{\citenamefont {{Espinoza}}\ \emph {et~al.}(2011)\citenamefont
  {{Espinoza}}, \citenamefont {{Lyne}}, \citenamefont {{Stappers}},\ and\
  \citenamefont {{Kramer}}}]{2011MNRAS.414.1679E}%
  \BibitemOpen
  \bibfield  {author} {\bibinfo {author} {\bibfnamefont {C.~M.}\ \bibnamefont
  {{Espinoza}}}, \bibinfo {author} {\bibfnamefont {A.~G.}\ \bibnamefont
  {{Lyne}}}, \bibinfo {author} {\bibfnamefont {B.~W.}\ \bibnamefont
  {{Stappers}}},\ and\ \bibinfo {author} {\bibfnamefont {M.}~\bibnamefont
  {{Kramer}}},\ }\bibfield  {title} {\bibinfo {title} {{A study of 315 glitches
  in the rotation of 102 pulsars}},\ }\href
  {https://doi.org/10.1111/j.1365-2966.2011.18503.x} {\bibfield  {journal}
  {\bibinfo  {journal} {Mon. Not. R. Astron. Soc.}\ }\textbf {\bibinfo {volume}
  {414}},\ \bibinfo {pages} {1679} (\bibinfo {year} {2011})},\ \Eprint
  {https://arxiv.org/abs/1102.1743} {arXiv:1102.1743 [astro-ph.HE]}
  \BibitemShut {NoStop}%
\bibitem [{\citenamefont {{Ellis}}\ \emph {et~al.}(2019)\citenamefont
  {{Ellis}}, \citenamefont {{Vallisneri}}, \citenamefont {{Taylor}},\ and\
  \citenamefont {{Baker}}}]{2019ascl.soft12015E}%
  \BibitemOpen
  \bibfield  {author} {\bibinfo {author} {\bibfnamefont {J.~A.}\ \bibnamefont
  {{Ellis}}}, \bibinfo {author} {\bibfnamefont {M.}~\bibnamefont
  {{Vallisneri}}}, \bibinfo {author} {\bibfnamefont {S.~R.}\ \bibnamefont
  {{Taylor}}},\ and\ \bibinfo {author} {\bibfnamefont {P.~T.}\ \bibnamefont
  {{Baker}}},\ }\href@noop {} {\bibinfo {title} {{ENTERPRISE: Enhanced
  Numerical Toolbox Enabling a Robust PulsaR Inference SuitE}}},\ \bibinfo
  {howpublished} {Astrophysics Source Code Library, record ascl:1912.015}
  (\bibinfo {year} {2019}),\ \Eprint {https://arxiv.org/abs/1912.015}
  {ascl:1912.015} \BibitemShut {NoStop}%
\bibitem [{\citenamefont {Taylor}\ \emph {et~al.}(2021)\citenamefont {Taylor},
  \citenamefont {Baker}, \citenamefont {Hazboun}, \citenamefont {Simon},\ and\
  \citenamefont {Vigeland}}]{enterprise}%
  \BibitemOpen
  \bibfield  {author} {\bibinfo {author} {\bibfnamefont {S.~R.}\ \bibnamefont
  {Taylor}}, \bibinfo {author} {\bibfnamefont {P.~T.}\ \bibnamefont {Baker}},
  \bibinfo {author} {\bibfnamefont {J.~S.}\ \bibnamefont {Hazboun}}, \bibinfo
  {author} {\bibfnamefont {J.}~\bibnamefont {Simon}},\ and\ \bibinfo {author}
  {\bibfnamefont {S.~J.}\ \bibnamefont {Vigeland}},\ }\href
  {https://github.com/nanograv/enterprise_extensions} {\bibinfo {title}
  {enterprise\_extensions}} (\bibinfo {year} {2021}),\ \bibinfo {note}
  {v2.4.3}\BibitemShut {NoStop}%
\bibitem [{\citenamefont {Detweiler}(1979)}]{Detweiler:1979wn}%
  \BibitemOpen
  \bibfield  {author} {\bibinfo {author} {\bibfnamefont {S.~L.}\ \bibnamefont
  {Detweiler}},\ }\bibfield  {title} {\bibinfo {title} {{Pulsar timing
  measurements and the search for gravitational waves}},\ }\href
  {https://doi.org/10.1086/157593} {\bibfield  {journal} {\bibinfo  {journal}
  {Astrophys. J.}\ }\textbf {\bibinfo {volume} {234}},\ \bibinfo {pages} {1100}
  (\bibinfo {year} {1979})}\BibitemShut {NoStop}%
\bibitem [{\citenamefont {Anholm}\ \emph {et~al.}(2009)\citenamefont {Anholm},
  \citenamefont {Ballmer}, \citenamefont {Creighton}, \citenamefont {Price},\
  and\ \citenamefont {Siemens}}]{Anholm:2008wy}%
  \BibitemOpen
  \bibfield  {author} {\bibinfo {author} {\bibfnamefont {M.}~\bibnamefont
  {Anholm}}, \bibinfo {author} {\bibfnamefont {S.}~\bibnamefont {Ballmer}},
  \bibinfo {author} {\bibfnamefont {J.~D.~E.}\ \bibnamefont {Creighton}},
  \bibinfo {author} {\bibfnamefont {L.~R.}\ \bibnamefont {Price}},\ and\
  \bibinfo {author} {\bibfnamefont {X.}~\bibnamefont {Siemens}},\ }\bibfield
  {title} {\bibinfo {title} {{Optimal strategies for gravitational wave
  stochastic background searches in pulsar timing data}},\ }\href
  {https://doi.org/10.1103/PhysRevD.79.084030} {\bibfield  {journal} {\bibinfo
  {journal} {Phys. Rev. D}\ }\textbf {\bibinfo {volume} {79}},\ \bibinfo
  {pages} {084030} (\bibinfo {year} {2009})},\ \Eprint
  {https://arxiv.org/abs/0809.0701} {arXiv:0809.0701 [gr-qc]} \BibitemShut
  {NoStop}%
\bibitem [{\citenamefont {Mingarelli}\ \emph {et~al.}(2013)\citenamefont
  {Mingarelli}, \citenamefont {Sidery}, \citenamefont {Mandel},\ and\
  \citenamefont {Vecchio}}]{Mingarelli:2013dsa}%
  \BibitemOpen
  \bibfield  {author} {\bibinfo {author} {\bibfnamefont {C.~M.~F.}\
  \bibnamefont {Mingarelli}}, \bibinfo {author} {\bibfnamefont
  {T.}~\bibnamefont {Sidery}}, \bibinfo {author} {\bibfnamefont
  {I.}~\bibnamefont {Mandel}},\ and\ \bibinfo {author} {\bibfnamefont
  {A.}~\bibnamefont {Vecchio}},\ }\bibfield  {title} {\bibinfo {title}
  {{Characterizing gravitational wave stochastic background anisotropy with
  pulsar timing arrays}},\ }\href {https://doi.org/10.1103/PhysRevD.88.062005}
  {\bibfield  {journal} {\bibinfo  {journal} {Phys. Rev. D}\ }\textbf {\bibinfo
  {volume} {88}},\ \bibinfo {pages} {062005} (\bibinfo {year} {2013})},\
  \Eprint {https://arxiv.org/abs/1306.5394} {arXiv:1306.5394 [astro-ph.HE]}
  \BibitemShut {NoStop}%
\bibitem [{\citenamefont {Gair}\ \emph {et~al.}(2014)\citenamefont {Gair},
  \citenamefont {Romano}, \citenamefont {Taylor},\ and\ \citenamefont
  {Mingarelli}}]{Gair:2014rwa}%
  \BibitemOpen
  \bibfield  {author} {\bibinfo {author} {\bibfnamefont {J.}~\bibnamefont
  {Gair}}, \bibinfo {author} {\bibfnamefont {J.~D.}\ \bibnamefont {Romano}},
  \bibinfo {author} {\bibfnamefont {S.}~\bibnamefont {Taylor}},\ and\ \bibinfo
  {author} {\bibfnamefont {C.~M.~F.}\ \bibnamefont {Mingarelli}},\ }\bibfield
  {title} {\bibinfo {title} {{Mapping gravitational-wave backgrounds using
  methods from CMB analysis: Application to pulsar timing arrays}},\ }\href
  {https://doi.org/10.1103/PhysRevD.90.082001} {\bibfield  {journal} {\bibinfo
  {journal} {Phys. Rev. D}\ }\textbf {\bibinfo {volume} {90}},\ \bibinfo
  {pages} {082001} (\bibinfo {year} {2014})},\ \Eprint
  {https://arxiv.org/abs/1406.4664} {arXiv:1406.4664 [gr-qc]} \BibitemShut
  {NoStop}%
\bibitem [{\citenamefont {Jenet}\ and\ \citenamefont
  {Romano}(2015)}]{Jenet:2014bea}%
  \BibitemOpen
  \bibfield  {author} {\bibinfo {author} {\bibfnamefont {F.~A.}\ \bibnamefont
  {Jenet}}\ and\ \bibinfo {author} {\bibfnamefont {J.~D.}\ \bibnamefont
  {Romano}},\ }\bibfield  {title} {\bibinfo {title} {{Understanding the
  gravitational-wave Hellings and Downs curve for pulsar timing arrays in terms
  of sound and electromagnetic waves}},\ }\href
  {https://doi.org/10.1119/1.4916358} {\bibfield  {journal} {\bibinfo
  {journal} {Am. J. Phys.}\ }\textbf {\bibinfo {volume} {83}},\ \bibinfo
  {pages} {635} (\bibinfo {year} {2015})},\ \Eprint
  {https://arxiv.org/abs/1412.1142} {arXiv:1412.1142 [gr-qc]} \BibitemShut
  {NoStop}%
\bibitem [{\citenamefont {Mingarelli}\ and\ \citenamefont
  {Sidery}(2014)}]{Mingarelli:2014xfa}%
  \BibitemOpen
  \bibfield  {author} {\bibinfo {author} {\bibfnamefont {C.~M.~F.}\
  \bibnamefont {Mingarelli}}\ and\ \bibinfo {author} {\bibfnamefont
  {T.}~\bibnamefont {Sidery}},\ }\bibfield  {title} {\bibinfo {title} {{Effect
  of small interpulsar distances in stochastic gravitational wave background
  searches with pulsar timing arrays}},\ }\href
  {https://doi.org/10.1103/PhysRevD.90.062011} {\bibfield  {journal} {\bibinfo
  {journal} {Phys. Rev. D}\ }\textbf {\bibinfo {volume} {90}},\ \bibinfo
  {pages} {062011} (\bibinfo {year} {2014})},\ \Eprint
  {https://arxiv.org/abs/1408.6840} {arXiv:1408.6840 [astro-ph.HE]}
  \BibitemShut {NoStop}%
\bibitem [{\citenamefont {Maggiore}(2018)}]{Maggiore:2018sht}%
  \BibitemOpen
  \bibfield  {author} {\bibinfo {author} {\bibfnamefont {M.}~\bibnamefont
  {Maggiore}},\ }\href@noop {} {\emph {\bibinfo {title} {{Gravitational Waves.
  Vol. 2: Astrophysics and Cosmology}}}}\ (\bibinfo  {publisher} {Oxford
  University Press},\ \bibinfo {year} {2018})\BibitemShut {NoStop}%
\bibitem [{\citenamefont {{Fermi}}(1922)}]{1922RendL..31...21F}%
  \BibitemOpen
  \bibfield  {author} {\bibinfo {author} {\bibfnamefont {E.}~\bibnamefont
  {{Fermi}}},\ }\bibfield  {title} {\bibinfo {title} {{Sopra i fenomeni che
  avvengono in vicinanza di una linea oraria}},\ }\href@noop {} {\bibfield
  {journal} {\bibinfo  {journal} {Rend. Lincei}\ }\textbf {\bibinfo {volume}
  {31}},\ \bibinfo {pages} {21} (\bibinfo {year} {1922})}\BibitemShut {NoStop}%
\bibitem [{\citenamefont {Manasse}\ and\ \citenamefont
  {Misner}(1963)}]{Manasse:1963zz}%
  \BibitemOpen
  \bibfield  {author} {\bibinfo {author} {\bibfnamefont {F.~K.}\ \bibnamefont
  {Manasse}}\ and\ \bibinfo {author} {\bibfnamefont {C.~W.}\ \bibnamefont
  {Misner}},\ }\bibfield  {title} {\bibinfo {title} {{Fermi Normal Coordinates
  and Some Basic Concepts in Differential Geometry}},\ }\href
  {https://doi.org/10.1063/1.1724316} {\bibfield  {journal} {\bibinfo
  {journal} {J. Math. Phys.}\ }\textbf {\bibinfo {volume} {4}},\ \bibinfo
  {pages} {735} (\bibinfo {year} {1963})}\BibitemShut {NoStop}%
\bibitem [{\citenamefont {Misner}\ \emph {et~al.}(1973)\citenamefont {Misner},
  \citenamefont {Thorne},\ and\ \citenamefont {Wheeler}}]{Misner:1973prb}%
  \BibitemOpen
  \bibfield  {author} {\bibinfo {author} {\bibfnamefont {C.~W.}\ \bibnamefont
  {Misner}}, \bibinfo {author} {\bibfnamefont {K.~S.}\ \bibnamefont {Thorne}},\
  and\ \bibinfo {author} {\bibfnamefont {J.~A.}\ \bibnamefont {Wheeler}},\
  }\href@noop {} {\emph {\bibinfo {title} {{Gravitation}}}}\ (\bibinfo
  {publisher} {W. H. Freeman},\ \bibinfo {address} {San Francisco},\ \bibinfo
  {year} {1973})\BibitemShut {NoStop}%
\bibitem [{\citenamefont {Ni}\ and\ \citenamefont
  {Zimmermann}(1978)}]{Ni:1978zz}%
  \BibitemOpen
  \bibfield  {author} {\bibinfo {author} {\bibfnamefont {W.-T.}\ \bibnamefont
  {Ni}}\ and\ \bibinfo {author} {\bibfnamefont {M.}~\bibnamefont
  {Zimmermann}},\ }\bibfield  {title} {\bibinfo {title} {{Inertial and
  gravitational effects in the proper reference frame of an accelerated,
  rotating observer}},\ }\href {https://doi.org/10.1103/PhysRevD.17.1473}
  {\bibfield  {journal} {\bibinfo  {journal} {Phys. Rev. D}\ }\textbf {\bibinfo
  {volume} {17}},\ \bibinfo {pages} {1473} (\bibinfo {year}
  {1978})}\BibitemShut {NoStop}%
\bibitem [{\citenamefont {{Li}}\ and\ \citenamefont
  {{Ni}}(1979)}]{1979JMP....20.1473L}%
  \BibitemOpen
  \bibfield  {author} {\bibinfo {author} {\bibfnamefont {W.-Q.}\ \bibnamefont
  {{Li}}}\ and\ \bibinfo {author} {\bibfnamefont {W.-T.}\ \bibnamefont
  {{Ni}}},\ }\bibfield  {title} {\bibinfo {title} {{Coupled inertial and
  gravitational effects in the proper reference frame of an accelerated
  rotating observer.}},\ }\href {https://doi.org/10.1063/1.524203} {\bibfield
  {journal} {\bibinfo  {journal} {Journal of Mathematical Physics}\ }\textbf
  {\bibinfo {volume} {20}},\ \bibinfo {pages} {1473} (\bibinfo {year}
  {1979})}\BibitemShut {NoStop}%
\bibitem [{\citenamefont {Marzlin}(1994)}]{Marzlin:1994ia}%
  \BibitemOpen
  \bibfield  {author} {\bibinfo {author} {\bibfnamefont {K.-P.}\ \bibnamefont
  {Marzlin}},\ }\bibfield  {title} {\bibinfo {title} {{Fermi coordinates for
  weak gravitational fields}},\ }\href
  {https://doi.org/10.1103/PhysRevD.50.888} {\bibfield  {journal} {\bibinfo
  {journal} {Phys. Rev. D}\ }\textbf {\bibinfo {volume} {50}},\ \bibinfo
  {pages} {888} (\bibinfo {year} {1994})},\ \Eprint
  {https://arxiv.org/abs/gr-qc/9403044} {arXiv:gr-qc/9403044} \BibitemShut
  {NoStop}%
\bibitem [{\citenamefont {Rakhmanov}(2014)}]{Rakhmanov:2014noa}%
  \BibitemOpen
  \bibfield  {author} {\bibinfo {author} {\bibfnamefont {M.}~\bibnamefont
  {Rakhmanov}},\ }\bibfield  {title} {\bibinfo {title} {{Fermi-normal, optical,
  and wave-synchronous coordinates for spacetime with a plane gravitational
  wave}},\ }\href {https://doi.org/10.1088/0264-9381/31/8/085006} {\bibfield
  {journal} {\bibinfo  {journal} {Class. Quant. Grav.}\ }\textbf {\bibinfo
  {volume} {31}},\ \bibinfo {pages} {085006} (\bibinfo {year} {2014})},\
  \Eprint {https://arxiv.org/abs/1409.4648} {arXiv:1409.4648 [gr-qc]}
  \BibitemShut {NoStop}%
\bibitem [{\citenamefont {Berlin}\ \emph {et~al.}(2022)\citenamefont {Berlin},
  \citenamefont {Blas}, \citenamefont {Tito~D'Agnolo}, \citenamefont {Ellis},
  \citenamefont {Harnik}, \citenamefont {Kahn},\ and\ \citenamefont
  {Sch{\"u}tte-Engel}}]{Berlin:2021txa}%
  \BibitemOpen
  \bibfield  {author} {\bibinfo {author} {\bibfnamefont {A.}~\bibnamefont
  {Berlin}}, \bibinfo {author} {\bibfnamefont {D.}~\bibnamefont {Blas}},
  \bibinfo {author} {\bibfnamefont {R.}~\bibnamefont {Tito~D'Agnolo}}, \bibinfo
  {author} {\bibfnamefont {S.~A.~R.}\ \bibnamefont {Ellis}}, \bibinfo {author}
  {\bibfnamefont {R.}~\bibnamefont {Harnik}}, \bibinfo {author} {\bibfnamefont
  {Y.}~\bibnamefont {Kahn}},\ and\ \bibinfo {author} {\bibfnamefont
  {J.}~\bibnamefont {Sch{\"u}tte-Engel}},\ }\bibfield  {title} {\bibinfo
  {title} {{Detecting high-frequency gravitational waves with microwave
  cavities}},\ }\href {https://doi.org/10.1103/PhysRevD.105.116011} {\bibfield
  {journal} {\bibinfo  {journal} {Phys. Rev. D}\ }\textbf {\bibinfo {volume}
  {105}},\ \bibinfo {pages} {116011} (\bibinfo {year} {2022})},\ \Eprint
  {https://arxiv.org/abs/2112.11465} {arXiv:2112.11465 [hep-ph]} \BibitemShut
  {NoStop}%
\bibitem [{\citenamefont {{Sazhin}}(1978)}]{1978SvA....22...36S}%
  \BibitemOpen
  \bibfield  {author} {\bibinfo {author} {\bibfnamefont {M.~V.}\ \bibnamefont
  {{Sazhin}}},\ }\bibfield  {title} {\bibinfo {title} {{Opportunities for
  detecting ultralong gravitational waves}},\ }\href@noop {} {\bibfield
  {journal} {\bibinfo  {journal} {Sov. Astron.}\ }\textbf {\bibinfo {volume}
  {22}},\ \bibinfo {pages} {36} (\bibinfo {year} {1978})}\BibitemShut {NoStop}%
\bibitem [{\citenamefont {Hellings}\ and\ \citenamefont
  {Downs}(1983)}]{Hellings:1983fr}%
  \BibitemOpen
  \bibfield  {author} {\bibinfo {author} {\bibfnamefont {R.~w.}\ \bibnamefont
  {Hellings}}\ and\ \bibinfo {author} {\bibfnamefont {G.~s.}\ \bibnamefont
  {Downs}},\ }\bibfield  {title} {\bibinfo {title} {{Upper Limits on the
  Isotropic Gravitational Radiation Background from Pulsar Timing Analysis}},\
  }\href {https://doi.org/10.1086/183954} {\bibfield  {journal} {\bibinfo
  {journal} {Astrophys. J. Lett.}\ }\textbf {\bibinfo {volume} {265}},\
  \bibinfo {pages} {L39} (\bibinfo {year} {1983})}\BibitemShut {NoStop}%
\bibitem [{\citenamefont {Allen}(2023)}]{Allen:2022dzg}%
  \BibitemOpen
  \bibfield  {author} {\bibinfo {author} {\bibfnamefont {B.}~\bibnamefont
  {Allen}},\ }\bibfield  {title} {\bibinfo {title} {{Variance of the
  Hellings-Downs correlation}},\ }\href
  {https://doi.org/10.1103/PhysRevD.107.043018} {\bibfield  {journal} {\bibinfo
   {journal} {Phys. Rev. D}\ }\textbf {\bibinfo {volume} {107}},\ \bibinfo
  {pages} {043018} (\bibinfo {year} {2023})},\ \Eprint
  {https://arxiv.org/abs/2205.05637} {arXiv:2205.05637 [gr-qc]} \BibitemShut
  {NoStop}%
\bibitem [{\citenamefont {Allen}\ and\ \citenamefont
  {Romano}(2023)}]{Allen:2022ksj}%
  \BibitemOpen
  \bibfield  {author} {\bibinfo {author} {\bibfnamefont {B.}~\bibnamefont
  {Allen}}\ and\ \bibinfo {author} {\bibfnamefont {J.~D.}\ \bibnamefont
  {Romano}},\ }\bibfield  {title} {\bibinfo {title} {{Hellings and Downs
  correlation of an arbitrary set of pulsars}},\ }\href
  {https://doi.org/10.1103/PhysRevD.108.043026} {\bibfield  {journal} {\bibinfo
   {journal} {Phys. Rev. D}\ }\textbf {\bibinfo {volume} {108}},\ \bibinfo
  {pages} {043026} (\bibinfo {year} {2023})},\ \Eprint
  {https://arxiv.org/abs/2208.07230} {arXiv:2208.07230 [gr-qc]} \BibitemShut
  {NoStop}%
\bibitem [{\citenamefont {Allen}\ and\ \citenamefont
  {Romano}(1999)}]{Allen:1997ad}%
  \BibitemOpen
  \bibfield  {author} {\bibinfo {author} {\bibfnamefont {B.}~\bibnamefont
  {Allen}}\ and\ \bibinfo {author} {\bibfnamefont {J.~D.}\ \bibnamefont
  {Romano}},\ }\bibfield  {title} {\bibinfo {title} {{Detecting a stochastic
  background of gravitational radiation: Signal processing strategies and
  sensitivities}},\ }\href {https://doi.org/10.1103/PhysRevD.59.102001}
  {\bibfield  {journal} {\bibinfo  {journal} {Phys. Rev. D}\ }\textbf {\bibinfo
  {volume} {59}},\ \bibinfo {pages} {102001} (\bibinfo {year} {1999})},\
  \Eprint {https://arxiv.org/abs/gr-qc/9710117} {arXiv:gr-qc/9710117}
  \BibitemShut {NoStop}%
\bibitem [{\citenamefont {Romano}\ and\ \citenamefont
  {Cornish}(2017)}]{Romano:2016dpx}%
  \BibitemOpen
  \bibfield  {author} {\bibinfo {author} {\bibfnamefont {J.~D.}\ \bibnamefont
  {Romano}}\ and\ \bibinfo {author} {\bibfnamefont {N.~J.}\ \bibnamefont
  {Cornish}},\ }\bibfield  {title} {\bibinfo {title} {{Detection methods for
  stochastic gravitational-wave backgrounds: a unified treatment}},\ }\href
  {https://doi.org/10.1007/s41114-017-0004-1} {\bibfield  {journal} {\bibinfo
  {journal} {Living Rev. Rel.}\ }\textbf {\bibinfo {volume} {20}},\ \bibinfo
  {pages} {2} (\bibinfo {year} {2017})},\ \Eprint
  {https://arxiv.org/abs/1608.06889} {arXiv:1608.06889 [gr-qc]} \BibitemShut
  {NoStop}%
\bibitem [{\citenamefont {Antypas}\ \emph {et~al.}(2022)\citenamefont {Antypas}
  \emph {et~al.}}]{Antypas:2022asj}%
  \BibitemOpen
  \bibfield  {author} {\bibinfo {author} {\bibfnamefont {D.}~\bibnamefont
  {Antypas}} \emph {et~al.},\ }\bibfield  {title} {\bibinfo {title} {{New
  Horizons: Scalar and Vector Ultralight Dark Matter}},\ }\href@noop {} {\
  (\bibinfo {year} {2022})},\ \Eprint {https://arxiv.org/abs/2203.14915}
  {arXiv:2203.14915 [hep-ex]} \BibitemShut {NoStop}%
\bibitem [{\citenamefont {Zel'dovich}\ and\ \citenamefont
  {Polnarev}(1974)}]{Zeldovich:1974gvh}%
  \BibitemOpen
  \bibfield  {author} {\bibinfo {author} {\bibfnamefont {Y.~B.}\ \bibnamefont
  {Zel'dovich}}\ and\ \bibinfo {author} {\bibfnamefont {A.~G.}\ \bibnamefont
  {Polnarev}},\ }\bibfield  {title} {\bibinfo {title} {{Radiation of
  gravitational waves by a cluster of superdense stars}},\ }\href@noop {}
  {\bibfield  {journal} {\bibinfo  {journal} {Sov. Astron.}\ }\textbf {\bibinfo
  {volume} {18}},\ \bibinfo {pages} {17} (\bibinfo {year} {1974})}\BibitemShut
  {NoStop}%
\bibitem [{\citenamefont {Braginsky}\ and\ \citenamefont
  {Grishchuk}(1985)}]{Braginsky:1985vlg}%
  \BibitemOpen
  \bibfield  {author} {\bibinfo {author} {\bibfnamefont {V.~B.}\ \bibnamefont
  {Braginsky}}\ and\ \bibinfo {author} {\bibfnamefont {L.~P.}\ \bibnamefont
  {Grishchuk}},\ }\bibfield  {title} {\bibinfo {title} {{Kinematic Resonance
  and Memory Effect in Free Mass Gravitational Antennas}},\ }\href@noop {}
  {\bibfield  {journal} {\bibinfo  {journal} {Sov. Phys. JETP}\ }\textbf
  {\bibinfo {volume} {62}},\ \bibinfo {pages} {427} (\bibinfo {year}
  {1985})}\BibitemShut {NoStop}%
\bibitem [{\citenamefont {Braginsky}\ and\ \citenamefont
  {Thorne}(1987)}]{Braginsky:1987kwo}%
  \BibitemOpen
  \bibfield  {author} {\bibinfo {author} {\bibfnamefont {V.~B.}\ \bibnamefont
  {Braginsky}}\ and\ \bibinfo {author} {\bibfnamefont {K.~S.}\ \bibnamefont
  {Thorne}},\ }\bibfield  {title} {\bibinfo {title} {{Gravitational-wave bursts
  with memory and experimental prospects}},\ }\href
  {https://doi.org/10.1038/327123a0} {\bibfield  {journal} {\bibinfo  {journal}
  {Nature}\ }\textbf {\bibinfo {volume} {327}},\ \bibinfo {pages} {123}
  (\bibinfo {year} {1987})}\BibitemShut {NoStop}%
\bibitem [{\citenamefont {Christodoulou}(1991)}]{Christodoulou:1991cr}%
  \BibitemOpen
  \bibfield  {author} {\bibinfo {author} {\bibfnamefont {D.}~\bibnamefont
  {Christodoulou}},\ }\bibfield  {title} {\bibinfo {title} {{Nonlinear nature
  of gravitation and gravitational wave experiments}},\ }\href
  {https://doi.org/10.1103/PhysRevLett.67.1486} {\bibfield  {journal} {\bibinfo
   {journal} {Phys. Rev. Lett.}\ }\textbf {\bibinfo {volume} {67}},\ \bibinfo
  {pages} {1486} (\bibinfo {year} {1991})}\BibitemShut {NoStop}%
\bibitem [{\citenamefont {Wiseman}\ and\ \citenamefont
  {Will}(1991)}]{Wiseman:1991ss}%
  \BibitemOpen
  \bibfield  {author} {\bibinfo {author} {\bibfnamefont {A.~G.}\ \bibnamefont
  {Wiseman}}\ and\ \bibinfo {author} {\bibfnamefont {C.~M.}\ \bibnamefont
  {Will}},\ }\bibfield  {title} {\bibinfo {title} {{Christodoulou's nonlinear
  gravitational wave memory: Evaluation in the quadrupole approximation}},\
  }\href {https://doi.org/10.1103/PhysRevD.44.R2945} {\bibfield  {journal}
  {\bibinfo  {journal} {Phys. Rev. D}\ }\textbf {\bibinfo {volume} {44}},\
  \bibinfo {pages} {R2945} (\bibinfo {year} {1991})}\BibitemShut {NoStop}%
\bibitem [{\citenamefont {{Blanchet}}\ and\ \citenamefont
  {{Damour}}(1992)}]{1992PhRvD..46.4304B}%
  \BibitemOpen
  \bibfield  {author} {\bibinfo {author} {\bibfnamefont {L.}~\bibnamefont
  {{Blanchet}}}\ and\ \bibinfo {author} {\bibfnamefont {T.}~\bibnamefont
  {{Damour}}},\ }\bibfield  {title} {\bibinfo {title} {{Hereditary effects in
  gravitational radiation}},\ }\href {https://doi.org/10.1103/PhysRevD.46.4304}
  {\bibfield  {journal} {\bibinfo  {journal} {\prd}\ }\textbf {\bibinfo
  {volume} {46}},\ \bibinfo {pages} {4304} (\bibinfo {year}
  {1992})}\BibitemShut {NoStop}%
\bibitem [{\citenamefont {Thorne}(1992)}]{Thorne:1992sdb}%
  \BibitemOpen
  \bibfield  {author} {\bibinfo {author} {\bibfnamefont {K.~S.}\ \bibnamefont
  {Thorne}},\ }\bibfield  {title} {\bibinfo {title} {{Gravitational-wave bursts
  with memory: The Christodoulou effect}},\ }\href
  {https://doi.org/10.1103/PhysRevD.45.520} {\bibfield  {journal} {\bibinfo
  {journal} {Phys. Rev. D}\ }\textbf {\bibinfo {volume} {45}},\ \bibinfo
  {pages} {520} (\bibinfo {year} {1992})}\BibitemShut {NoStop}%
\bibitem [{\citenamefont {Bieri}\ and\ \citenamefont
  {Garfinkle}(2014)}]{Bieri:2013ada}%
  \BibitemOpen
  \bibfield  {author} {\bibinfo {author} {\bibfnamefont {L.}~\bibnamefont
  {Bieri}}\ and\ \bibinfo {author} {\bibfnamefont {D.}~\bibnamefont
  {Garfinkle}},\ }\bibfield  {title} {\bibinfo {title} {{Perturbative and gauge
  invariant treatment of gravitational wave memory}},\ }\href
  {https://doi.org/10.1103/PhysRevD.89.084039} {\bibfield  {journal} {\bibinfo
  {journal} {Phys. Rev. D}\ }\textbf {\bibinfo {volume} {89}},\ \bibinfo
  {pages} {084039} (\bibinfo {year} {2014})},\ \Eprint
  {https://arxiv.org/abs/1312.6871} {arXiv:1312.6871 [gr-qc]} \BibitemShut
  {NoStop}%
\bibitem [{\citenamefont {Favata}(2009)}]{Favata:2009ii}%
  \BibitemOpen
  \bibfield  {author} {\bibinfo {author} {\bibfnamefont {M.}~\bibnamefont
  {Favata}},\ }\bibfield  {title} {\bibinfo {title} {{Nonlinear
  gravitational-wave memory from binary black hole mergers}},\ }\href
  {https://doi.org/10.1088/0004-637X/696/2/L159} {\bibfield  {journal}
  {\bibinfo  {journal} {Astrophys. J. Lett.}\ }\textbf {\bibinfo {volume}
  {696}},\ \bibinfo {pages} {L159} (\bibinfo {year} {2009})},\ \Eprint
  {https://arxiv.org/abs/0902.3660} {arXiv:0902.3660 [astro-ph.SR]}
  \BibitemShut {NoStop}%
\bibitem [{\citenamefont {Pollney}\ and\ \citenamefont
  {Reisswig}(2011)}]{Pollney:2010hs}%
  \BibitemOpen
  \bibfield  {author} {\bibinfo {author} {\bibfnamefont {D.}~\bibnamefont
  {Pollney}}\ and\ \bibinfo {author} {\bibfnamefont {C.}~\bibnamefont
  {Reisswig}},\ }\bibfield  {title} {\bibinfo {title} {{Gravitational memory in
  binary black hole mergers}},\ }\href
  {https://doi.org/10.1088/2041-8205/732/1/L13} {\bibfield  {journal} {\bibinfo
   {journal} {Astrophys. J. Lett.}\ }\textbf {\bibinfo {volume} {732}},\
  \bibinfo {pages} {L13} (\bibinfo {year} {2011})},\ \Eprint
  {https://arxiv.org/abs/1004.4209} {arXiv:1004.4209 [gr-qc]} \BibitemShut
  {NoStop}%
\bibitem [{\citenamefont {Madison}\ \emph {et~al.}(2014)\citenamefont
  {Madison}, \citenamefont {Cordes},\ and\ \citenamefont
  {Chatterjee}}]{Madison:2014vca}%
  \BibitemOpen
  \bibfield  {author} {\bibinfo {author} {\bibfnamefont {D.~R.}\ \bibnamefont
  {Madison}}, \bibinfo {author} {\bibfnamefont {J.~M.}\ \bibnamefont
  {Cordes}},\ and\ \bibinfo {author} {\bibfnamefont {S.}~\bibnamefont
  {Chatterjee}},\ }\bibfield  {title} {\bibinfo {title} {{Assessing Pulsar
  Timing Array Sensitivity to Gravitational Wave Bursts with Memory}},\ }\href
  {https://doi.org/10.1088/0004-637X/788/2/141} {\bibfield  {journal} {\bibinfo
   {journal} {Astrophys. J.}\ }\textbf {\bibinfo {volume} {788}},\ \bibinfo
  {pages} {141} (\bibinfo {year} {2014})},\ \Eprint
  {https://arxiv.org/abs/1404.5682} {arXiv:1404.5682 [astro-ph.HE]}
  \BibitemShut {NoStop}%
\bibitem [{\citenamefont {{Epstein}}(1978)}]{1978ApJ...223.1037E}%
  \BibitemOpen
  \bibfield  {author} {\bibinfo {author} {\bibfnamefont {R.}~\bibnamefont
  {{Epstein}}},\ }\bibfield  {title} {\bibinfo {title} {{The generation of
  gravitational radiation by escaping supernova neutrinos.}},\ }\href
  {https://doi.org/10.1086/156337} {\bibfield  {journal} {\bibinfo  {journal}
  {\apj}\ }\textbf {\bibinfo {volume} {223}},\ \bibinfo {pages} {1037}
  (\bibinfo {year} {1978})}\BibitemShut {NoStop}%
\bibitem [{\citenamefont {Turner}(1978)}]{Turner:1978jj}%
  \BibitemOpen
  \bibfield  {author} {\bibinfo {author} {\bibfnamefont {M.~S.}\ \bibnamefont
  {Turner}},\ }\bibfield  {title} {\bibinfo {title} {{Gravitational Radiation
  from Supernova Neutrino Bursts}},\ }\href {https://doi.org/10.1038/274565a0}
  {\bibfield  {journal} {\bibinfo  {journal} {Nature}\ }\textbf {\bibinfo
  {volume} {274}},\ \bibinfo {pages} {565} (\bibinfo {year}
  {1978})}\BibitemShut {NoStop}%
\bibitem [{\citenamefont {{Burrows}}\ and\ \citenamefont
  {{Hayes}}(1996)}]{1996PhRvL..76..352B}%
  \BibitemOpen
  \bibfield  {author} {\bibinfo {author} {\bibfnamefont {A.}~\bibnamefont
  {{Burrows}}}\ and\ \bibinfo {author} {\bibfnamefont {J.}~\bibnamefont
  {{Hayes}}},\ }\bibfield  {title} {\bibinfo {title} {{Pulsar Recoil and
  Gravitational Radiation Due to Asymmetrical Stellar Collapse and
  Explosion}},\ }\href {https://doi.org/10.1103/PhysRevLett.76.352} {\bibfield
  {journal} {\bibinfo  {journal} {\prl}\ }\textbf {\bibinfo {volume} {76}},\
  \bibinfo {pages} {352} (\bibinfo {year} {1996})},\ \Eprint
  {https://arxiv.org/abs/astro-ph/9511106} {arXiv:astro-ph/9511106 [astro-ph]}
  \BibitemShut {NoStop}%
\bibitem [{\citenamefont {Strominger}\ and\ \citenamefont
  {Zhiboedov}(2016)}]{Strominger:2014pwa}%
  \BibitemOpen
  \bibfield  {author} {\bibinfo {author} {\bibfnamefont {A.}~\bibnamefont
  {Strominger}}\ and\ \bibinfo {author} {\bibfnamefont {A.}~\bibnamefont
  {Zhiboedov}},\ }\bibfield  {title} {\bibinfo {title} {{Gravitational Memory,
  BMS Supertranslations and Soft Theorems}},\ }\href
  {https://doi.org/10.1007/JHEP01(2016)086} {\bibfield  {journal} {\bibinfo
  {journal} {JHEP}\ }\textbf {\bibinfo {volume} {01}},\ \bibinfo {pages}
  {086}},\ \Eprint {https://arxiv.org/abs/1411.5745} {arXiv:1411.5745 [hep-th]}
  \BibitemShut {NoStop}%
\bibitem [{\citenamefont {Strominger}(2018)}]{Strominger:2017zoo}%
  \BibitemOpen
  \bibfield  {author} {\bibinfo {author} {\bibfnamefont {A.}~\bibnamefont
  {Strominger}},\ }\href@noop {} {\emph {\bibinfo {title} {{Lectures on the
  Infrared Structure of Gravity and Gauge Theory}}}}\ (\bibinfo  {publisher}
  {Princeton University Press},\ \bibinfo {year} {2018})\ \Eprint
  {https://arxiv.org/abs/1703.05448} {arXiv:1703.05448 [hep-th]} \BibitemShut
  {NoStop}%
\bibitem [{\citenamefont {He}\ \emph {et~al.}(2024)\citenamefont {He},
  \citenamefont {Raclariu},\ and\ \citenamefont {Zurek}}]{He:2023qha}%
  \BibitemOpen
  \bibfield  {author} {\bibinfo {author} {\bibfnamefont {T.}~\bibnamefont
  {He}}, \bibinfo {author} {\bibfnamefont {A.-M.}\ \bibnamefont {Raclariu}},\
  and\ \bibinfo {author} {\bibfnamefont {K.~M.}\ \bibnamefont {Zurek}},\
  }\bibfield  {title} {\bibinfo {title} {{From shockwaves to the gravitational
  memory effect}},\ }\href {https://doi.org/10.1007/JHEP01(2024)006} {\bibfield
   {journal} {\bibinfo  {journal} {JHEP}\ }\textbf {\bibinfo {volume} {01}},\
  \bibinfo {pages} {006}},\ \Eprint {https://arxiv.org/abs/2305.14411}
  {arXiv:2305.14411 [hep-th]} \BibitemShut {NoStop}%
\bibitem [{\citenamefont {Verlinde}\ and\ \citenamefont
  {Zurek}(2022)}]{Verlinde:2022hhs}%
  \BibitemOpen
  \bibfield  {author} {\bibinfo {author} {\bibfnamefont {E.}~\bibnamefont
  {Verlinde}}\ and\ \bibinfo {author} {\bibfnamefont {K.~M.}\ \bibnamefont
  {Zurek}},\ }\bibfield  {title} {\bibinfo {title} {{Modular fluctuations from
  shockwave geometries}},\ }\href {https://doi.org/10.1103/PhysRevD.106.106011}
  {\bibfield  {journal} {\bibinfo  {journal} {Phys. Rev. D}\ }\textbf {\bibinfo
  {volume} {106}},\ \bibinfo {pages} {106011} (\bibinfo {year} {2022})},\
  \Eprint {https://arxiv.org/abs/2208.01059} {arXiv:2208.01059 [hep-th]}
  \BibitemShut {NoStop}%
\bibitem [{\citenamefont {He}\ \emph {et~al.}(2025)\citenamefont {He},
  \citenamefont {Raclariu},\ and\ \citenamefont {Zurek}}]{He:2024vlp}%
  \BibitemOpen
  \bibfield  {author} {\bibinfo {author} {\bibfnamefont {T.}~\bibnamefont
  {He}}, \bibinfo {author} {\bibfnamefont {A.-M.}\ \bibnamefont {Raclariu}},\
  and\ \bibinfo {author} {\bibfnamefont {K.~M.}\ \bibnamefont {Zurek}},\
  }\bibfield  {title} {\bibinfo {title} {{An infrared on-shell action and its
  implications for soft charge fluctuations in asymptotically flat
  spacetimes}},\ }\href {https://doi.org/10.1088/1751-8121/adc4a2} {\bibfield
  {journal} {\bibinfo  {journal} {J. Phys. A}\ }\textbf {\bibinfo {volume}
  {58}},\ \bibinfo {pages} {165402} (\bibinfo {year} {2025})},\ \Eprint
  {https://arxiv.org/abs/2408.01485} {arXiv:2408.01485 [hep-th]} \BibitemShut
  {NoStop}%
\bibitem [{\citenamefont {Wang}\ \emph {et~al.}(2015)\citenamefont {Wang} \emph
  {et~al.}}]{Wang:2014zls}%
  \BibitemOpen
  \bibfield  {author} {\bibinfo {author} {\bibfnamefont {J.~B.}\ \bibnamefont
  {Wang}} \emph {et~al.},\ }\bibfield  {title} {\bibinfo {title} {{Searching
  for gravitational wave memory bursts with the Parkes Pulsar Timing Array}},\
  }\href {https://doi.org/10.1093/mnras/stu2137} {\bibfield  {journal}
  {\bibinfo  {journal} {Mon. Not. Roy. Astron. Soc.}\ }\textbf {\bibinfo
  {volume} {446}},\ \bibinfo {pages} {1657} (\bibinfo {year} {2015})},\ \Eprint
  {https://arxiv.org/abs/1410.3323} {arXiv:1410.3323 [astro-ph.GA]}
  \BibitemShut {NoStop}%
\bibitem [{\citenamefont {Lasky}\ \emph {et~al.}(2016)\citenamefont {Lasky},
  \citenamefont {Thrane}, \citenamefont {Levin}, \citenamefont {Blackman},\
  and\ \citenamefont {Chen}}]{Lasky:2016knh}%
  \BibitemOpen
  \bibfield  {author} {\bibinfo {author} {\bibfnamefont {P.~D.}\ \bibnamefont
  {Lasky}}, \bibinfo {author} {\bibfnamefont {E.}~\bibnamefont {Thrane}},
  \bibinfo {author} {\bibfnamefont {Y.}~\bibnamefont {Levin}}, \bibinfo
  {author} {\bibfnamefont {J.}~\bibnamefont {Blackman}},\ and\ \bibinfo
  {author} {\bibfnamefont {Y.}~\bibnamefont {Chen}},\ }\bibfield  {title}
  {\bibinfo {title} {{Detecting gravitational-wave memory with LIGO:
  implications of GW150914}},\ }\href
  {https://doi.org/10.1103/PhysRevLett.117.061102} {\bibfield  {journal}
  {\bibinfo  {journal} {Phys. Rev. Lett.}\ }\textbf {\bibinfo {volume} {117}},\
  \bibinfo {pages} {061102} (\bibinfo {year} {2016})},\ \Eprint
  {https://arxiv.org/abs/1605.01415} {arXiv:1605.01415 [astro-ph.HE]}
  \BibitemShut {NoStop}%
\bibitem [{\citenamefont {H{\"u}bner}\ \emph {et~al.}(2021)\citenamefont
  {H{\"u}bner}, \citenamefont {Lasky},\ and\ \citenamefont
  {Thrane}}]{Hubner:2021amk}%
  \BibitemOpen
  \bibfield  {author} {\bibinfo {author} {\bibfnamefont {M.}~\bibnamefont
  {H{\"u}bner}}, \bibinfo {author} {\bibfnamefont {P.}~\bibnamefont {Lasky}},\
  and\ \bibinfo {author} {\bibfnamefont {E.}~\bibnamefont {Thrane}},\
  }\bibfield  {title} {\bibinfo {title} {{Memory remains undetected: Updates
  from the second LIGO/Virgo gravitational-wave transient catalog}},\ }\href
  {https://doi.org/10.1103/PhysRevD.104.023004} {\bibfield  {journal} {\bibinfo
   {journal} {Phys. Rev. D}\ }\textbf {\bibinfo {volume} {104}},\ \bibinfo
  {pages} {023004} (\bibinfo {year} {2021})},\ \Eprint
  {https://arxiv.org/abs/2105.02879} {arXiv:2105.02879 [gr-qc]} \BibitemShut
  {NoStop}%
\bibitem [{\citenamefont {Cheung}\ \emph {et~al.}(2024)\citenamefont {Cheung},
  \citenamefont {Lasky},\ and\ \citenamefont {Thrane}}]{Cheung:2024zow}%
  \BibitemOpen
  \bibfield  {author} {\bibinfo {author} {\bibfnamefont {S.~Y.}\ \bibnamefont
  {Cheung}}, \bibinfo {author} {\bibfnamefont {P.~D.}\ \bibnamefont {Lasky}},\
  and\ \bibinfo {author} {\bibfnamefont {E.}~\bibnamefont {Thrane}},\
  }\bibfield  {title} {\bibinfo {title} {{Does spacetime have memories?
  Searching for gravitational-wave memory in the third LIGO-Virgo-KAGRA
  gravitational-wave transient catalogue}},\ }\href
  {https://doi.org/10.1088/1361-6382/ad3ffe} {\bibfield  {journal} {\bibinfo
  {journal} {Class. Quant. Grav.}\ }\textbf {\bibinfo {volume} {41}},\ \bibinfo
  {pages} {115010} (\bibinfo {year} {2024})},\ \Eprint
  {https://arxiv.org/abs/2404.11919} {arXiv:2404.11919 [gr-qc]} \BibitemShut
  {NoStop}%
\bibitem [{\citenamefont {Arzoumanian}\ \emph {et~al.}(2015)\citenamefont
  {Arzoumanian} \emph {et~al.}}]{NANOGrav:2015xuc}%
  \BibitemOpen
  \bibfield  {author} {\bibinfo {author} {\bibfnamefont {Z.}~\bibnamefont
  {Arzoumanian}} \emph {et~al.} (\bibinfo {collaboration} {NANOGrav}),\
  }\bibfield  {title} {\bibinfo {title} {{NANOGrav Constraints on Gravitational
  Wave Bursts with Memory}},\ }\href
  {https://doi.org/10.1088/0004-637X/810/2/150} {\bibfield  {journal} {\bibinfo
   {journal} {Astrophys. J.}\ }\textbf {\bibinfo {volume} {810}},\ \bibinfo
  {pages} {150} (\bibinfo {year} {2015})},\ \Eprint
  {https://arxiv.org/abs/1501.05343} {arXiv:1501.05343 [astro-ph.GA]}
  \BibitemShut {NoStop}%
\bibitem [{\citenamefont {Aggarwal}\ \emph {et~al.}(2020)\citenamefont
  {Aggarwal} \emph {et~al.}}]{NANOGrav:2019vto}%
  \BibitemOpen
  \bibfield  {author} {\bibinfo {author} {\bibfnamefont {K.}~\bibnamefont
  {Aggarwal}} \emph {et~al.} (\bibinfo {collaboration} {NANOGrav}),\ }\bibfield
   {title} {\bibinfo {title} {{The NANOGrav 11 yr Data Set: Limits on
  Gravitational Wave Memory}},\ }\href
  {https://doi.org/10.3847/1538-4357/ab6083} {\bibfield  {journal} {\bibinfo
  {journal} {Astrophys. J.}\ }\textbf {\bibinfo {volume} {889}},\ \bibinfo
  {pages} {38} (\bibinfo {year} {2020})},\ \Eprint
  {https://arxiv.org/abs/1911.08488} {arXiv:1911.08488 [astro-ph.HE]}
  \BibitemShut {NoStop}%
\bibitem [{\citenamefont {Agazie}\ \emph {et~al.}(2024)\citenamefont {Agazie}
  \emph {et~al.}}]{NANOGrav:2023vfo}%
  \BibitemOpen
  \bibfield  {author} {\bibinfo {author} {\bibfnamefont {G.}~\bibnamefont
  {Agazie}} \emph {et~al.} (\bibinfo {collaboration} {NANOGrav}),\ }\bibfield
  {title} {\bibinfo {title} {{The NANOGrav 12.5 yr Data Set: Search for
  Gravitational Wave Memory}},\ }\href
  {https://doi.org/10.3847/1538-4357/ad0726} {\bibfield  {journal} {\bibinfo
  {journal} {Astrophys. J.}\ }\textbf {\bibinfo {volume} {963}},\ \bibinfo
  {pages} {61} (\bibinfo {year} {2024})},\ \Eprint
  {https://arxiv.org/abs/2307.13797} {arXiv:2307.13797 [gr-qc]} \BibitemShut
  {NoStop}%
\bibitem [{\citenamefont {Agazie}\ \emph {et~al.}(2025)\citenamefont {Agazie}
  \emph {et~al.}}]{Agazie:2025oug}%
  \BibitemOpen
  \bibfield  {author} {\bibinfo {author} {\bibfnamefont {G.}~\bibnamefont
  {Agazie}} \emph {et~al.},\ }\bibfield  {title} {\bibinfo {title} {{The
  NANOGrav 15 yr Data Set: Search for Gravitational-wave Memory}},\ }\href
  {https://doi.org/10.3847/1538-4357/add874} {\bibfield  {journal} {\bibinfo
  {journal} {Astrophys. J.}\ }\textbf {\bibinfo {volume} {987}},\ \bibinfo
  {pages} {5} (\bibinfo {year} {2025})},\ \Eprint
  {https://arxiv.org/abs/2502.18599} {arXiv:2502.18599 [gr-qc]} \BibitemShut
  {NoStop}%
\bibitem [{\citenamefont {Islo}\ \emph {et~al.}(2019)\citenamefont {Islo},
  \citenamefont {Simon}, \citenamefont {Burke-Spolaor},\ and\ \citenamefont
  {Siemens}}]{Islo:2019qht}%
  \BibitemOpen
  \bibfield  {author} {\bibinfo {author} {\bibfnamefont {K.}~\bibnamefont
  {Islo}}, \bibinfo {author} {\bibfnamefont {J.}~\bibnamefont {Simon}},
  \bibinfo {author} {\bibfnamefont {S.}~\bibnamefont {Burke-Spolaor}},\ and\
  \bibinfo {author} {\bibfnamefont {X.}~\bibnamefont {Siemens}},\ }\bibfield
  {title} {\bibinfo {title} {{Prospects for Memory Detection with Low-Frequency
  Gravitational Wave Detectors}},\ }\href@noop {} {\  (\bibinfo {year}
  {2019})},\ \Eprint {https://arxiv.org/abs/1906.11936} {arXiv:1906.11936
  [astro-ph.HE]} \BibitemShut {NoStop}%
\bibitem [{\citenamefont {Johnson}\ \emph {et~al.}(2019)\citenamefont
  {Johnson}, \citenamefont {Kapadia}, \citenamefont {Osborne}, \citenamefont
  {Hixon},\ and\ \citenamefont {Kennefick}}]{Johnson:2018xly}%
  \BibitemOpen
  \bibfield  {author} {\bibinfo {author} {\bibfnamefont {A.~D.}\ \bibnamefont
  {Johnson}}, \bibinfo {author} {\bibfnamefont {S.~J.}\ \bibnamefont
  {Kapadia}}, \bibinfo {author} {\bibfnamefont {A.}~\bibnamefont {Osborne}},
  \bibinfo {author} {\bibfnamefont {A.}~\bibnamefont {Hixon}},\ and\ \bibinfo
  {author} {\bibfnamefont {D.}~\bibnamefont {Kennefick}},\ }\bibfield  {title}
  {\bibinfo {title} {{Prospects of detecting the nonlinear gravitational wave
  memory}},\ }\href {https://doi.org/10.1103/PhysRevD.99.044045} {\bibfield
  {journal} {\bibinfo  {journal} {Phys. Rev. D}\ }\textbf {\bibinfo {volume}
  {99}},\ \bibinfo {pages} {044045} (\bibinfo {year} {2019})},\ \Eprint
  {https://arxiv.org/abs/1810.09563} {arXiv:1810.09563 [gr-qc]} \BibitemShut
  {NoStop}%
\bibitem [{\citenamefont {Grant}\ and\ \citenamefont
  {Nichols}(2023)}]{Grant:2022bla}%
  \BibitemOpen
  \bibfield  {author} {\bibinfo {author} {\bibfnamefont {A.~M.}\ \bibnamefont
  {Grant}}\ and\ \bibinfo {author} {\bibfnamefont {D.~A.}\ \bibnamefont
  {Nichols}},\ }\bibfield  {title} {\bibinfo {title} {{Outlook for detecting
  the gravitational-wave displacement and spin memory effects with current and
  future gravitational-wave detectors}},\ }\href
  {https://doi.org/10.1103/PhysRevD.107.064056} {\bibfield  {journal} {\bibinfo
   {journal} {Phys. Rev. D}\ }\textbf {\bibinfo {volume} {107}},\ \bibinfo
  {pages} {064056} (\bibinfo {year} {2023})},\ \bibinfo {note} {[Erratum:
  Phys.Rev.D 108, 029901 (2023)]},\ \Eprint {https://arxiv.org/abs/2210.16266}
  {arXiv:2210.16266 [gr-qc]} \BibitemShut {NoStop}%
\bibitem [{\citenamefont {Inchausp{\'e}}\ \emph {et~al.}(2025)\citenamefont
  {Inchausp{\'e}}, \citenamefont {Gasparotto}, \citenamefont {Blas},
  \citenamefont {Heisenberg}, \citenamefont {Zosso},\ and\ \citenamefont
  {Tiwari}}]{Inchauspe:2024ibs}%
  \BibitemOpen
  \bibfield  {author} {\bibinfo {author} {\bibfnamefont {H.}~\bibnamefont
  {Inchausp{\'e}}}, \bibinfo {author} {\bibfnamefont {S.}~\bibnamefont
  {Gasparotto}}, \bibinfo {author} {\bibfnamefont {D.}~\bibnamefont {Blas}},
  \bibinfo {author} {\bibfnamefont {L.}~\bibnamefont {Heisenberg}}, \bibinfo
  {author} {\bibfnamefont {J.}~\bibnamefont {Zosso}},\ and\ \bibinfo {author}
  {\bibfnamefont {S.}~\bibnamefont {Tiwari}},\ }\bibfield  {title} {\bibinfo
  {title} {{Measuring gravitational wave memory with LISA}},\ }\href
  {https://doi.org/10.1103/PhysRevD.111.044044} {\bibfield  {journal} {\bibinfo
   {journal} {Phys. Rev. D}\ }\textbf {\bibinfo {volume} {111}},\ \bibinfo
  {pages} {044044} (\bibinfo {year} {2025})},\ \Eprint
  {https://arxiv.org/abs/2406.09228} {arXiv:2406.09228 [gr-qc]} \BibitemShut
  {NoStop}%
\bibitem [{\citenamefont {Favata}(2010)}]{Favata:2010zu}%
  \BibitemOpen
  \bibfield  {author} {\bibinfo {author} {\bibfnamefont {M.}~\bibnamefont
  {Favata}},\ }\bibfield  {title} {\bibinfo {title} {{The gravitational-wave
  memory effect}},\ }\href {https://doi.org/10.1088/0264-9381/27/8/084036}
  {\bibfield  {journal} {\bibinfo  {journal} {Class. Quant. Grav.}\ }\textbf
  {\bibinfo {volume} {27}},\ \bibinfo {pages} {084036} (\bibinfo {year}
  {2010})},\ \Eprint {https://arxiv.org/abs/1003.3486} {arXiv:1003.3486
  [gr-qc]} \BibitemShut {NoStop}%
\bibitem [{\citenamefont {Mitman}\ \emph {et~al.}(2024)\citenamefont {Mitman}
  \emph {et~al.}}]{Mitman:2024uss}%
  \BibitemOpen
  \bibfield  {author} {\bibinfo {author} {\bibfnamefont {K.}~\bibnamefont
  {Mitman}} \emph {et~al.},\ }\bibfield  {title} {\bibinfo {title} {{A review
  of gravitational memory and BMS frame fixing in numerical relativity}},\
  }\href {https://doi.org/10.1088/1361-6382/ad83c2} {\bibfield  {journal}
  {\bibinfo  {journal} {Class. Quant. Grav.}\ }\textbf {\bibinfo {volume}
  {41}},\ \bibinfo {pages} {223001} (\bibinfo {year} {2024})},\ \Eprint
  {https://arxiv.org/abs/2405.08868} {arXiv:2405.08868 [gr-qc]} \BibitemShut
  {NoStop}%
\bibitem [{\citenamefont {Aichelburg}\ and\ \citenamefont
  {Sexl}(1971)}]{Aichelburg:1970dh}%
  \BibitemOpen
  \bibfield  {author} {\bibinfo {author} {\bibfnamefont {P.~C.}\ \bibnamefont
  {Aichelburg}}\ and\ \bibinfo {author} {\bibfnamefont {R.~U.}\ \bibnamefont
  {Sexl}},\ }\bibfield  {title} {\bibinfo {title} {{On the Gravitational field
  of a massless particle}},\ }\href {https://doi.org/10.1007/BF00758149}
  {\bibfield  {journal} {\bibinfo  {journal} {Gen. Rel. Grav.}\ }\textbf
  {\bibinfo {volume} {2}},\ \bibinfo {pages} {303} (\bibinfo {year}
  {1971})}\BibitemShut {NoStop}%
\bibitem [{\citenamefont {Dray}\ and\ \citenamefont
  {'t~Hooft}(1985)}]{Dray:1984ha}%
  \BibitemOpen
  \bibfield  {author} {\bibinfo {author} {\bibfnamefont {T.}~\bibnamefont
  {Dray}}\ and\ \bibinfo {author} {\bibfnamefont {G.}~\bibnamefont
  {'t~Hooft}},\ }\bibfield  {title} {\bibinfo {title} {{The Gravitational Shock
  Wave of a Massless Particle}},\ }\href
  {https://doi.org/10.1016/0550-3213(85)90525-5} {\bibfield  {journal}
  {\bibinfo  {journal} {Nucl. Phys. B}\ }\textbf {\bibinfo {volume} {253}},\
  \bibinfo {pages} {173} (\bibinfo {year} {1985})}\BibitemShut {NoStop}%
\bibitem [{\citenamefont {Tolish}\ and\ \citenamefont
  {Wald}(2014)}]{Tolish:2014bka}%
  \BibitemOpen
  \bibfield  {author} {\bibinfo {author} {\bibfnamefont {A.}~\bibnamefont
  {Tolish}}\ and\ \bibinfo {author} {\bibfnamefont {R.~M.}\ \bibnamefont
  {Wald}},\ }\bibfield  {title} {\bibinfo {title} {{Retarded Fields of Null
  Particles and the Memory Effect}},\ }\href
  {https://doi.org/10.1103/PhysRevD.89.064008} {\bibfield  {journal} {\bibinfo
  {journal} {Phys. Rev. D}\ }\textbf {\bibinfo {volume} {89}},\ \bibinfo
  {pages} {064008} (\bibinfo {year} {2014})},\ \Eprint
  {https://arxiv.org/abs/1401.5831} {arXiv:1401.5831 [gr-qc]} \BibitemShut
  {NoStop}%
\bibitem [{\citenamefont {Zhang}\ \emph {et~al.}(2017)\citenamefont {Zhang},
  \citenamefont {Duval}, \citenamefont {Gibbons},\ and\ \citenamefont
  {Horvathy}}]{Zhang:2017rno}%
  \BibitemOpen
  \bibfield  {author} {\bibinfo {author} {\bibfnamefont {P.~M.}\ \bibnamefont
  {Zhang}}, \bibinfo {author} {\bibfnamefont {C.}~\bibnamefont {Duval}},
  \bibinfo {author} {\bibfnamefont {G.~W.}\ \bibnamefont {Gibbons}},\ and\
  \bibinfo {author} {\bibfnamefont {P.~A.}\ \bibnamefont {Horvathy}},\
  }\bibfield  {title} {\bibinfo {title} {{The Memory Effect for Plane
  Gravitational Waves}},\ }\href
  {https://doi.org/10.1016/j.physletb.2017.07.050} {\bibfield  {journal}
  {\bibinfo  {journal} {Phys. Lett. B}\ }\textbf {\bibinfo {volume} {772}},\
  \bibinfo {pages} {743} (\bibinfo {year} {2017})},\ \Eprint
  {https://arxiv.org/abs/1704.05997} {arXiv:1704.05997 [gr-qc]} \BibitemShut
  {NoStop}%
\bibitem [{\citenamefont {Zhang}\ \emph {et~al.}(2018)\citenamefont {Zhang},
  \citenamefont {Duval},\ and\ \citenamefont {Horvathy}}]{Zhang:2017jma}%
  \BibitemOpen
  \bibfield  {author} {\bibinfo {author} {\bibfnamefont {P.~M.}\ \bibnamefont
  {Zhang}}, \bibinfo {author} {\bibfnamefont {C.}~\bibnamefont {Duval}},\ and\
  \bibinfo {author} {\bibfnamefont {P.~A.}\ \bibnamefont {Horvathy}},\
  }\bibfield  {title} {\bibinfo {title} {{Memory effect for impulsive
  gravitational waves}},\ }\href {https://doi.org/10.1088/1361-6382/aaa987}
  {\bibfield  {journal} {\bibinfo  {journal} {Class. Quant. Grav.}\ }\textbf
  {\bibinfo {volume} {35}},\ \bibinfo {pages} {065011} (\bibinfo {year}
  {2018})},\ \Eprint {https://arxiv.org/abs/1709.02299} {arXiv:1709.02299
  [gr-qc]} \BibitemShut {NoStop}%
\bibitem [{\citenamefont {van Haasteren}\ and\ \citenamefont
  {Levin}(2010)}]{vanHaasteren:2009fy}%
  \BibitemOpen
  \bibfield  {author} {\bibinfo {author} {\bibfnamefont {R.}~\bibnamefont {van
  Haasteren}}\ and\ \bibinfo {author} {\bibfnamefont {Y.}~\bibnamefont
  {Levin}},\ }\bibfield  {title} {\bibinfo {title} {{Gravitational-wave memory
  and pulsar timing arrays}},\ }\href
  {https://doi.org/10.1111/j.1365-2966.2009.15885.x} {\bibfield  {journal}
  {\bibinfo  {journal} {Mon. Not. Roy. Astron. Soc.}\ }\textbf {\bibinfo
  {volume} {401}},\ \bibinfo {pages} {2372} (\bibinfo {year} {2010})},\ \Eprint
  {https://arxiv.org/abs/0909.0954} {arXiv:0909.0954 [astro-ph.IM]}
  \BibitemShut {NoStop}%
\end{thebibliography}%

\end{document}